\documentclass[a4paper,fleqn,usenatbib,useAMS]{mnras}

\usepackage{graphicx}	
\usepackage{amsmath}	
\usepackage{amssymb}	
\usepackage{multicol}        
\usepackage{bm}	
\usepackage{pdflscape}	
\usepackage{subfigure}
\usepackage{longtable}

\usepackage[T1]{fontenc}
\usepackage{ae,aecompl}

\title[Jet Propagation in Neutron Star Mergers]{Jet Propagation in Neutron Star Mergers and GW170817}

\author[Hamidani, Kiuchi, \& Ioka]{Hamid Hamidani$^1$\thanks{E-mail: hamidani.hamid@yukawa.kyoto-u.ac.jp}, Kenta Kiuchi$^2$, and Kunihito Ioka$^1$
\\
$^{1}$ Yukawa Institute for Theoretical Physics, Kyoto University, Kyoto 606-8502, Japan\\
$^{2}$ Max Planck Institute for Gravitational Physics (Albert Einstein Institute), Am Muehlenberg, Potsdam-Golm, D-14476, Germany}

\date{Last updated 2019 November 13}

\pubyear{2019}

\begin{document}
\label{firstpage}
\pagerange{\pageref{firstpage}--\pageref{lastpage}}
\maketitle

\begin{abstract}
The gravitational wave event from the binary neutron star (BNS) merger GW170817 and the following multi-messenger observations present strong evidence for i) merger ejecta expanding with substantial velocities and ii) a relativistic jet which had to propagate through the merger ejecta. The ejecta's expansion velocity is not negligible for the jet head motion, which is a fundamental difference from the other systems like collapsars and active galactic nuclei. Here we present an analytic model of the jet propagation in an expanding medium. In particular, we notice a new term in the expression of the breakout time and velocity. In parallel, we perform a series of over a hundred 2D numerical simulations of jet propagation. The BNS merger ejecta is prepared based on numerical relativity simulations of a BNS merger with the highest-resolution to date. We show that our analytic results agree with numerical simulations over a wide parameter space. Then we apply our analytic model to GW170817, and obtain two solid constraints on: i) the central engine luminosity as $L_{iso,0} \sim 3\times10^{49}-2.5\times10^{52}$ erg s$^{-1}$, and on ii) the delay time between the merger and engine activation $t_0-t_m < 1.3$ s. The engine power implies that the apparently-faint \textit{short} gamma-ray burst (\textit{s}GRB) \textit{s}GRB 170817A is similar to typical \textit{s}GRBs if observed on-axis.

\end{abstract}



\section{Introduction}
\begin{figure}
 \includegraphics[width=0.99\linewidth]{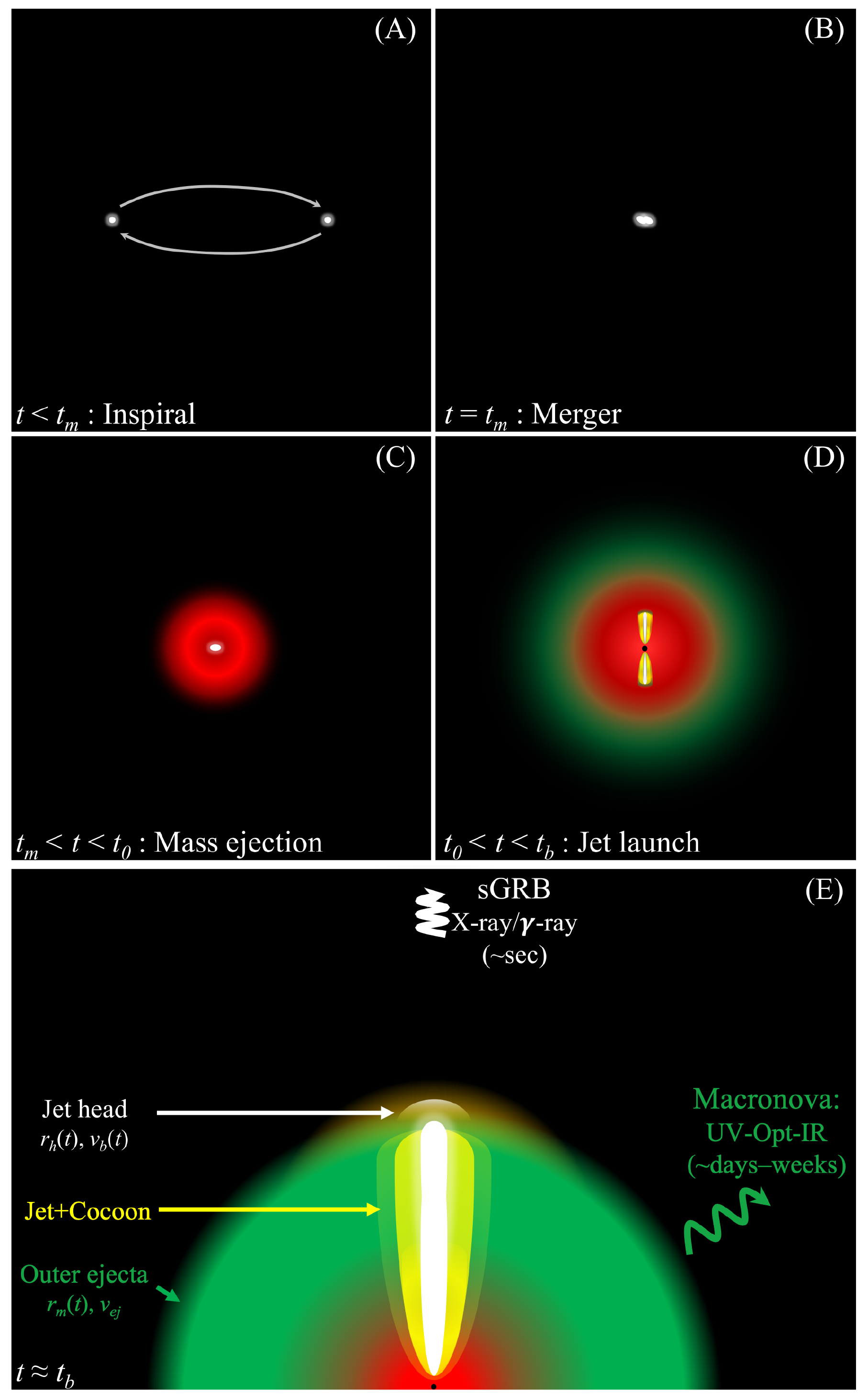}
 \caption{A schematic description of the main phases in a BNS merger event. (A) A BNS system before the merger. (B) The merger taking place at $t=t_m$. (C) Just after the merger, the by-product of the merger is surrounded by expanding ejecta of $M_{ej}\sim0.01 M_\odot$. (D) The expansion reduces the ejecta's density, and at $t=t_0$, the engine is activated and polar jets are launched. (E) The jet breaks out at $t=t_b$, as the jet head catches up with the ejecta's outer radius: $r_h(t=t_b)=r_m(t=t_b)$.}
 \label{fig:P4a-F00.pdf}
\end{figure}

Gravitational wave (GW) observation of GW170817 by the Laser Interferometer Gravitational-Wave Observatory (LIGO) and the Virgo Consortium (LVC), and the follow-up observations across the electromagnetic (EM) spectrum were historical discoveries. 
The binary neutron star (BNS) merger event GW170817 was associated with the short gamma-ray burst \textit{s}GRB 170817A (\citealt{2017PhRvL.119p1101A}; \citealt{2017ApJ...848L..13A}), which gave the first direct observational clue to the scenario of BNS mergers for \textit{s}GRBs (\citealt{1986ApJ...308L..43P}; \citealt{1986ApJ...308L..47G}; \citealt{1989Natur.340..126E}). 
In this scenario, the merger produces a central engine surrounded by an accretion disk and ejecta of $\sim 10^{-2}-10^{-3}M_\odot$ (\citealt{1999PhRvD..60j4052S}; \citealt{2000PhRvD..61f4001S}).
Accretion of matter fuels the central engine to power a relativistic jet. 
As illustrated in figure \ref{fig:P4a-F00.pdf}, eventually, high-energy photons can be released as the prompt emission, once the jet breaks out of the ejecta (\citealt{2014ApJ...784L..28N}; \citealt{2014ApJ...788L...8M}).

The observation of GW170817/\textit{s}GRB 170817A (hereafter referred to as GW170817) was a major turning point for the multi-messenger astrophysics.
As one of the best \textit{s}GRB events observed ever, GW170817 is rich in new data; such as, the mass of the BNS ($\sim 2.7 M_\odot$; \citealt{2017PhRvL.119p1101A}), the viewing angle ($\sim20^\circ-30^\circ$; \citealt{2017PhRvL.119p1101A}; \citealt{2018arXiv180806617T}), and the delay between the GW and the EM signal ($\sim 1.7$s; \citealt{2017PhRvL.119p1101A}; \citealt{2017ApJ...848L..13A}), all inferred for the first time.
Superluminal motion in late radio afterglow observations \citep{2018Natur.561..355M}, and macronova/kilonova (hereafter macronova) emission with a strong support for r-process nucleosynthesis (\citealt{Arcavi:2017vbi}; \citealt{Chornock:2017sdf}; \citealt{Coulter:2017wya}; \citealt{2017ApJ...848L..29D}; \citealt{2017Sci...358.1570D}; \citealt{Kilpatrick:2017mhz}; \citealt{2017Sci...358.1559K}; \citealt{Nicholl:2017ahq}; \citealt{Pian:2017gtc}; \citealt{Smartt:2017fuw}; \citealt{Shappee:2017zly}; \citealt{Soares-Santos:2017lru}; \citealt{2017PASJ...69..102T}; \citealt{Utsumi:2017cti}; \citealt{Valenti:2017ngx}) are also new revelations. 
This event also had impacts on the equation of state of neutron stars, relativity, cosmology, etc. (e.g. \citealt{2017Natur.551...85A}).


Although GW170817 did answer several fundamental questions related to \textit{s}GRBs, it also did prompt new questions. Because \textit{s}GRB 170817A is several orders-of-magnitude fainter than ordinary \textit{s}GRBs \citep{2017ApJ...848L..13A}, the jet associated with GW170817 is not fully understood yet.
This peculiar faintness has been interpreted as most likely due to the large off-axis viewing angle (see: \citealt{2017ApJ...848L..13A}; \citealt{2017Sci...358.1559K}; \citealt{2017ApJ...850L..24G}; \citealt{2018PTEP.2018d3E02I}).
With radio afterglow observations showing evidence for superluminal motion, and the light curve behavior after the peak, we are forced to conclude that a relativistic jet did exist and the faintness is due to the off-axis viewing angle.
The peculiar spectral properties of \textit{s}GRB 170817A (\citealt{2019MNRAS.486.1563M}; \citealt{2019MNRAS.483.1247M}) are also interpreted within the framework of the off-axis jet model by considering the jet structure \citep{2019MNRAS.487.4884I}. 
Still there remains an open question whether the emission comes from the jet or not (for example the emission could come from the cocoon breakout from the ejecta; \citealt{2017Sci...358.1559K}; \citealt{2018MNRAS.479..588G}) mainly because the jet properties are not fully determined yet.

A breakout has to take place for the high-energy photons of the prompt emission to be released.
Jet propagation is determined by the properties of the ejecta, and also by the parameters of the engine (e.g., its luminosity).
Hence, the jet propagation is a key process in order to tackle the open questions concerning GW170817.  
Numerical relativity simulations, with the help of GW observations (i.e. mass measurement; \citealt{2017PhRvL.119p1101A}), provide insights into the properties of the dynamical ejecta and the post-merger wind (\citealt{2013PhRvD..87b4001H}; \citealt{2013MNRAS.435..502F}; \citealt{2013ApJ...773...78B}; \citealt{2015MNRAS.448..541J}; \citealt{2018ApJ...860...64F}; \citealt{2019MNRAS.482.3373F}; also see \citealt{2019ARNPS..6901918S} for a review). 
Observations of the macronova emission in combination with theoretical works provide additional information; e.g., limits on the ejecta mass and velocity (\citealt{2017PASJ...69..102T}; \citealt{2017Sci...358.1570D}; \citealt{2017Sci...358.1559K}; \citealt{2017ApJ...851L..21V}; \citealt{2018MNRAS.481.3423W}; \citealt{2018ApJ...865L..21K}; \citealt{2019arXiv190805815K}; etc.).
Therefore, with these properties of the ejecta, we can infer the jet and central engine of GW170817 by studying the jet propagation in the merger ejecta.

Jet propagation has been intensively studied in the context of active galactic nuclei (AGNs) and collapsars (\citealt{1989ApJ...345L..21B}; \citealt{1997ApJ...479..151M}; \citealt{1999ApJ...524..262M}; \citealt{2003MNRAS.345..575M}; \citealt{2013ApJ...777..162M}).
In both AGNs and collapsars, the medium is static.
Jet propagation in a static medium is well-understood thanks to several analytic and numerical studies (\citealt{2003MNRAS.345..575M}; \citealt{2011ApJ...740..100B}; \citealt{2013ApJ...777..162M}; \citealt{ 2018MNRAS.477.2128H}).

Jet propagation in the ejecta of a BNS merger is different in many aspects, in particular the outward expansion of the merger ejecta with substantial velocities ($\sim 0.2c$; \citealt{2013PhRvD..87b4001H}; \citealt{2013ApJ...773...78B}; \citealt{2015MNRAS.448..541J}).  
There are two important consequences of the outward expansion in the merger ejecta. 
First, the jet head must reach a higher velocity than that of the outer ejecta in order to catch it up and ensure the breakout.
Second, as the outer radius of the ejecta continuously expands with time, the ejecta's volume increases with time. 
As a result, unlike the collapsar case, the density (and the pressure) is time dependent.
This profoundly affects jet propagation. 

Until very recently, the jet propagation in the merger ejecta was not studied as much as that in the collapsar case. However, after GRB 130603B with the first indications of a macronova \citep{2013ApJ...774L..23B} and later GW170817 \citep{2017PhRvL.119p1101A}, the interests grew. So far, most studies have been based on numerical hydrodynamical simulations (\citealt{2014ApJ...784L..28N}; \citealt{2015ApJ...813...64D}; \citealt{2018ApJ...866....3D}; \citealt{2017ApJ...848L...6L}; \citealt{2018MNRAS.473..576G}; \citealt{2018MNRAS.479..588G}; \citealt{2018ApJ...863...58X}). 
While numerical simulations give important insights on the phenomenology of jet propagation, it is very hard to cover a wide parameter-space with simulations alone.
Hence, analytic modeling is indispensable, especially as several key parameters remain beyond reach (e.g., parameters of the central engine).
Analytic modeling of the jet propagation in a BNS merger has been presented using ideas from the collapsar case (\citealt{2017ApJ...835L..34M}; \citealt{2018PTEP.2018d3E02I}; \citealt{2018ApJ...866L..16M}; \citealt{2019ApJ...876..139G}; \citealt{2019ApJ...877L..40G}; \citealt{2019arXiv190408425L}).
These studies tried to make constraints on the equation of state of neutron stars \citep{2019arXiv190408425L}, on the properties of the ejecta (or the wind) (\citealt{2017ApJ...835L..34M}; \citealt{2019arXiv190408425L}), and on the timescale between the merger and BH formation (\citealt{2017ApJ...835L..34M}; \citealt{2018ApJ...866L..16M}; \citealt{2019ApJ...876..139G}; \citealt{2019ApJ...877L..40G}).
However, previous analytic models of the jet propagation did not properly take the expansion of the medium into account; also, there has been no evidence of their consistency with numerical simulations.

Here we present a work that combines high-resolution hydrodynamical simulations of jet propagation in the ejecta of BNS mergers with proper analytic modeling for the jet propagation.
We use refined numerical relativity simulations of BNS merger by \citet{2017PhRvD..96h4060K} to understand the early BNS merger ejecta down to the vicinity of the central engine.
We analytically derive the jet head motion, taking the expansion of the ejecta correctly into account. 
In parallel, we carry out a series of about $100$ numerical simulations using a 2D hydrodynamic relativistic code, in order to compare analytical and numerical results over a wide parameter space.
Our motivations are to: i) understand the jet propagation in the BNS merger ejecta and the difference with the jet propagation in the collapsar case; ii) understand more about the properties of the central engine in the event GW170817/\textit{s}GRB 170817A; and iii) understand the nature of \textit{s}GRB 170817A relative to typical \textit{s}GRBs.

This paper is organized as follows. 
In $\S$ \ref{sec:The Analytical Model}, the analytical model for jet propagation is presented, and two cases are presented: the collapsar case (static medium) and the BNS merger case (expanding medium). 
Analytical results and numerical simulations' results are presented, compared and discussed in $\S$ \ref{sec:Comparison with Numerical Simulations}. 
In $\S$ \ref{sec:4.NR} we show and discuss numerical relativity simulations' results about BNS merger ejecta. 
In $\S$ \ref{sec:GW170817}, we apply our analytical model to GW170817/\textit{s}GRB 170817A, and present several constraints on its central engine. 
And in $\S$ \ref{sec:GRB 170817A} we discuss the implications of our findings on the nature of \textit{s}GRB 170817A in comparison to typical \textit{s}GRBs. 
A conclusion is given at the end of this paper ($\S$ \ref{sec:Conclusion}).
We present the analytic calculations in Appendix \ref{app:A} and \ref{app:B}, for the static medium case and the expanding medium case, respectively.
And in Appendix \ref{sec:app C}, we present the analytical model for jet propagation taking the evolution of the jet opening angle into account.

\section{The Analytical Model}
\label{sec:The Analytical Model}
\subsection{Jump conditions and the equation of motion for the jet head}

Let's consider a jet launched through an ambient medium with a total mass $M_{ej}$. We consider two cases. First, a static medium case, as in the collapsar case, where the medium is the stellar envelope of a dying massive star. Second, an expanding medium case, as in the case of a BNS merger, where the medium is the dynamical ejecta (refer to $\S$ \ref{sec:4.NR} for a full explanation of the dynamical ejecta). 

In both cases, a jet head and a cocoon are formed. The jet head is composed of the shocked jet and the shocked ambient medium, both of which are pressure balanced. The jet head is continuously pushed forward. Therefore, the pressure in the jet head is very high, which results in matter expanding sideways to form the cocoon surrounding the jet.

The jet is assumed to be launched at the vicinity of the central engine $\sim 10^6-10^7$ cm (depending on the engine and its mechanism). The jet head position at a given time, $t$, is $r_h(t)$. The velocity of the jet head in the lab frame (i.e., the central engine frame) is $c\beta_h=dr_h(t)/dt$. This velocity is determined by the ram pressure balance (\citealt{1989ApJ...345L..21B}; \citealt{1997ApJ...479..151M}; \citealt{2003MNRAS.345..575M}; \citealt{2011ApJ...740..100B}; \citealt{2013ApJ...777..162M}):
\begin{eqnarray}
h_j \rho_j c^2 (\Gamma\beta)_{jh}^2 + P_j = h_a \rho_a c^2 (\Gamma\beta)_{ha}^2 + P_a, 
\end{eqnarray}
where $h$, $\rho$, and $P$ are enthalpy, density, and pressure of each fluid element, all measured in the fluid's rest frame. The subscripts $j$, $h$, and $a$ refer to the three domains: the jet, jet head, and ambient medium, respectively. $(\Gamma\beta)_{jh}$ is the four-velocity of the jet relative to the jet head, and $(\Gamma\beta)_{ha}$ is the four-velocity of the jet head relative to the ambient medium. As $P_a$ and $P_j$ are cold, they can be neglected. Hence, the jet head velocity can be written as:
\begin{eqnarray}
\beta_h  =  \frac{\beta_j - \beta_a}{1 + \tilde{L}^{-1/2}} +  \beta_a,
\label{eq:beta_h_1}
\end{eqnarray}
where $\tilde{L}$ is the ratio of the energy density between the jet and the ambient medium \citep{2003MNRAS.345..575M,2011ApJ...740..100B,2018PTEP.2018d3E02I,2018MNRAS.477.2128H}:
\begin{eqnarray}
\tilde{L}  =  \frac{h_j \rho_j \Gamma_j^2}{h_a \rho_a \Gamma_a^2} \simeq \frac{L_j}{\Sigma_j \rho_a c^3}.
\label{eq:L expression}
\end{eqnarray}
$\Sigma_j$ is the cross section of the jet.  With $\theta_j$ as the jet opening angle, $\Sigma_j=\pi\theta_j^2 r_h^2(t)$. 
$L_j$ is the jet luminosity (one sided). 
We can usually take  $\Gamma_a\simeq 1$ even for the case of BNS merger ejecta.

\subsubsection{Approximations}
\label{sec:model assumptions}
\paragraph{A roughly constant opening angle $\theta_j$}
\label{sec:A roughly constant cross section}
A close look at the jet opening angle $\theta_j(t)$ shows that it does not vary significantly throughout most of the jet head propagation. At first, let's approximate it as constant $\theta_j(t)\equiv\theta_j$ (see Appendix \ref{sec:app C} for the case of a time-dependent opening angle). That is, the jet opening angle can be written as:
\begin{eqnarray}
    \theta_j \approx \theta_0/f_j,    
    \label{eq:f_j}
\end{eqnarray}
where $\theta_0$ is the initial opening angle\footnote{The initial opening angle, accounting for relativistic spreading of the jet, is given by $\theta_0 \approx \theta_{inj} + 1/\Gamma_0$, where $\theta_{inj}$ and $\Gamma_0$ are the opening angle and the Lorentz factor of the injected outflow. In our simulations, $\theta_{inj}$ is typically set as $\theta_{inj}\approx1/\Gamma_0$ (see $\S$ \ref{sec:Comparison with Numerical Simulations}).}, and $f_j$ (>1) is a constant that accounts for the average degree of collimation. The slow evolution of the opening angle is based on results from the analytic calculations [see Appendix \ref{sec:app C} and equation (\ref{eq:theta_j/theta_0 app})] and confirmed by results from numerical simulations
(see figure \ref{fig:P4a-rh2}, and figure 3 in \citealt{2014ApJ...784L..28N}). This approximation is useful as a first step because it simplifies the complexity of the analytic formula, but still keeps reasonable accuracy before going to the complex equations for the evolution of the jet collimation in Appendix \ref{sec:app C}.

\paragraph{The collimation factor $f_j =\theta_0/\theta_j$}
\label{sec:f_j}
The numerical results within our explored parameter space show that the collimation factor $f_j$ takes values in the range $\sim 2-10$ (depending on the engine luminosity, the initial opening angle, and the ejecta mass), with $f_j\approx 5$ being a typical value for the case of a non-relativistic jet head (i.e. $\tilde{L}\ll (1-\beta_a)^2$; see figure \ref{fig:P4a-rh2}). Hence, within the current parameter space, we take $f_j=5$ as a fiducial case in our analytic modeling. We stress that the approximation $f_j\approx 5$ is valid only within our current parameter space for the typical BNS merger case, and should not be understood as a universally valid value. 

\paragraph{A homologous expansion}
The ejecta is approximated to follow a homologous expansion. That is, the profile of velocity satisfies $v_a(r)\propto r$. In the case of BNS dynamical ejecta, numerical relativity simulations confirm this approximation (see figure \ref{fig:H_HB_135_135}).   

\paragraph{Definition of the ejecta}
The term ``ejecta'' commonly refers to the part of material ejected from the system of the central engine. Here we use this term to refer to the ejected part (gravitationally unbound) in addition to the inner part (gravitationally bound). The reason is that the jet is launched at the vicinity of the central engine $10^6-10^7$ cm, which is gravitationally bound to the central engine. Hence, the density in the inner region (although gravitationally bound) is also relevant for jet propagation. Therefore, we define the mass of the ejecta as $M_{ej}=\int_{r_0}^{r_{m,0}}4\pi r^2 \rho_a(r) dr$; where $r_0\sim10^6-10^7$ cm, $r_{m,0}$ is the outer radius of the ejecta when the jet is launched, and $\rho_a(r)$ is the density of the ambient medium in the polar region when the jet is launched\footnote{Note that the $\rho_a(r)$ is the density in the polar region, and the ejecta mass is not isotropically distributed (for more information see $\S$ \ref{sec:4.NR}). The total mass of the ejecta is larger than $M_{ej}$, but, as far as the jet propagation is concerned, $M_{ej}$ is the relevant mass.}. Note that, as the volume of the inner part is very small in comparison to the ejecta's total volume, and as the density profile in this inner part can be approximated to a power-law with an index $n\approx 2$ (see figure \ref{fig:H_HB_135_135} and the discussion in $\S$ \ref{sec:NSs}), the inclusion of this inner part will not significantly affect the total mass of the ejecta (contributes with a change of only $\sim 10\%$).

\paragraph{A power-law density profile}
The ejecta's density profile, as found in numerical relativity simulations, can be approximated by a combination of power-law functions; where the power-law index in the inner region of the ejecta is different (smaller) from the index in the outer region of the ejecta (see $\S$ \ref{sec:4.NR} and figure \ref{fig:H_HB_135_135}). 
Here, this is simplified by approximating the entire ejecta's density profile to one power-law function with one index $n$. 
We adopt the power-law index of the inner region, $n=2$, for the whole ejecta, although in reality the outer part of the ejecta (excluding the fast tail) shows a higher index $n\sim 3 -3.5$ (see $\S$ \ref{sec:4.NR} and figure \ref{fig:H_HB_135_135}).
The validity of this approximation is discussed in $\S$ \ref{sec:Density profile of the ejecta}.
Such an approximation is also reasonable for the stellar envelope in the collapsar case; although the index $n$ differs (based on the stellar calculations by \citealt{2006ApJ...637..914W}; see \citealt{2013ApJ...777..162M} figure 2 for an illustration of $n$).

\paragraph{A negligible ambient velocity}
Since $\beta_j \gg \beta_a$, we approximately take that $\beta_j - \beta_a \simeq \beta_j \simeq 1$. This is a very good approximation for slowly expanding ejecta. Also, even for fast ejecta ($\sim0.4c$), this is a good approximation during most of the jet propagation time, until the jet head starts interacting with the outer region where $\beta_a$ is substantial, because the velocity structure is homologous. Hence, equation (\ref{eq:beta_h_1}) is simplified to the following form:
\begin{eqnarray}
\beta_h  \simeq \frac{1}{1 + \tilde{L}^{-1/2}} + \beta_a .
\label{eq:beta_h_15}
\end{eqnarray}
A particular case where this approximation was avoided is presented in Appendix \ref{app:exact solution n=2} for reference.

\paragraph{A non-relativistic jet head $\tilde{L}\ll (1-\beta_a)^2$}
This study is limited to the case of a non-relativistic jet head. This is guaranteed by the following requirement $\tilde{L}\ll (1-\beta_a)^2$; this condition allows equation (\ref{eq:beta_h_15}) to be simplified as:
\begin{eqnarray}
\beta_h  \simeq \tilde{L}^{1/2} + \beta_a.
\label{eq:beta_h_2}
\end{eqnarray}
The above condition also ensures that $\beta_h$ is less than unity. This condition can be understood as analogous to $\tilde{L}\ll 1$ in the collapsar case ($\beta_a=0$) \citep{2003MNRAS.345..575M,2011ApJ...740..100B}.

\paragraph{A calibration coefficient for the analytical $\tilde{L}$}
\label{sec:calibration}
Comparison of numerical simulations (of collapsar jets) shows that, in reality, the analytical modeling does not capture all the physics of fluid dynamics and the jet head propagation (e.g., oblique motion).
As a result, as firstly found in \citet{2013ApJ...777..162M} and examined in detail in \citet{2018MNRAS.477.2128H}, the analytical modeling overestimates the jet head radius (effectively $\tilde{L}$) in comparison with simulations.
In \citet{2018MNRAS.477.2128H}, the intensive comparison with simulations shows that, in the non-relativistic domain, the analytical equations give $\sim 2.5 - 3$ times faster jet head velocity (see figure 12 in \citealt{2018MNRAS.477.2128H}).
That is, for more accurate results, $\tilde{L}^{1/2}$ has to be corrected by a calibration coefficient $N_s \sim \frac{1}{2.5}$ to $\frac{1}{3}$.
So far, there has been no estimation of this calibration coefficient for the case of an expanding medium.
\citet{2018ApJ...866L..16M,2019arXiv190707599S} used the same calibration coefficient for the case of an expanding medium as an assumption, but offered no evidence.
Here, after intense comparison with numerical simulations (in $\S$ \ref{sec:Comparison with Numerical Simulations}), we found that \citet{2013ApJ...777..162M} and \citet{2018MNRAS.477.2128H} findings can be generalized for the case of an expanding medium; for the case of non-relativistic jet head propagation in an expanding medium, we show for the first time that this calibration coefficient is i) necessary and that ii) it takes roughly the same range of values as it does in the collapsar case.
Hence, hereafter we adopt a constant calibration coefficient $N_s$ for our analytical $\tilde{L}^{1/2}$ as:
\begin{eqnarray}
N_s\approx 2/5.
\label{eq:N_s}
\end{eqnarray}

\subsection{Case I: Static medium}
\label{subsubsec:Case I: Static medium}
\subsubsection{Equation of motion}
In this case I, the medium is assumed to remain static during the timescale of jet propagation (i.e., $\beta_a=0$), as in the collapsar case. Hence, the density profile of the ambient medium (assumed as a power-law) is not time dependent. The motion of the jet head derived from equation (\ref{eq:beta_h_2}) is given by the following first-order differential equation (for a detailed calculation, refer to Appendix \ref{app subsec:Static ambient medium}):

\begin{eqnarray}
\frac{dr_h(t)}{dt}  =  A\:r_h(t)^\frac{n-2}{2} .
\label{eq:diff static case}
\end{eqnarray}
With the jet opening angle $\theta_j$ approximated as constant (see $\S$ \ref{sec:A roughly constant cross section}), the solution can be written with:

\begin{align}
r_h(t)  =  \left[\left(\frac{4-n}{2}\right)A\:(t-t_0) +  r_0^\frac{4-n}{2}\right]^{\frac{2}{4-n}},
\label{eq:r stat}
\end{align}

\begin{align}
v_h(t)  =  A\:\left[ \left(\frac{4-n}{2}\right)\:A\:(t-t_0)    + r_0^\frac{4-n}{2}            \right]^\frac{n-2}{4-n},
\label{eq:v stat}
\end{align}
where $t_0$ is the time at which the engine activates, and $A$ is a constant that can be written as:
\begin{equation*}
A=\sqrt{ \left(\frac{r_m^{3-n}-r_0^{3-n}}{3-n}\right)\left(\frac{4\:L_j}{\theta_j^2\:M_{ej}\:c}\right)    },
\end{equation*}
with $\tilde{L}^{1/2} \propto A$.
Since $\tilde{L}^{1/2}$ needs to be corrected using the calibration coefficient $N_s$ [see $\S$ \ref{sec:calibration} and equation (\ref{eq:N_s})], the above $A$ is corrected to $A_c$ as:
\begin{equation}
A \to A_c=N_s\sqrt{ \left(\frac{r_m^{3-n}-r_0^{3-n}}{3-n}\right)\left(\frac{4\:L_j}{\theta_j^2\:M_{ej}\:c}\right)    }.
\label{eq:A_c static}
\end{equation}
The opening angle of the jet, $\theta_j=\theta_0/f_j$, is unknown here. With numerical simulations $f_j$ can be determined. For typical collapsar parameters, we get $f_j\approx 10$ [see equation (\ref{eq:tb stat sim 2}) and the discussion that follows]. For a small opening angle $\theta_0$, we can write the isotropic equivalent luminosity at the base of the jet as:
\begin{eqnarray}
L_{iso,0}\simeq4L_j/\theta_0^2   . 
\end{eqnarray}
We later refer to this luminosity as the isotropic equivalent luminosity of the central engine. Finally, with the jet launch taking place at the vicinity of the central engine $r_0\approx10^6-10^7$ cm\footnote{Note that as long as $r_0\ll r_{m}$, the exact value of $r_0$ is not relevant to the analytic results or the breakout time.}, the jet head motion is determined by the following macroscopic parameters: 
\begin{enumerate}
    \item The engine power: $L_j$ and the jet initial opening angle $\theta_0$; or the jet isotropic luminosity $L_{iso,0}$.
    \item The ejecta: its total mass $M_{ej}$, and the way this mass is distributed with the density profile's power-law index $n$.
    \item The outer radius: $r_{m}$, constant in this case.
\end{enumerate}

\subsubsection{The breakout}
The jet breaks out of the ejecta at $t = t_b$, where $t_b-t_0$ is the necessary time for the jet to break out of the ejecta since the jet launch. At the very instant of the breakout $t=t_b$, $r_h(t_b)=r_m$; hence, the breakout time and velocity can be found analytically as:
\begin{align}
\label{eq:tb stat 1}
t_b - t_0 =& \frac{2}{A\:(4-n)}\left[r_m^\frac{4-n}{2} - r_0^\frac{4-n}{2} \right], \\
\label{eq:vb stat 1}
v_b =& A\:r_m^\frac{n-2}{2}.
\end{align}
Note that $A$ should be replaced by $A_c$ [in equation (\ref{eq:A_c static})] for calibrated analytic results. 

For $r_{m}\gg r_0$, and after replacing $\theta_j$ with $\theta_j = \theta_0/f_j$, $A_c$ in equation (\ref{eq:A_c static}) can be simplified to the following:
\begin{equation}
A_{c}\simeq 
    \begin{cases}
      N_s\sqrt{\frac{r_m^{3-n} L_{iso,0}  f_j^2}{(3-n)\:M_{ej}\:c}   }\:\:\:& (\text{for}\: n<3),\\
      N_s\sqrt{\frac{r_0^{3-n} L_{iso,0} f_j^2}{(n-3)\:M_{ej}\:c}   }\:\:\:& (\text{for}\: n>3).
    \end{cases}       
\label{eq:A_c static sim}
\end{equation}
$t_b - t_0$ [in equation (\ref{eq:tb stat 1})] and $v_b$ [in equation (\ref{eq:vb stat 1})] can be also simplified to the following:
\begin{align}
\label{eq:tb stat sim}
t_b - t_0 &\simeq
\begin{cases}
\frac{1}{N_s}\sqrt{ \frac{4(3-n)r_m M_{ej} c}{(4-n)^2 f_j^2 L_{iso,0}} }&(\text{for}\: n<3),\\
 \\
\frac{1}{N_s}\sqrt{ \frac{4(n-3)r_0^{n-3} r_m^{4-n} M_{ej} c}{(4-n)^2 f_j^2 L_{iso,0}} }&(\text{for}\: 4>n>3),\\
\end{cases}\\
\label{eq:vb stat sim}
v_b &\simeq 
\begin{cases}
N_s\sqrt{\frac{r_m L_{iso,0} f_j^2}{(3-n)\:M_{ej}\:c}   }&(\text{for}\: n<3),\\
 \\
N_s\sqrt{\frac{r_0^{3-n} r_m^{n-2}  L_{iso,0}  f_j^2}{(n-3)\:M_{ej}\:c}   }&(\text{for}\: 4>n>3).\\
\end{cases}
\end{align}
In this case I, the medium's outer radius and the breakout radius are equal: $r_m = R_b$. Hence, for typical collapsar parameters (see model G5.0 in \citealt{2013ApJ...777..162M}), taking $n\approx2$, the breakout time can be written as:
\begin{equation}
\begin{split}
& t_b - t_0 \simeq 4.61\:\text{s} \left(\frac{M_{ej}}{14 M_\odot}\right)^\frac{1}{2}\left(\frac{r_{m}}{4\times 10^{10} \text{cm}}\right)^\frac{1}{2}\\ 
&\left(\frac{L_{iso,0}}{5\times10^{52}\: \text{erg s}^{-1}}\right)^{-\frac{1}{2}} \left(\frac{\frac{3-n}{(4-n)^2}}{1/4}\right)^\frac{1}{2}
\left(\frac{N_{s}}{2/5}\right)^{-1}\left(\frac{f_{j}}{14}\right)^{-1}.
\end{split}
\label{eq:tb stat sim 2}
\end{equation}
Simulation of the model G5.0 in \citet{2013ApJ...777..162M} gives a breakout time of $4.5$ s, closely consistent with to the above $t_b - t_0\simeq 4.61$ s (with $f_j = 14$). 

\subsection{Case II. Expanding medium}
\label{subsubsec:Case II. Expanding medium}
In this case II, the difference is that we consider an ambient medium expanding outward with a non-negligible radial velocity, as in the case of BNS merger ejecta. The ejecta's outer radius $r_m(t)$, the expansion velocity at a given radius $r$, $v_a(r,t)$, and the density at a given radius $r$, $\rho_a(r,t)$, are all dependent on the time $t$. The equations are:
\begin{eqnarray}
    r_m(t) =& v_{ej}\: (t-t_0) + r_{m,0}, \\
    v_a(r,t) =& v_{ej}\left(\frac{r}{r_{m}(t)}\right), \\ 
    \rho_a(r,t) =& \rho_0 \left(\frac{r_{0}}{r}\right)^n\left(\frac{r_{m,0}}{r_m(t)}\right)^{3-n},
    \label{eq:the ejecta spec}
\end{eqnarray}
with $r_{m,0}$, $r_0$, $\rho_0$ are the ejecta outer radius, the jet head position, and the density at $r=r_0$, all at $t=t_0$ when the jet is launched (see figure \ref{fig:P4a-F00.pdf}). $v_{ej}$ is the ejecta's maximum velocity and $n$ is the power-law index of the density profile. Note that assuming that matter is ejected at the very moment of the merger $t=t_m$, $r_{m,0}$ can be approximated to $r_{m,0}\approx v_{ej}\:(t_0-t_m)+r_0$. 

\subsubsection{Equation of motion}
The motion of the jet head in an expanding medium is derived from equation (\ref{eq:beta_h_2}) as the following first-order differential equation (detailed calculations are presented in Appendix \ref{app subsec:Dynamic ambient medium: BNS merger ejecta}
):

\begin{eqnarray}
    \frac{dr_h(t)}{dt}  + \left( -\frac{v_{ej}}{r_m(t)}\right)r_h(t) = {A}\:{r_m(t)}^\frac{3-n}{2}{r_h(t)}^\frac{n-2}{2},
\end{eqnarray}
where $\tilde{L}^{1/2} \propto A$, and $A$ is a constant that depends only on the parameters of the ejecta and the engine: 
\begin{equation*}
A=\sqrt{ \left(\frac{r_{m,0}^{3-n}-r_0^{3-n}}{(3-n)\:r_{m,0}^{3-n}}\right)\left(\frac{4\:L_j}{\theta_j^2\:M_{ej}\:c}\right)    },
\end{equation*}

In comparison with the equation of motion in the case of the static medium, there are two differences. First, the additional term  $(-v_{ej}/r_m(t))r_h(t)$ and its negative sign can be understood as the comoving speed that the jet head gains from the expanding ejecta as a background. Second, the term in the right-hand side is different, which is due to the evolution of the density.

With the jet opening angle $\theta_j$ approximated as constant over time (see $\S$ \ref{sec:A roughly constant cross section}), $A$ is also constant and the analytic solution can be found as (more details are given in Appendix \ref{app subsec:Dynamic ambient medium: BNS merger ejecta}):

\begin{equation}
\begin{split}
    r_h(t) =& \left[ (4-n){\frac{A}{v_{ej}}}[\sqrt{r_m(t)} - \sqrt{r_{m,0}}] + \left(\frac{r_0}{r_{m,0}}\right)^{\frac{4-n}{2}} \right]^\frac{2}{4-n} \\ 
    & \times r_m(t),
\end{split}
\label{eq:r dyn}
\end{equation}

\begin{equation}
v_h(t) = A \left[\frac{r_h(t)}{r_m(t)}\right]^\frac{n-2}{2}\sqrt{r_m(t)} + v_{ej}\:\left[ \frac{r_h(t)}{r_m(t)}\right].
\label{eq:v dyn}
\end{equation}

Through comparison with numerical simulations, we find that the analytic $\tilde{L}^{1/2}$ is overestimated in the same way as previously found for the static medium case \citep{2013ApJ...777..162M,2018MNRAS.477.2128H}. 
Hence, correcting $\tilde{L}^{1/2}$ gives the corrected form of $A$ as:
\begin{equation}
A \to A_c =N_s\sqrt{ \left(\frac{r_{m,0}^{3-n}-r_0^{3-n}}{(3-n)\:r_{m,0}^{3-n}}\right)\left(\frac{4\:L_j}{\theta_j^2\:M_{ej}\:c}\right)    },
\label{eq:A_c dyn}
\end{equation}
where we take $N_s\approx 2/5$ [see $\S$ \ref{sec:calibration} and equation (\ref{eq:N_s})]. 
The opening angle of the jet $\theta_j=\theta_0/f_j$ is unknown. 
Within the explored parameter space for the BNS merger case, we find that taking $f_j\approx 5$ is a reasonably good approximation for the case of a non-relativistic jet head. Hence, with the central engine's isotropic luminosity $L_{iso,0}\simeq4L_j/\theta_0^2$ and $r_0\approx10^6-10^7$ cm, the jet head motion is determined by the following macroscopic parameters: 
\begin{enumerate}
    \item The engine power: $L_j$ and the jet initial opening angle $\theta_0$; or the jet isotropic luminosity $L_{iso,0}$.
    \item The ejecta: its total mass $M_{ej}$, and the way this mass is distributed with the density profile's power-law index $n$.
    \item The outer radius: outer radius at the moment the jet is launched $r_{m,0}$, and its expansion velocity $v_{ej}$.
\end{enumerate}

\subsubsection{The breakout}
As shown in figure \ref{fig:P4a-F00.pdf}, at the breakout $t=t_b$, which is $t_b-t_0$ after the jet launch, the following equation is fulfilled: $r_h(t=t_b)=r_m(t=t_b)=R_b$. Hence, the breakout time since the jet launch $t_b-t_0$ and the breakout velocity $v_b$ can be analytically found as:

\begin{align}
\label{eq:tb dyn}
    t_b - t_0 =&\left[ \frac{ r_{m,0}^\frac{4-n}{2} - r_{0}^\frac{4-n}{2}  }{ r_{m,0}^\frac{4-n}{2}} \frac{\sqrt{v_{ej}}}{(4-n)A} + \sqrt{\frac{r_{m,0}}{v_{ej}}}         \right]^2 - \frac{r_{m,0}}{v_{ej}},\\
    \label{eq:v_b dyn}
    v_b =& A \:\sqrt{R_b} + v_{ej} .
\end{align}
Here too, $A$ should be replaced by $A_c$ in equation (\ref{eq:A_c dyn}).

For $r_{m,0}\gg r_0$, and $\theta_j = \theta_0/f_j$, $A_c$ in equation (\ref{eq:A_c dyn}) can be simplified to the following:
\begin{equation}
A_{c} \simeq  
\begin{cases}
N_s\sqrt{\frac{L_{iso,0} f_j^2}{(3-n)\:M_{ej}\:c}} &(\text{for}\:n<3),\\
N_s\sqrt{\frac{r_{m,0}^{n-3} L_{iso,0} f_j^2}{(n-3)r_0^{n-3}\:M_{ej}\:c}} &(\text{for}\:n>3).
\end{cases}
\label{eq:A_c dyn sim}
\end{equation}
For $n<4$, equation (\ref{eq:tb dyn}) can be also simplified to the following:
\begin{align}
\label{eq:tb dyn sim}
t_b - t_0 \simeq & \frac{2\sqrt{r_{m,0}}}{(4-n)A_c} + \frac{v_{ej}}{(4-n)^2 A_c^2}.
\end{align}
Inserting $A_c$ [equation (\ref{eq:A_c dyn sim})] in $t_b-t_0$ [equation (\ref{eq:tb dyn sim})] and in $v_b$ [in equation (\ref{eq:v_b dyn})] gives:
\begin{equation}
\label{eq:tb dyn sim 3}
\begin{split}
& t_b - t_0 \simeq\\
& 
\begin{cases}
\frac{1}{N_s}\sqrt{ \frac{4(3-n)r_{m,0} M_{ej} c}{(4-n)^2 f_j^2 L_{iso,0}} } + \frac{1}{N_s^2}\frac{(3-n)M_{ej} c \:v_{ej}}{(4-n)^2 L_{iso,0} f_j^2} &(n<3),\\
\frac{1}{N_s}\sqrt{ \frac{4(n-3)r_0^{n-3} r_{m,0}^{4-n} M_{ej} c}{(4-n)^2 f_j^2 L_{iso,0}} } + \frac{1}{N_s^2}\frac{(n-3)r_0^{n-3}M_{ej} c \:v_{ej}}{(4-n)^2r_{m,0}^{n-3} L_{iso,0} f_j^2}&(4>n>3),\\
\end{cases}
\end{split}
\end{equation}
and:
\begin{equation}
\label{eq:v_b dyn sim}
\begin{split}
& v_b \simeq \\
&
\begin{cases}
N_s\sqrt{\frac{r_{m,0} L_{iso,0} f_j^2}{(3-n)\:M_{ej}\:c}   }\sqrt{R_b/r_{m,0}}  + v_{ej} &(\text{for}\: n<3),\\
 \\
N_s\sqrt{\frac{r_0^{3-n} r_{m,0}^{n-2}  L_{iso,0}  f_j^2}{(n-3)\:M_{ej}\:c}   }\sqrt{R_b/r_{m,0}}+ v_{ej}
&(\text{for}\: 4>n>3).\\
\end{cases}
\end{split}
\end{equation}

From the comparison of the expression of the breakout time in an expanding medium [equations (\ref{eq:tb dyn sim 3}) and (\ref{eq:v_b dyn sim})] with the expression of the breakout time in a static medium [equations (\ref{eq:tb stat sim}) and (\ref{eq:vb stat sim})], we can identify: i) the very same expressions found in the static case (in the first term, for both cases $n<3$ and $4>n>3$) which is independent of $v_{ej}$, and ii) an additional term linearly dependent on $v_{ej}$. 
And the same can be mentioned for the breakout velocity from equations (\ref{eq:vb stat sim}) and (\ref{eq:v_b dyn sim}).

The second term proportional to $v_{ej}$ in the breakout time expression [equations (\ref{eq:tb dyn sim 3}) and (\ref{eq:v_b dyn sim})] reflects the expansion of the ejecta. 
For a very small expansion velocity ($v_{ej}\simeq 0$), this analytic breakout time [equations (\ref{eq:tb dyn sim 3}) and (\ref{eq:v_b dyn sim})] converges to the very same analytic form found in the static medium case [equation (\ref{eq:tb stat sim}) in $\S$ \ref{subsubsec:Case I: Static medium}]. 
The same can be mentioned about the breakout velocity ($v_b$) when comparing equation (\ref{eq:v_b dyn sim}) with equation (\ref{eq:vb stat sim}). 

For typical parameters of GW170817's ejecta (later explained in $\S$ \ref{sec:4.NR}: refer to $\S$ \ref{sec:Density profile of the ejecta}; $\S$ \ref{sec:Velocity profile of the ejecta}; and $\S$ \ref{sec:Angular dependance}) and engine (later discussed in $\S$ \ref{sec:GW170817}; see figure \ref{fig:P4a-L=dt}), the breakout time can be written as: 
\begin{equation}
\begin{split}
& t_b - t_0 \\ 
& \simeq 0.173\:\text{s}\left(\frac{M_{ej}}{0.002M_\odot}\right)^{\frac{1}{2}}\left(\frac{r_{m,0}}{10^9\text{cm}}\right)^{\frac{1}{2}}\left(\frac{L_{iso,0}}{10^{51} \text{erg s}^{-1}}\right)^{-\frac{1}{2}}\\ 
& \left(\frac{\frac{(3-n)}{(4-n)^2}}{1/4}\right)^{\frac{1}{2}}
\left(\frac{N_{s}}{2/5}\right)^{-1}\left(\frac{f_{j}}{5}\right)^{-1}\\
& + 0.078\:\text{s}\left(\frac{M_{ej}}{0.002 M_\odot}\right)\left(\frac{v_{ej}}{0.35c}\right)\left(\frac{L_{iso,0}}{10^{51} \text{erg s}^{-1}}\right)^{-1}\\ 
& \left(\frac{\frac{(3-n)}{(4-n)^2}}{1/4}\right)
\left(\frac{N_{s}}{2/5}\right)^{-2}\left(\frac{f_{j}}{5}\right)^{-2}\\
& (\text{for}\:n < 3),
\end{split}
\label{eq:tb dyn sim n<3}
\end{equation}

\begin{equation}
\begin{split}
& t_b - t_0 \\ 
& \simeq 0.087\:\text{s}\left(\frac{M_{ej}}{0.002M_\odot}\right)^{\frac{1}{2}}\left(\frac{r_{0}}{10^6\text{cm}}\right)^{\frac{1}{4}}\left(\frac{r_{m,0}}{10^9\text{cm}}\right)^{\frac{1}{4}}\\
& \left(\frac{L_{iso,0}}{10^{51} \text{erg s}^{-1}}\right)^{-\frac{1}{2}} \left(\frac{\frac{(n-3)}{(4-n)^2}}{2}\right)^{\frac{1}{2}}
\left(\frac{N_{s}}{2/5}\right)^{-1}\left(\frac{f_{j}}{5}\right)^{-1}\\
& + 0.020\:\text{s}\left(\frac{M_{ej}}{0.002 M_\odot}\right)\left(\frac{r_{0}}{10^6\text{cm}}\right)^{\frac{1}{2}}\left(\frac{r_{m,0}}{10^9\text{cm}}\right)^{-\frac{1}{2}}\left(\frac{v_{ej}}{0.35c}\right)\\ 
& \left(\frac{L_{iso,0}}{10^{51} \text{erg s}^{-1}}\right)^{-1}\left(\frac{\frac{(n-3)}{(4-n)^2}}{2}\right)
\left(\frac{N_{s}}{2/5}\right)^{-2}\left(\frac{f_{j}}{5}\right)^{-2}\\
& (\text{for}\:4>n>3).
\end{split}
\label{eq:tb dyn sim n>3}
\end{equation}

First, in both cases ($n<3$ and $4>n>3$), the first term in the expression of the breakout time (independent of $v_{ej}$) is roughly of the same order as the second term, for the parameters chosen above. 
Second, comparison of these two analytic breakout times for different density profile indices, $n=2$ [equation (\ref{eq:tb dyn sim n<3})] and $n=3.5$ [equation (\ref{eq:tb dyn sim n>3})], shows that the results vary within a factor of $\sim 2$\footnote{Note that, for simplicity, we took the same ejecta velocity $v_{ej} = 0.346c$ (and the same $r_{m,0}$) independently of the density profile index $n$. Ideally, $v_{ej}$ should be set in accordance with the value $n$, so that the average velocity (and the kinetic energy) of the ejecta is kept the same for a fair comparison.}.
This breakout time will be further discussed in comparison with numerical simulations (see $\S$ \ref{sec:Comparison with Numerical Simulations} or table \ref{table:models}).

\subsubsection{Comparaison with other analytical estimations of the breakout time}
The breakout time of a relativistic jet head in an expanding medium has been derived analytically in several recent works.
\citet{2018MNRAS.475.2659M}, \citet{2018ApJ...866L..16M} and \citet{2019ApJ...876..139G} have presented analytic modeling of jet propagation in an expanding medium.

In \citet{2018MNRAS.475.2659M}, the context is jet propagation in the expanding ejecta of superluminous SN associated with GRB.
\citet{2018MNRAS.475.2659M} presented an analytical solution using a set of reasonable approximations within the context of collapsar jets (\citealt{2011ApJ...740..100B}). However, since the jet head motion has been derived with the approximation of neglecting the expansion of the ejecta [unlike the case here; compare equation (\ref{eq:beta_h_2}) here with equation (17) in \citealt{2018MNRAS.475.2659M}], their model cannot be applied in the limit of the BNS merger case where the ejecta's expansion is substantial ($\sim 0.2c$) and comparable to the jet head velocity.

In \citet{2018ApJ...866L..16M} and \citet{2019ApJ...876..139G} the context is jet propagation in an expanding medium, the same as here.
In \citet{2018ApJ...866L..16M} and \citet{2019ApJ...876..139G}, the analytic modeling of the jet propagation follows the same arguments of \citet{2011ApJ...740..100B}.
Both studies presented numerically solved jet head equation of motion, for a certain set of parameters.
This is in contrast with the results presented here, where jet propagation is solved analytically (see Appendices \ref{app:B} and \ref{sec:app C}). 
As a result, we demonstrate that the breakout time in the case of an expanding medium is determined by the sum of two components: the same component as in the case of a static medium; and another component which has different dependence on the parameters of the engine and the ejecta [refer to equations (\ref{eq:tb dyn sim n<3}) and (\ref{eq:tb dyn sim n>3}) and the discussion that follows].
This is an important new finding which was found to be valid for a large parameter space in consistence with numerical simulations.
In addition, an extensive comparison with numerical simulations presented here and others in the literature have been carried out.

Comparison with numerical simulations show that our analytical model gives the best results (later discussed in $\S$ \ref{sec:Comparison with Numerical Simulations}).
It also shows that estimations from previous studies (\citealt{2018MNRAS.475.2659M}; \citealt{2018ApJ...866L..16M}; \citealt{2019ApJ...876..139G}); which did relay on arguments made for the case of a static medium (\citealt{2011ApJ...740..100B}) and did not present a comparison of the numerical/analytical results with numerical simulations; did not properly account for the expansion of the ejecta, in particular its effect on the lateral radius of the cocoon $r_c$ [i.e., the parameter $\chi$ in equation (\ref{eq:c1}); see Appendix \ref{sec:app C}]. The lack of this treatment, combined with assuming $\eta\approx 1$ is found to dramatically (and unphysically) enhance the pressure of the cocoon in the case where the medium is expanding, as in BNS merger's ejecta.
As a result, their estimations give an unphysically high jet collimation, and hence the value of $\tilde{L}$ is found higher by roughly a factor of $\sim 4$ (see Appendix \ref{sec:app C}).

\citet{2018ApJ...866....3D} is a similar study to the one presented here, where the jet propagation was analytically modeled and extensively compared with numerical simulations.
However, there are several noticeable differences. 
While our study is limited to the case of successful jets (which is most likely the case for GW170817; see \citealt{2018Natur.561..355M}), \citet{2018ApJ...866....3D} did cover both the case of successful jets and failed jets (referred to as the early breakout and the late breakout, respectively), where in the case of failed jets a cocoon breakout takes place after the engine is turned off.
On the other hand, the analytic treatment presented here is more extensive than in \citet{2018ApJ...866....3D} in several points. 
As it can be seen from a comparison of the derived breakout time, the parameter dependence is different [compare equations (\ref{eq:tb dyn sim n<3}) and (\ref{eq:tb dyn sim n>3}) with equation (21) in \citealt{2018ApJ...866....3D}], with our analytic model incorporating a larger set of parameters (of the ejecta and the engine), which makes it more extensive and scalable for different models. 
For instance, in our model, no assumption was taken for the delay time between the merger and the jet launch $t_0-t_m$ and it was taken to vary freely, while in \citet{2018ApJ...866....3D} jet launch was assumed to take place at the moment the merger takes place (i.e. $t_0-t_m\sim 0s$; which might not be always the case depending on the type of the collapse; see \citealt{2019arXiv190703790K}).
In addition, the treatment of the jet collimation here is more rigorous, where approximations were taken based on the fundamental equations (see Appendix \ref{sec:app C}).



\section{Comparison to Numerical Simulations}
\label{sec:Comparison with Numerical Simulations}
We present a large series of 2D relativistic hydrodynamical simulations. Our numerical study is unique as: i) it is the largest collection of numerical simulations of BNS mergers' jet launch to our best knowledge, with nearly a hundred models computed; ii) the sample as a whole investigates all the key parameters that are relevant for jet propagation, and hence, it explores a very wide parameter space.

\subsection{Numerical simulations}
\label{sec:NSs}
\subsubsection{Setup}
We carry out numerical simulations using a two-dimensional relativistic hydrodynamical code, which was previously used for core-collapse simulations in \citet{2014MNRAS.438.3119Y} and \citet{2012AIPC.1484..418O} (see \citealt{2017MNRAS.469.2361H} for more details about the code method). 

We follow the same setup as in \citet{2014ApJ...784L..28N}. Based on the approximations presented in $\S$ \ref{sec:model assumptions}, the initial conditions are set. The ejecta is spherically symmetric.
The injected jet is hot, with its initial enthalpy set as $h=20$.  
The jet is injected with an opening angle $\theta_{inj}$ and an initial Lorentz factor $\Gamma_0\simeq 1/\theta_{inj}$. Hence, accounting for the fact that the jet is hot and for the relativistic spreading, the opening angle of the jet is $\theta_0 \sim \theta_{inj} + \Gamma_0^{-1} \sim 2 \theta_{inj}$. The jet is injected with a constant power $L_j$ (per one polar jet) throughout the simulation. 

We use the Polytropic equation of state (EOS) with $\gamma=4/3$. The pressure in the ejecta is scaled to density as $P = K_{ef}\: \rho^{4/3}$ where $K_{ef} = 2.6 \times 10^{15} $g$^{-1/3}$ cm$^{3}$ s$^{-2}$ \citep{2014ApJ...784L..28N}. Such a scaling factor may give a colder temperature than that of the actual radioactive ejecta. However, this is not expected to affect the dynamics of the jet propagation. The circumstellar medium through which the ejecta expands (hereafter, CSM) is assumed to have a much lower density than the ejecta (we use $\rho_{CSM}=10^{-10}$ g cm$^{-3}$). The CSM is also assumed to be static ($v=0$). 

The coordinate system $(r, \theta)$ is spherical, with axisymmetry and equatorial plane symmetry. The inner boundary of the computation domain is $r_{in} = 1.2 \times 10^8 $ cm\footnote{Ideally, the inner boundary $r_{in}$ should be in the order of $r_0\sim 10^6-10^7$ cm, however, as calculations with such a small inner boundary are numerically very challenging and extremely time consuming, a larger inner boundary is considered. The artificial effects from such a larger inner boundary are minor as long as $r_{m,0}\gg r_{in}$.} (as in \citealt{2014ApJ...784L..28N}). The mesh is allocated using an adaptive mesh refinement (AMR) manner. The initial number of grid is $\sim 8000 \times 512$ for high-resolution calculation, and $\sim 2500\times 512$ for the lowest resolution (with angular grid $N_\theta=512$ in both cases). The radial resolution is the highest at the inner boundary, and decreases logarithmically in the domain $r_{in} < r <10^{10}$ cm. For $r > 10^{10}$ cm the resolution is kept constant, unless the jet is detected; in which case the AMR algorithm allocates more grids accordingly. For high-resolution calculations, the highest resolution is $\Delta r_{min} = 10^5$ cm. For low-resolution calculations, the highest resolution is $\Delta r_{min} = 10^6$ cm. Angular resolution is also distributed logarithmically, with the angular resolution around the on-axis being 10 times the angular resolution near the equator. The angular resolution around the on-axis is always high ($\Delta \theta_{min} = 0.04^\circ$).

Considering the timescale of the simulations (< 1s) and its goal of studying the jet head dynamics, many effects are neglected. Magnetism, gravity, neutrino pressure, the ejecta's fast tail, and general relativistic effects are all ignored.

\subsubsection{Simulated models}
\label{sec:Simulated models}
As shown in table \ref{table:models}, we present the list of the simulated models. Models are classified into different groups of simulations; each group is intended to explore certain space of parameters. These series of models are set so that all the relevant parameters for jet propagation are covered. 

The first group (labeled by ``T'') focuses on exploring the time delay between the merger and the jet launch $t_0-t_m$; which can affect the jet structure \citep{2019ApJ...877L..40G}. 
We present models varying in $t_0-t_m$, logarithmically, from 20 ms to 320 ms. Other parameters varies in this group, such as the opening angle $\theta_0$, engine luminosity $L_{iso,0}$, and maximum resolution. This group is aimed to take parameters similar to GW170817. For this group we take ejecta parameters based on numerical relativity simulations for a $1.35-1.35 M_\odot$ BNS merger: the density profile power-law index $n=2$, the ejecta maximum velocity $v_{ej}=0.2\sqrt{3}c$, and the ejecta's total mass $M_{ej}=0.002 M_\odot$ (refer to $\S$ \ref{sec:Density profile of the ejecta}; $\S$ \ref{sec:Velocity profile of the ejecta}; and $\S$ \ref{sec:Angular dependance}; respectively). 
The jet's initial opening angle $\theta_0$ takes two values; the ``narrow'' jet case ($6.8^\circ$) and the ``wide'' jet case ($18.0^\circ$). This choice of values for $\theta_0$ is based on the assumption that the final jet opening angle, accounting for its spreading after the breakout, is taken as $\theta_f \sim\theta_0/2$\footnote{From \citet{2013ApJ...777..162M} we know that for a typical collapsar, the jet opening angle after the breakout is $\sim \theta_0/5$. However, as there are no such studies for jet opening angle evolution after the breakout for the expanding ejecta case, we assume a conservative jet opening angle after the breakout of $\theta_0/2$.}. Hence, our choice of $\theta_0$ values is so that it covers the expected opening angle for GW170817's jet as suggested by observations (see $\S$ \ref{subsec:Late observations}  for a rigorous explanation). 

The second group (labeled by ``N'') is meant to explore density profiles (varying in $n$), and test the consistency of our analytic model's breakout time. Also, it allows to compare our simulations to others in the literature. It takes ejecta parameters similar to those in \citet{2014ApJ...784L..28N}. Resolution does vary from high to low within this group.

The third group of models (labeled by ``V'') presents models varying in the ejecta's maximum velocity $v_{ej}$. It aims to test our analytic equations of jet head motion, in a general way, independently of GW170817 typical parameters. $v_{ej}$ takes the values $0$, $0.1c$, $0.2c$ and $0.4c$. The density profile also varies from the flat case $n=0$ to a very steep case $n=5$.

For reference, in figure \ref{fig:P4a-snap} two snapshots showing the density and the velocity are presented for two models (T02-H and N30-H).

Finally, in figure \ref{fig:P4a-rh2} we show the average opening angle of the jet head in numerical simulations, from $t_0$ to the breakout time $t_b$. We show four different models (T02-H, T07-H, T12-H and T17-H; all with $t_0-t_m=80$ ms). The average opening angle is derived as:
\begin{eqnarray}
      \theta_{j,av} = \frac{\int_{r_h(t)/2}^{r_h(t)}\theta_j(r)dr}{r_h(t)/2},
      \label{eq:theta_av}
\end{eqnarray}
where $r_h(t)$ is the jet head radius and $\theta_j(r)$ is the opening angle of the jet outflow at the radius $r$. The jet outflow was defined by the following two requirements: $\Gamma > 3$ and $h\Gamma > 10$; this ensures the exclusion of the ejecta and the cocoon material. This definition of the average jet opening angle is similar to that in \citet{2014ApJ...784L..28N} [see fig 3 and equation (9) in \citealt{2014ApJ...784L..28N}]. However, in our case, the integration starts from half the jet head radius $r_h(t)/2$. This modification is added to focus on the outer part of the collimated jet, so that the average opening angle represents the jet head's opening angle. Note that, it is the jet head's opening angle which is relevant to the jet head cross-section $\Sigma_j$ and $\tilde{L}$ [as discussed in $\S$ \ref{sec:The Analytical Model}; see equation (\ref{eq:L expression})]. 

From figure \ref{fig:P4a-rh2}, it can be seen that our approximations in $\S$ \ref{sec:A roughly constant cross section} are quite reasonable. 
First, figure \ref{fig:P4a-rh2} shows that the jet head opening angle is roughly constant over time [the time dependence is weak; see equation (\ref{eq:theta_j^2}) in Appendix \ref{sec:app C}].
Second, figure \ref{fig:P4a-rh2} shows that the collimation factor $f_j=\theta_0/\theta_j$ takes values as $f_j \sim 2 - 10$ depending on the model. The typical value of $f_j$ in the case of a non-relativistic jet head ($\tilde{L}\ll (1-\beta_a)^2$) is $f_j\sim 5$.

\begin{table*}
\caption {Simulated models.}
\label{table:models}
\begin{tabular}{l|lllllll|lll}
  \hline
        & $M_{ej}$ & $n$$^a$ & $L_{iso,0}$$^b$ &  $\theta_0$$^c$ & $r_{in}$ & $t_0-t_m$$^d$ (or $r_m$)$^e$ & $v_{ej}$   & Resolution$^f$ & $t_b-t_0$ [s] & $t_b-t_0$ $^g$ [s] \\
    Model  &  [$10^{-2} M_\odot$] & & [erg s$^{-1}$] &  [deg] & [cm] & [ms]     ([cm]) & [c]   &  & (Simulation) & (Analytic) \\
        \hline
    T00-H & $0.2$ & 2 & $5\times10^{50}$ & 6.8 & $1.2\times 10^8$ & 20 & 0.346 & High  & 0.155 & 0.146\\ 
    T01-H & $0.2$ & 2 & $5\times10^{50}$ & 6.8 & $1.2\times 10^8$ & 40 & 0.346 & High  & 0.255 & 0.246\\
    T02-H & $0.2$ & 2 & $5\times10^{50}$ & 6.8 & $1.2\times 10^8$ & 80 & 0.346 & High  & 0.227 & 0.340\\
    T03-H & $0.2$ & 2 & $5\times10^{50}$ & 6.8 & $1.2\times 10^8$ & 160 & 0.346 & High  & 0.221 & 0.448\\
    T04-H & $0.2$ & 2 & $5\times10^{50}$ & 6.8 & $1.2\times 10^8$ & 320 & 0.346 & High  & 0.326 & 0.587\\
        \hline
    T05-H & $0.2$ & 2 & $5\times10^{51}$ & 6.8 & $1.2\times 10^8$ & 20 & 0.346 & High  & 0.051 & 0.031\\
    T06-H & $0.2$ & 2 & $5\times10^{51}$ & 6.8 & $1.2\times 10^8$ & 40 & 0.346 & High  & 0.084 & 0.054\\
    T07-H & $0.2$ & 2 & $5\times10^{51}$ & 6.8 & $1.2\times 10^8$ & 80 & 0.346 & High  & 0.096 & 0.079$^*$\\
    T08-H & $0.2$ & 2 & $5\times10^{51}$ & 6.8 & $1.2\times 10^8$ & 160 & 0.346 & High  & 0.154 & 0.111$^*$\\
    T09-H & $0.2$ & 2 & $5\times10^{51}$ & 6.8 & $1.2\times 10^8$ & 320 & 0.346 & High  & 0.257 & 0.153$^*$\\
        \hline
    T10-H & $0.2$ & 2 & $5\times10^{50}$ & 18 & $1.2\times 10^8$ & 20 & 0.346 & High  & 0.127 & 0.146\\
    T11-H & $0.2$ & 2 & $5\times10^{50}$ & 18 & $1.2\times 10^8$ & 40 & 0.346 & High  & 0.356 & 0.246\\
    T12-H & $0.2$ & 2 & $5\times10^{50}$ & 18 & $1.2\times 10^8$ & 80 & 0.346 & High  & 0.403 & 0.340\\
    T13-H & $0.2$ & 2 & $5\times10^{50}$ & 18 & $1.2\times 10^8$ & 160 & 0.346 & High  & 0.429 & 0.448\\
    T14-H & $0.2$ & 2 & $5\times10^{50}$ & 18 & $1.2\times 10^8$ & 320 & 0.346 & High  & $\gtrsim$0.650 & 0.587\\
        \hline
    T15-H & $0.2$ & 2 & $5\times10^{51}$ & 18 & $1.2\times 10^8$ & 20 & 0.346 & High  & 0.048 & 0.031\\
    T16-H & $0.2$ & 2 & $5\times10^{51}$ & 18 & $1.2\times 10^8$ & 40 & 0.346 & High  & 0.074 & 0.054\\
    T17-H & $0.2$ & 2 & $5\times10^{51}$ & 18 & $1.2\times 10^8$ & 80 & 0.346 & High  & 0.164 & 0.079$^*$\\
    T18-H & $0.2$ & 2 & $5\times10^{51}$ & 18 & $1.2\times 10^8$ & 160 & 0.346 & High  & 0.234 & 0.111$^*$\\
    T19-H & $0.2$ & 2 & $5\times10^{51}$ & 18 & $1.2\times 10^8$ & 320 & 0.346 & High  & 0.378 & 0.153$^*$\\
        \hline
    T00-L & $0.2$ & 2 & $5\times10^{50}$ & 6.8 & $1.2\times 10^8$ & 20 & 0.346 & Low  & 0.190	& 0.146\\
    T01-L & $0.2$ & 2 & $5\times10^{50}$ & 6.8 & $1.2\times 10^8$ & 40 & 0.346 & Low  & 0.145	& 0.246\\
    T02-L & $0.2$ & 2 & $5\times10^{50}$ & 6.8 & $1.2\times 10^8$ & 80 & 0.346 & Low  & 0.278	& 0.340\\
    T03-L & $0.2$ & 2 & $5\times10^{50}$ & 6.8 & $1.2\times 10^8$ & 160 & 0.346 & Low  & 0.297	& 0.448\\
    T04-L & $0.2$ & 2 & $5\times10^{50}$ & 6.8 & $1.2\times 10^8$ & 320 & 0.346 & Low  & 0.375	& 0.587\\
        \hline
    T05-L & $0.2$ & 2 & $5\times10^{51}$ & 6.8 & $1.2\times 10^8$ & 20 & 0.346 & Low  & 0.067	& 0.031\\
    T06-L & $0.2$ & 2 & $5\times10^{51}$ & 6.8 & $1.2\times 10^8$ & 40 & 0.346 & Low  & 0.100	& 0.054\\
    T07-L & $0.2$ & 2 & $5\times10^{51}$ & 6.8 & $1.2\times 10^8$ & 80 & 0.346 & Low  & 0.080	& 0.079$^*$\\
    T08-L & $0.2$ & 2 & $5\times10^{51}$ & 6.8 & $1.2\times 10^8$ & 160 & 0.346 & Low  & 0.145	& 0.111$^*$\\
    T09-L & $0.2$ & 2 & $5\times10^{51}$ & 6.8 & $1.2\times 10^8$ & 320 & 0.346 & Low  & 0.240	& 0.153$^*$\\
        \hline
    T10-L & $0.2$ & 2 & $5\times10^{50}$ & 18 & $1.2\times 10^8$ & 20 & 0.346 & Low  & 0.180	& 0.146\\
    T11-L & $0.2$ & 2 & $5\times10^{50}$ & 18 & $1.2\times 10^8$ & 40 & 0.346 & Low  & 0.470	& 0.246\\
    T12-L & $0.2$ & 2 & $5\times10^{50}$ & 18 & $1.2\times 10^8$ & 80 & 0.346 & Low  & 0.405	& 0.340\\
    T13-L & $0.2$ & 2 & $5\times10^{50}$ & 18 & $1.2\times 10^8$ & 160 & 0.346 & Low  & 0.675	& 0.448\\
    T14-L & $0.2$ & 2 & $5\times10^{50}$ & 18 & $1.2\times 10^8$ & 320 & 0.346 & Low  & 0.640	& 0.587\\
        \hline
    T15-L & $0.2$ & 2 & $5\times10^{51}$ & 18 & $1.2\times 10^8$ & 20 & 0.346 & Low  & 0.050	& 0.031\\
    T16-L & $0.2$ & 2 & $5\times10^{51}$ & 18 & $1.2\times 10^8$ & 40 & 0.346 & Low  & 0.076	& 0.054\\
    T17-L & $0.2$ & 2 & $5\times10^{51}$ & 18 & $1.2\times 10^8$ & 80 & 0.346 & Low  & 0.138	& 0.079$^*$\\
    T18-L & $0.2$ & 2 & $5\times10^{51}$ & 18 & $1.2\times 10^8$ & 160 & 0.346 & Low  & 0.243	& 0.111$^*$\\
    T19-L & $0.2$ & 2 & $5\times10^{51}$ & 18 & $1.2\times 10^8$ & 320 & 0.346 & Low  & 0.360	& 0.153$^*$\\
        \hline
        \hline
    N10-H & $1$ & 1 & $1.46\times10^{51}$ & 30 & $1.2\times 10^8$ & $60$ & 0.2 & High  & 0.196	& 0.259\\
    N15-H & $1$ & 1.5 & $1.46\times10^{51}$ & 30 & $1.2\times 10^8$ & $60$ & 0.2 & High  & 0.226	& 0.262\\
    N20-H & $1$ & 2   & $1.46\times10^{51}$ & 30 & $1.2\times 10^8$ & $60$ & 0.2 & High  & 0.240	& 0.263\\
    N22-H & $1$ & 2.25& $1.46\times10^{51}$ & 30 & $1.2\times 10^8$ & $60$ & 0.2 & High  & 0.227	& 0.263\\
    N25-H & $1$ & 2.5 & $1.46\times10^{51}$ & 30 & $1.2\times 10^8$ & $60$ & 0.2 & High  & 0.220	& 0.262\\
    N30-H & $1$ & 3   & $1.46\times10^{51}$ & 30 & $1.2\times 10^8$ & $60$ & 0.2 & High  & 0.229	& 0.258\\
    N35-H & $1$ & 3.5 & $1.46\times10^{51}$ & 30 & $1.2\times 10^8$ & $60$ & 0.2 & High  & 0.280	& 0.253\\
    N40-H & $1$ & 4   & $1.46\times10^{51}$ & 30 & $1.2\times 10^8$ & $60$ & 0.2 & High  & 0.242	& 0.245\\
    ...\\
  \hline
  \\
  \\
  \\
  \\
 \end{tabular}
\end{table*}
\begin{table*}
\addtocounter{table}{-1}
\caption{Continued...}
\begin{tabular}{l|lllllll|lll}
  \hline
        & $M_{ej}$ & $n$$^a$ & $L_{iso,0}$$^b$ &  $\theta_0$$^c$ & $r_{in}$ & $t_0-t_m$$^d$ (or $r_m$)$^e$ & $v_{ej}$   & Resolution$^f$ & $t_b-t_0$ [s] & $t_b-t_0$ $^g$ [s] \\
    Model  &  [$10^{-2} M_\odot$] & & [erg s$^{-1}$] &  [deg] & [cm] & [ms]     ([cm]) & [c]   &  & (Simulation) & (Analytic) \\
        \hline
    N00-L & $1$ & 0   & $1.46\times10^{51}$ & 30 & $1.2\times 10^8$ & $60$ & 0.2 & Low  & 0.190	& 0.248\\
    N10-L & $1$ & 1   & $1.46\times10^{51}$ & 30 & $1.2\times 10^8$ & $60$ & 0.2 & Low  & 0.211	& 0.259\\
    N12-L & $1$ & 1.25& $1.46\times10^{51}$ & 30 & $1.2\times 10^8$ & $60$ & 0.2 & Low  & 0.231	& 0.261\\
    N15-L & $1$ & 1.5 & $1.46\times10^{51}$ & 30 & $1.2\times 10^8$ & $60$ & 0.2 & Low  & 0.251	& 0.262\\
    N17-L & $1$ & 1.75& $1.46\times10^{51}$ & 30 & $1.2\times 10^8$ & $60$ & 0.2 & Low  & 0.259	& 0.263\\
    N20-L & $1$ & 2   & $1.46\times10^{51}$ & 30 & $1.2\times 10^8$ & $60$ & 0.2 & Low  & 0.271	& 0.263\\
    N22-L & $1$ & 2.25& $1.46\times10^{51}$ & 30 & $1.2\times 10^8$ & $60$ & 0.2 & Low  & 0.282	& 0.263\\
    N25-L & $1$ & 2.5 & $1.46\times10^{51}$ & 30 & $1.2\times 10^8$ & $60$ & 0.2 & Low  & 0.285	& 0.262\\
    N27-L & $1$ & 2.75& $1.46\times10^{51}$ & 30 & $1.2\times 10^8$ & $60$ & 0.2 & Low  & 0.283	& 0.261\\
    N30-L & $1$ & 3   & $1.46\times10^{51}$ & 30 & $1.2\times 10^8$ & $60$ & 0.2 & Low  & 0.301	& 0.258\\
    N32-L & $1$ & 3.25& $1.46\times10^{51}$ & 30 & $1.2\times 10^8$ & $60$ & 0.2 & Low  & 0.267	& 0.256\\
    N35-L & $1$ & 3.5 & $1.46\times10^{51}$ & 30 & $1.2\times 10^8$ & $60$ & 0.2 & Low  & 0.267	& 0.253\\
    N37-L & $1$ & 3.75& $1.46\times10^{51}$ & 30 & $1.2\times 10^8$ & $60$ & 0.2 & Low  & 0.257	& 0.249\\
    N40-L & $1$ & 4   & $1.46\times10^{51}$ & 30 & $1.2\times 10^8$ & $60$ & 0.2 & Low  & 0.272	& 0.245\\
    N50-L & $1$ & 5   & $1.46\times10^{51}$ & 30 & $1.2\times 10^8$ & $60$ & 0.2 & Low  & 0.260	& 0.225\\
  \hline\hline
    V00-L & $1$ & 0 & $3.28\times10^{51}$ & 20 & $1.2\times 10^8$ & ($1.2\times 10^9$) & 0.0 & Low  & 0.160	& 0.200\\
    V01-L & $1$ & 1 & $3.28\times10^{51}$ & 20 & $1.2\times 10^8$ & ($1.2\times 10^9$) & 0.0 & Low  & 0.163	& 0.214\\
    V02-L & $1$ & 2 & $3.28\times10^{51}$ & 20 & $1.2\times 10^8$ & ($1.2\times 10^9$) & 0.0 & Low  & 0.193	& 0.221\\
    V03-L & $1$ & 3 & $3.28\times10^{51}$ & 20 & $1.2\times 10^8$ & ($1.2\times 10^9$) & 0.0 & Low  & 0.183	& 0.210\\
    V04-L & $1$ & 4 & $3.28\times10^{51}$ & 20 & $1.2\times 10^8$ & ($1.2\times 10^9$) & 0.0 & Low  & 0.210	& 0.179\\
    V05-L & $1$ & 5 & $3.28\times10^{51}$ & 20 & $1.2\times 10^8$ & ($1.2\times 10^9$) & 0.0 & Low  & 0.177	& 0.143\\
  \hline
    V10-L & $1$ & 0 & $3.28\times10^{51}$ & 20 & $1.2\times 10^8$ & $400$ & 0.1 & Low  & 0.210	& 0.225\\
    V11-L & $1$ & 1 & $3.28\times10^{51}$ & 20 & $1.2\times 10^8$ & $400$ & 0.1 & Low  & 0.215	& 0.243\\
    V12-L & $1$ & 2 & $3.28\times10^{51}$ & 20 & $1.2\times 10^8$ & $400$ & 0.1 & Low  & 0.240	& 0.252\\
    V13-L & $1$ & 3 & $3.28\times10^{51}$ & 20 & $1.2\times 10^8$ & $400$ & 0.1 & Low  & 0.226	& 0.238\\
    V14-L & $1$ & 4 & $3.28\times10^{51}$ & 20 & $1.2\times 10^8$ & $400$ & 0.1 & Low  & 0.272	& 0.199\\
    V15-L & $1$ & 5 & $3.28\times10^{51}$ & 20 & $1.2\times 10^8$ & $400$ & 0.1 & Low  & 0.228	& 0.156\\
  \hline
    V20-L & $1$ & 0 & $3.28\times10^{51}$ & 20 & $1.2\times 10^8$ & $200$ & 0.2 & Low  & 0.267	& 0.250\\
    V21-L & $1$ & 1 & $3.28\times10^{51}$ & 20 & $1.2\times 10^8$ & $200$ & 0.2 & Low  & 0.267	& 0.271\\
    V22-L & $1$ & 2 & $3.28\times10^{51}$ & 20 & $1.2\times 10^8$ & $200$ & 0.2 & Low  & 0.310	& 0.283\\
    V23-L & $1$ & 3 & $3.28\times10^{51}$ & 20 & $1.2\times 10^8$ & $200$ & 0.2 & Low  & 0.347	& 0.265\\
    V24-L & $1$ & 4 & $3.28\times10^{51}$ & 20 & $1.2\times 10^8$ & $200$ & 0.2 & Low  & 0.270	& 0.219\\
    V25-L & $1$ & 5 & $3.28\times10^{51}$ & 20 & $1.2\times 10^8$ & $200$ & 0.2 & Low  & 0.257	& 0.169\\
  \hline
    V40-L & $1$ & 0 & $3.28\times10^{51}$ & 20 & $1.2\times 10^8$ & $100$ & 0.4 & Low  & 0.525	& 0.300\\
    V41-L & $1$ & 1 & $3.28\times10^{51}$ & 20 & $1.2\times 10^8$ & $100$ & 0.4 & Low  & 0.605	& 0.329\\
    V42-L & $1$ & 2 & $3.28\times10^{51}$ & 20 & $1.2\times 10^8$ & $100$ & 0.4 & Low  & 0.603	& 0.344\\
    V43-L & $1$ & 3 & $3.28\times10^{51}$ & 20 & $1.2\times 10^8$ & $100$ & 0.4 & Low  & 0.560	& 0.320\\
    V44-L & $1$ & 4 & $3.28\times10^{51}$ & 20 & $1.2\times 10^8$ & $100$ & 0.4 & Low  & 0.538	& 0.259\\
    V45-L & $1$ & 5 & $3.28\times10^{51}$ & 20 & $1.2\times 10^8$ & $100$ & 0.4 & Low  & 0.533	& 0.195\\  
  \hline\hline
 \end{tabular}
 \\
 \vspace{1ex}
 \raggedright {\footnotesize 
     $^a$ Density profile's power-law index. Note that as $n=3$ (and $n=4$) cannot be calculated with our set of equations, we effectively use $3.01$  (and $4.01$) instead.\\
     $^b$ The engine isotropic equivalent luminosity. Conversion to the jet luminosity $L_j$ (one sided) can be done using: $L_j \simeq L_{iso,0}\: \theta_0^2/4$.\\
     $^c$ The opening angle at the base of the jet accounting for relativistic spreading of the jet is defined as $\theta_0=\theta_{inj}+1/\Gamma_0$. Note that real values may vary slightly due to the approximation $\sim 1/\Gamma_0$.\\
     $^d$ The delay between the merger time $t_m$ and the jet launch time $t_0$. The outer radius of the ejecta is determined using: $r_{m,0}\simeq v_{ej}(t_0-t_m)+r_{0}$, with $r_{0}\sim 10^7$ cm. \\
     $^e$ The outer radius of the medium (in the case of a static outer medium $v_{ej} = 0$).\\
     $^f$ Relates to resolution used in the simulation.\\ 
     $^g$ Analytic breakout times are calculated using equation (\ref{eq:tb dyn}) [and equation (\ref{eq:tb stat 1}) in the static medium case: $v_{ej}=0$] and assuming $\theta_j\approx \theta_0/f_j$, where $f_j\approx 5$, and a calibration coefficient of $N_s = 2/5$. Note that for these values are calculated using a non-relativistic model for the jet head, and since the jet head in models T08-H, T09-H, T18-H, T19-H, T08-L, T09-L, T18-L, and T19-L reach relativistic velocities, the analytic breakout times for these models are not fully reliable.\\ 
     $^*$ The analytic results are questionable, because the requirement $\tilde{L}\ll (1-\beta_a)^2$ is not always fulfilled (or is at its limit). In other words, the approximation of a non-relativistic jet head and the analytic results are questionable.\\
     }
\end{table*}

\begin{figure*}
  \begin{subfigure}
    \centering
    \includegraphics[width=0.87\linewidth]{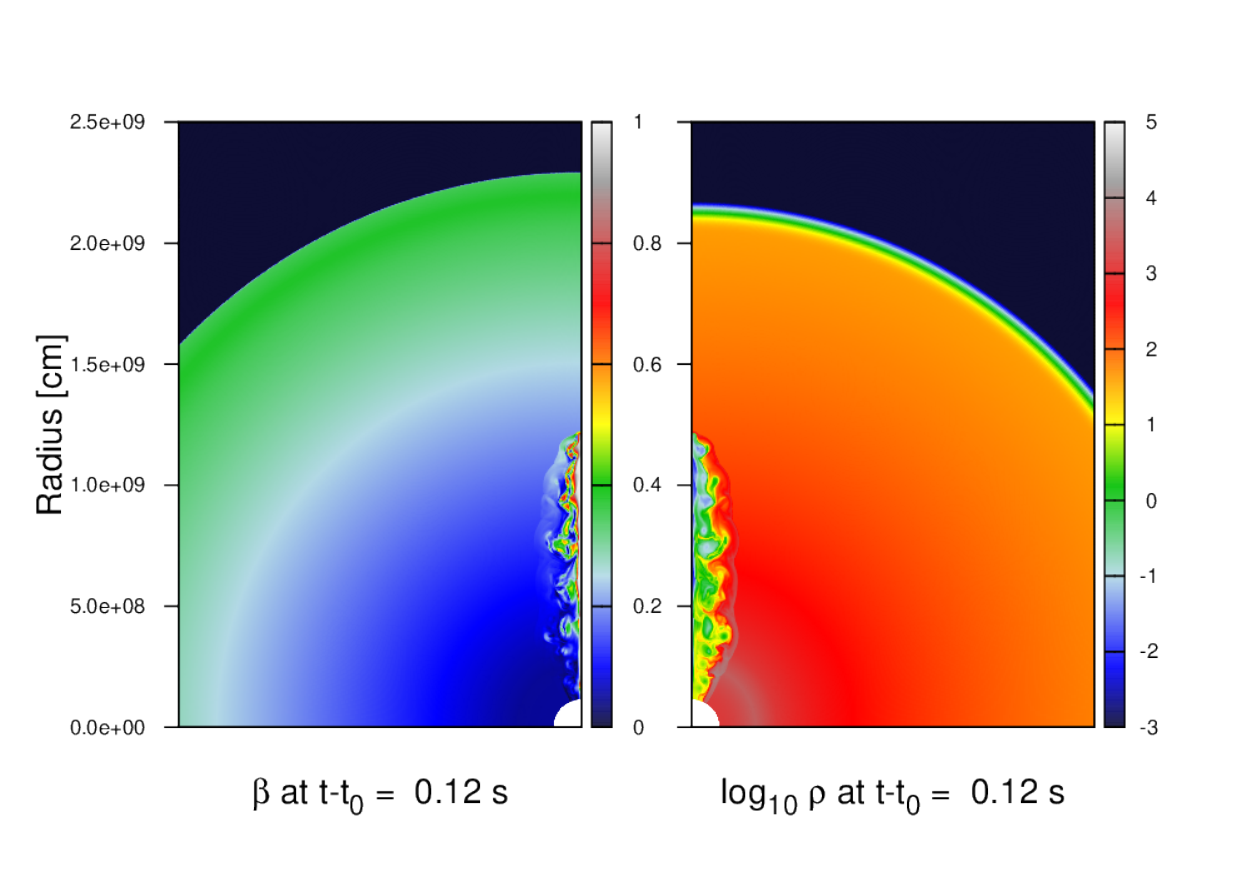}
  \end{subfigure}
  \begin{subfigure}
    \centering
    \includegraphics[width=0.87\linewidth]{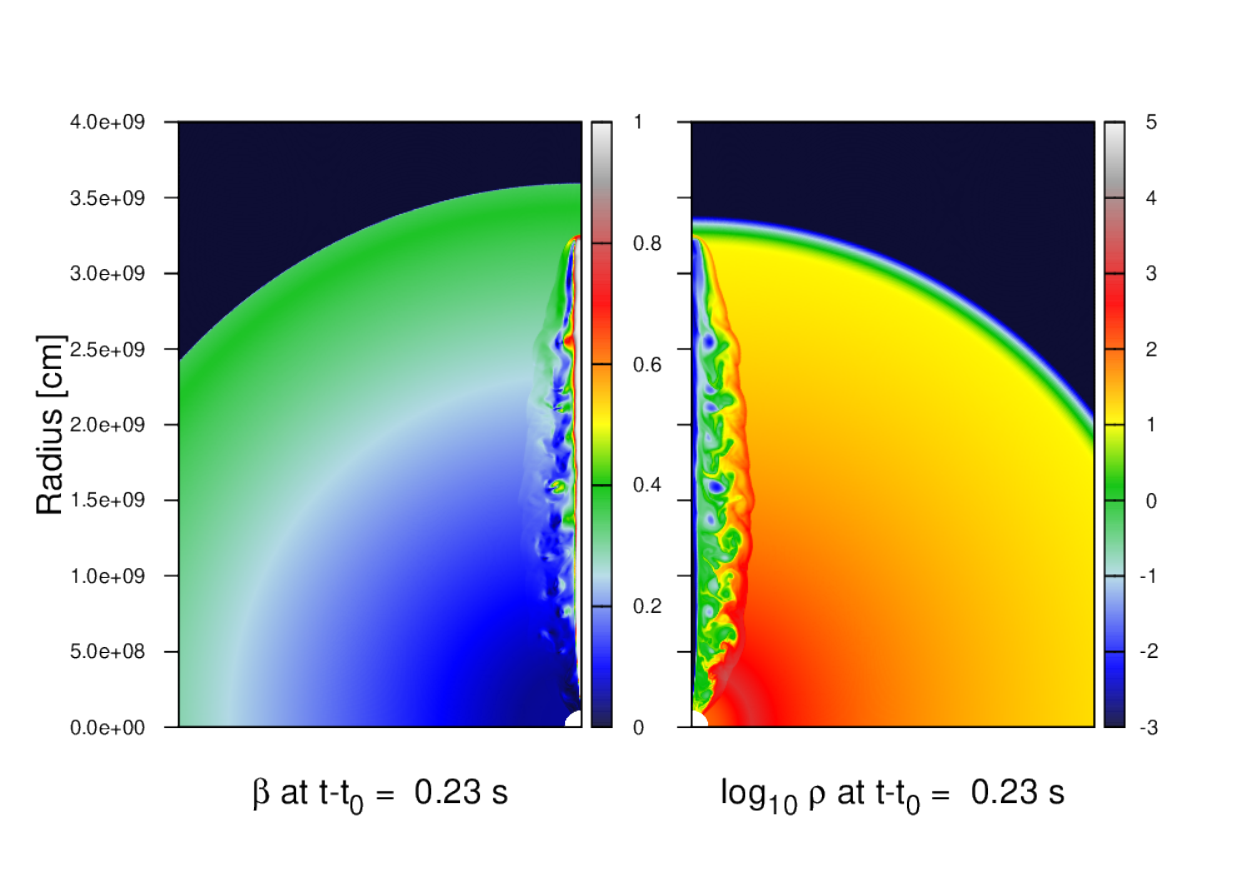}
    \caption{Four panels showing the velocity map (on the left) and the density map (on the right) in each panel. The first two panels are for the model T02-H, and the last two (in the next page) are for N30-H. The jet is injected at $t=t_0$. For each of the two models, the first panel shows the jet at $t\simeq t_0 + (t_b-t_0)/2$, and the second panel shows the jet at the moment of the breakout $t \approx t_b$. Note that, as we assume a relatively low density for the CSM, the interaction of the outer edge of the ejecta with the CSM widens the ejecta's outer edge and produces a slightly faster component (note that its density and mass are very small to affect the jet propagation).}
  \end{subfigure} 
  \end{figure*}
  
  \begin{figure*}
  \setcounter{figure}{1}
  \clearpage   
  \begin{subfigure}
    \centering
    \includegraphics[width=0.87\linewidth]{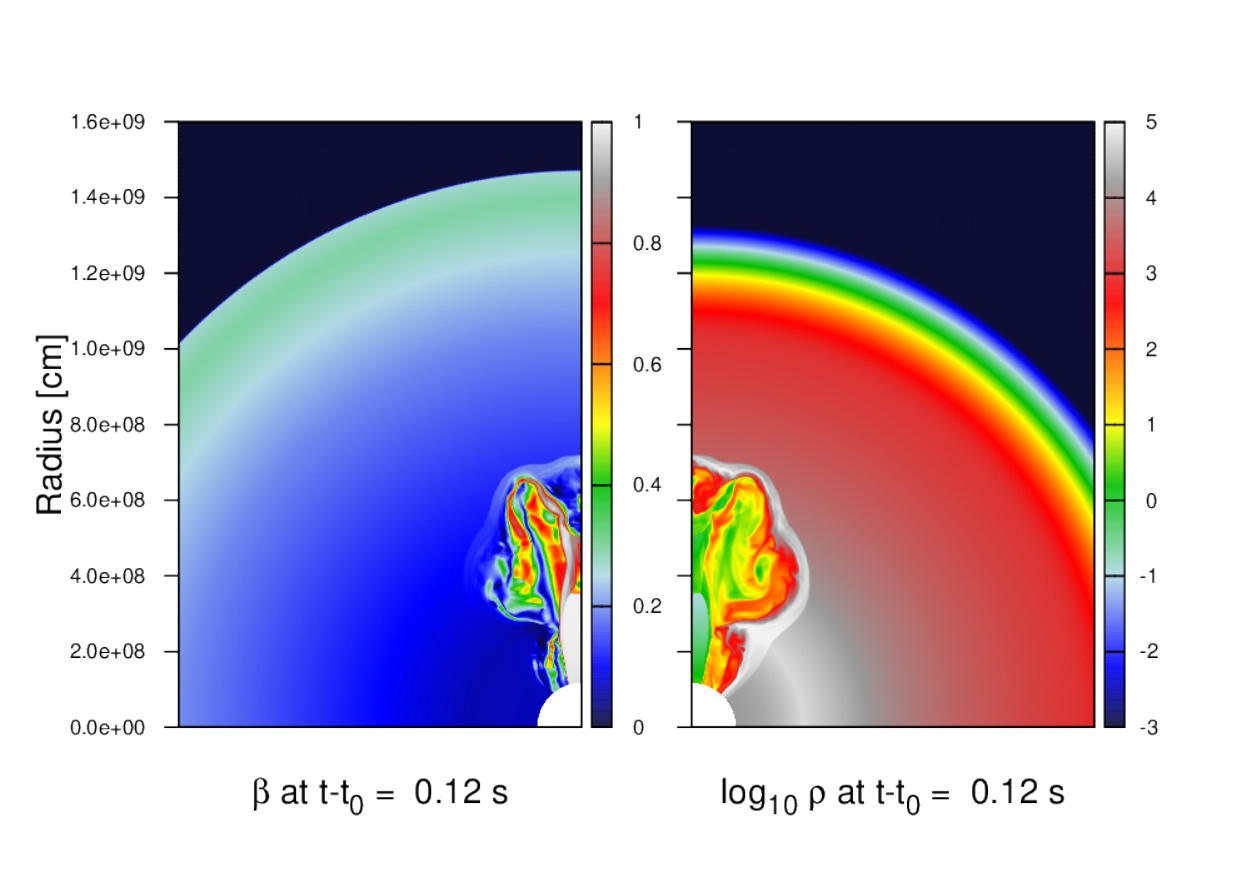}
  \end{subfigure}
  \begin{subfigure}
    \centering
    \includegraphics[width=0.87\linewidth]{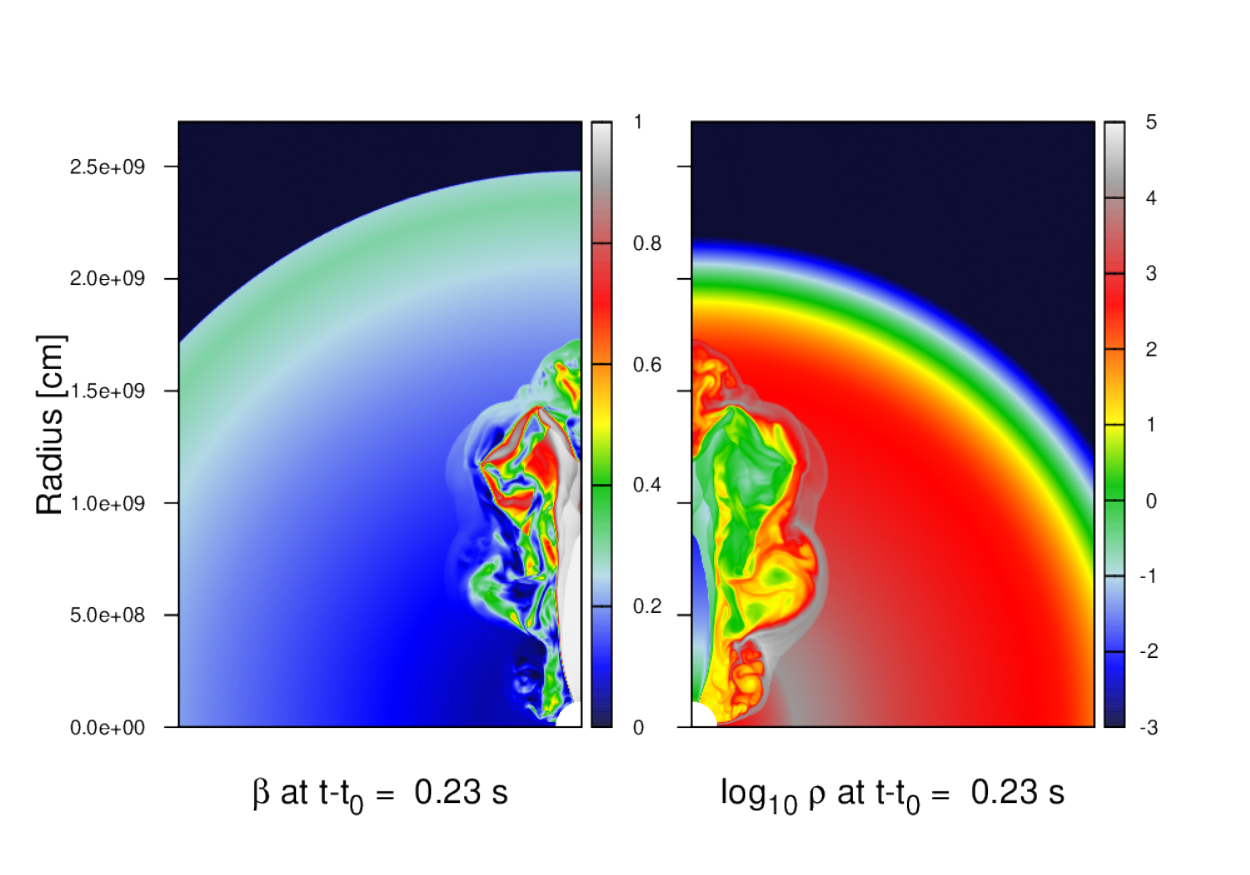}
  \end{subfigure} 
    \caption{(Continued.)
    }
  \label{fig:P4a-snap} 
\end{figure*}

\begin{figure}
    \centering
    \includegraphics[width=0.95\linewidth]{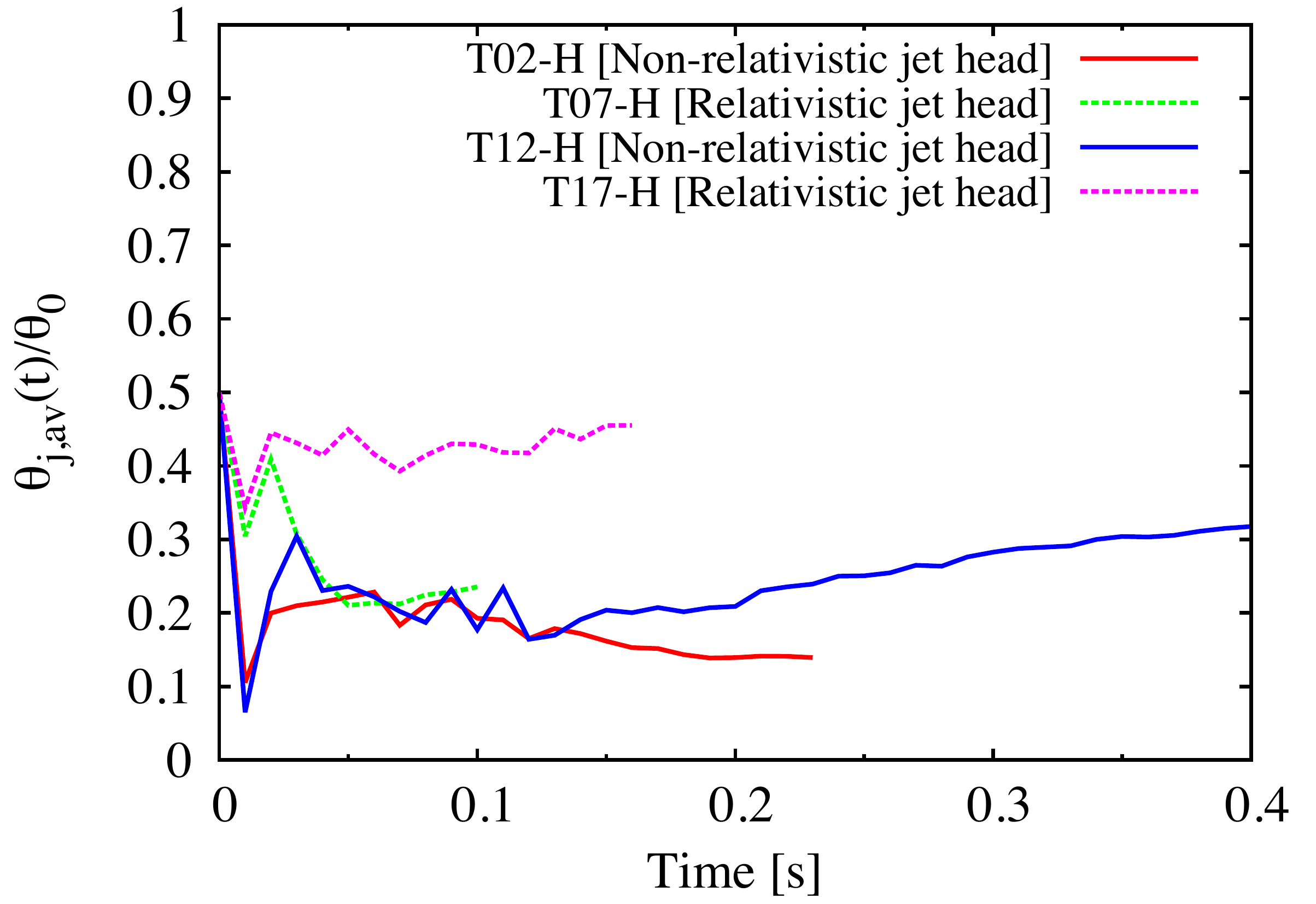} 
  \caption{Evolution of the average opening angle $\theta_{j,av}$ of the jet head in numerical simulations, normalized to the initial opening angle $\theta_0 = \theta_{inj}+1/\Gamma_0$, as a function of the time since the jet launch, up to the breakout time. The average opening angle is defined as presented in equation (\ref{eq:theta_av}). We show the results for four models (see table \ref{table:models} for the parameters of the models). The variation of the average opening angle over time is not significant during most of the jet head path up to the breakout [see also Appendix \ref{sec:app C} in particular equation (\ref{eq:theta_j/theta_0 app})]. In the case of a non-relativistic jet head, we roughly get $\theta_j(t)\approx \theta_0/5$. For a comparison, see figure 3 in \citet{2014ApJ...784L..28N}.}
  \label{fig:P4a-rh2} 
\end{figure}

\subsubsection{Jet head motion}
\label{subsec:jet head motion}
We track the jet head position in simulations using an algorithm that accurately detects the sharp changes; such as fluid's density and fluid's velocity (in particular sharp changes in $\beta_\theta$ and $\beta_r$); between the unshocked ejecta and the jet head.

\begin{figure*}
    \vspace{4ex}
  \begin{subfigure}
    \centering
    \includegraphics[width=0.49\linewidth]{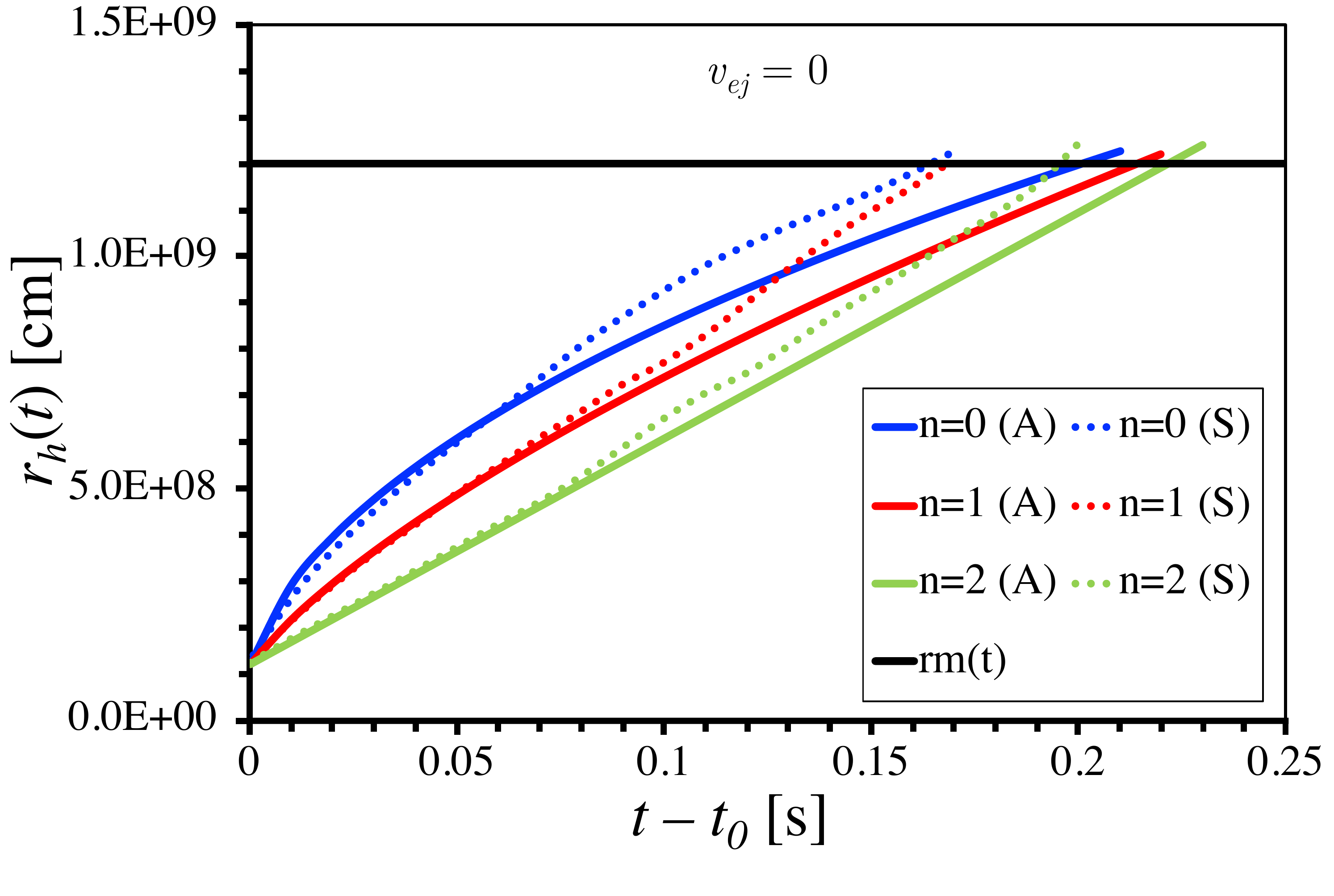} 
  \end{subfigure}
  \begin{subfigure}
    \centering
    \includegraphics[width=0.49\linewidth]{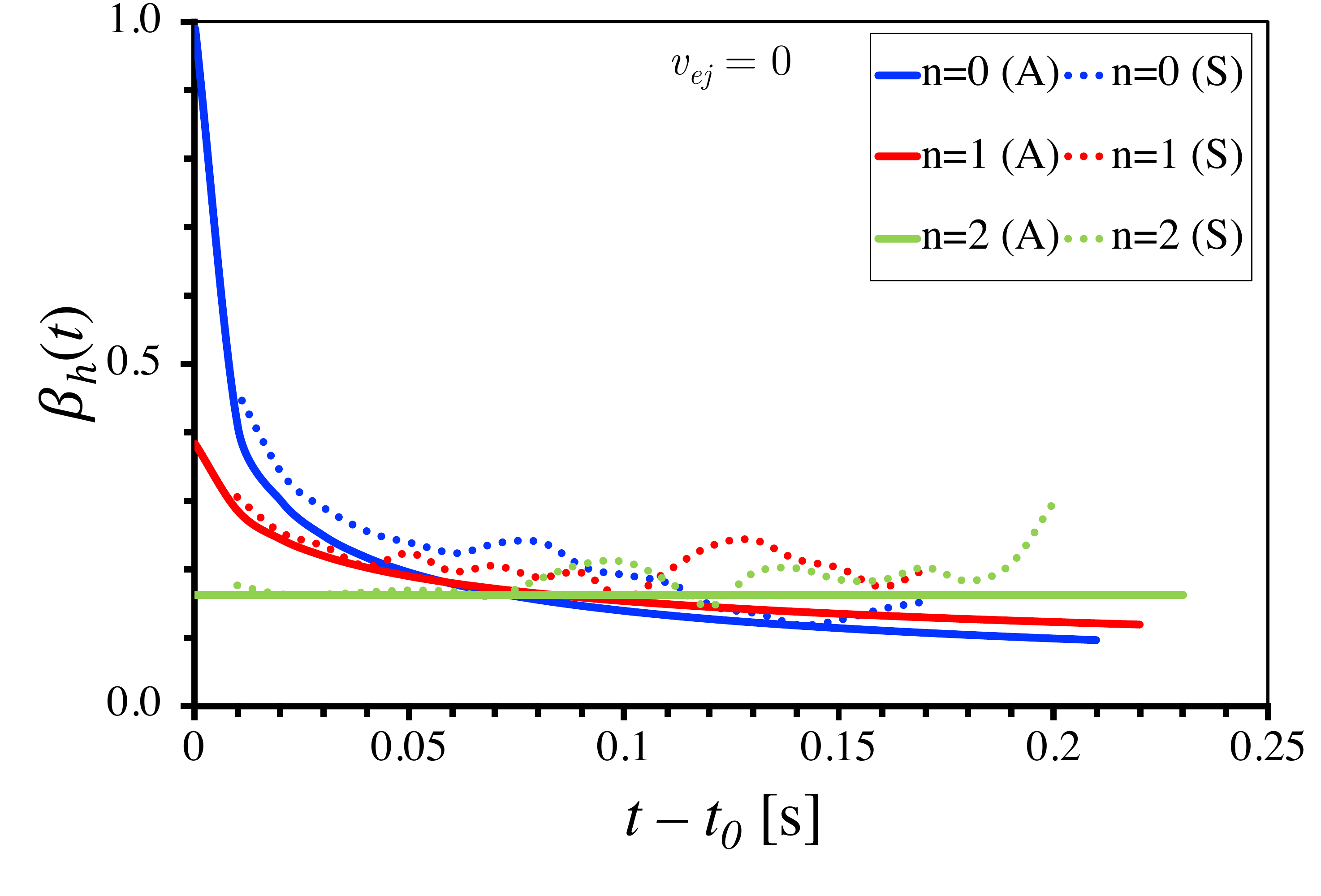} 
  \end{subfigure} 
  \begin{subfigure}
    \centering
    \includegraphics[width=0.49\linewidth]{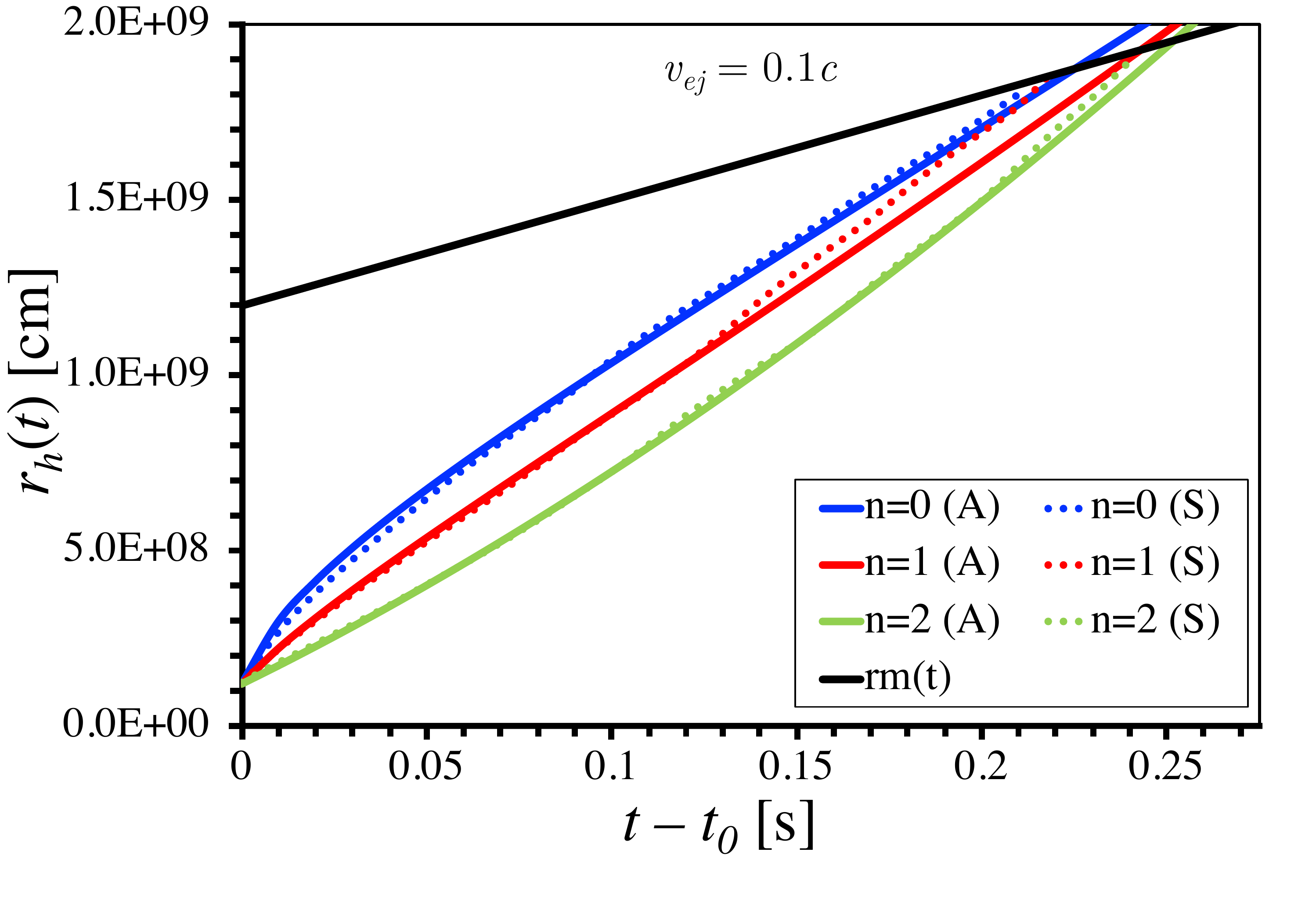} 
  \end{subfigure}
  \begin{subfigure}
    \centering
    \includegraphics[width=0.49\linewidth]{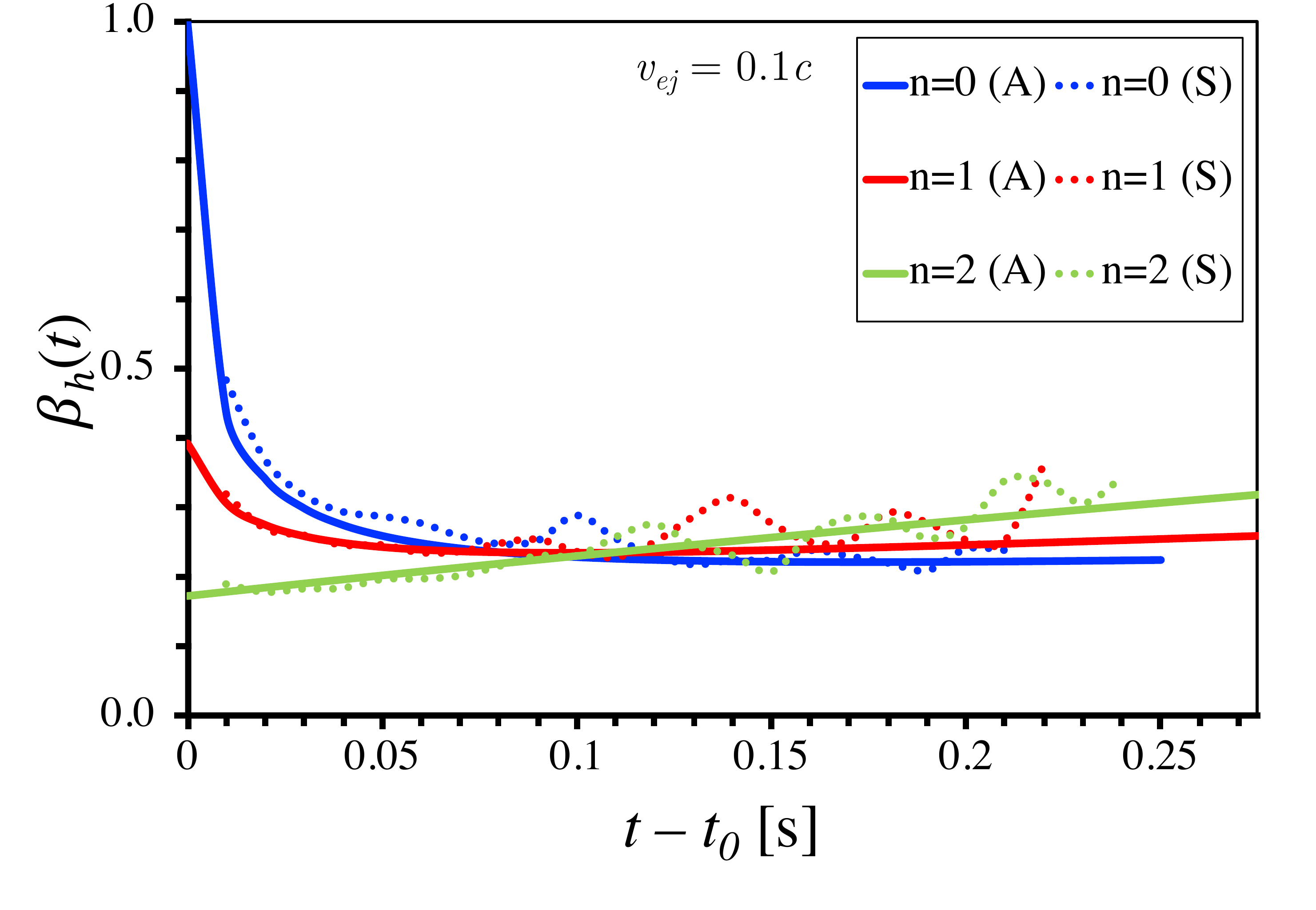}
  \end{subfigure} 
  \begin{subfigure}
    \centering
    \includegraphics[width=0.49\linewidth]{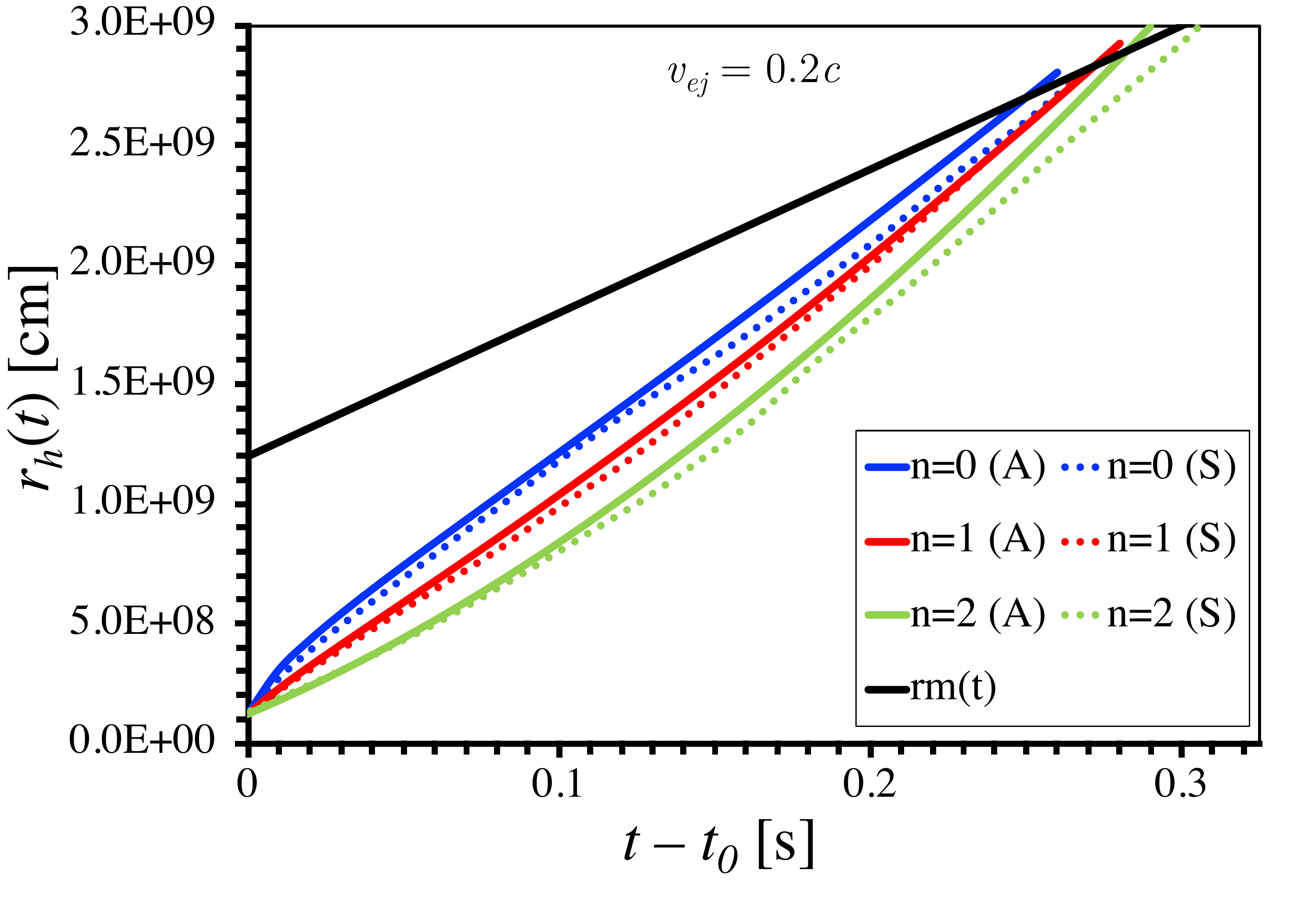}
  \end{subfigure}
  \begin{subfigure}
    \centering
    \includegraphics[width=0.49\linewidth]{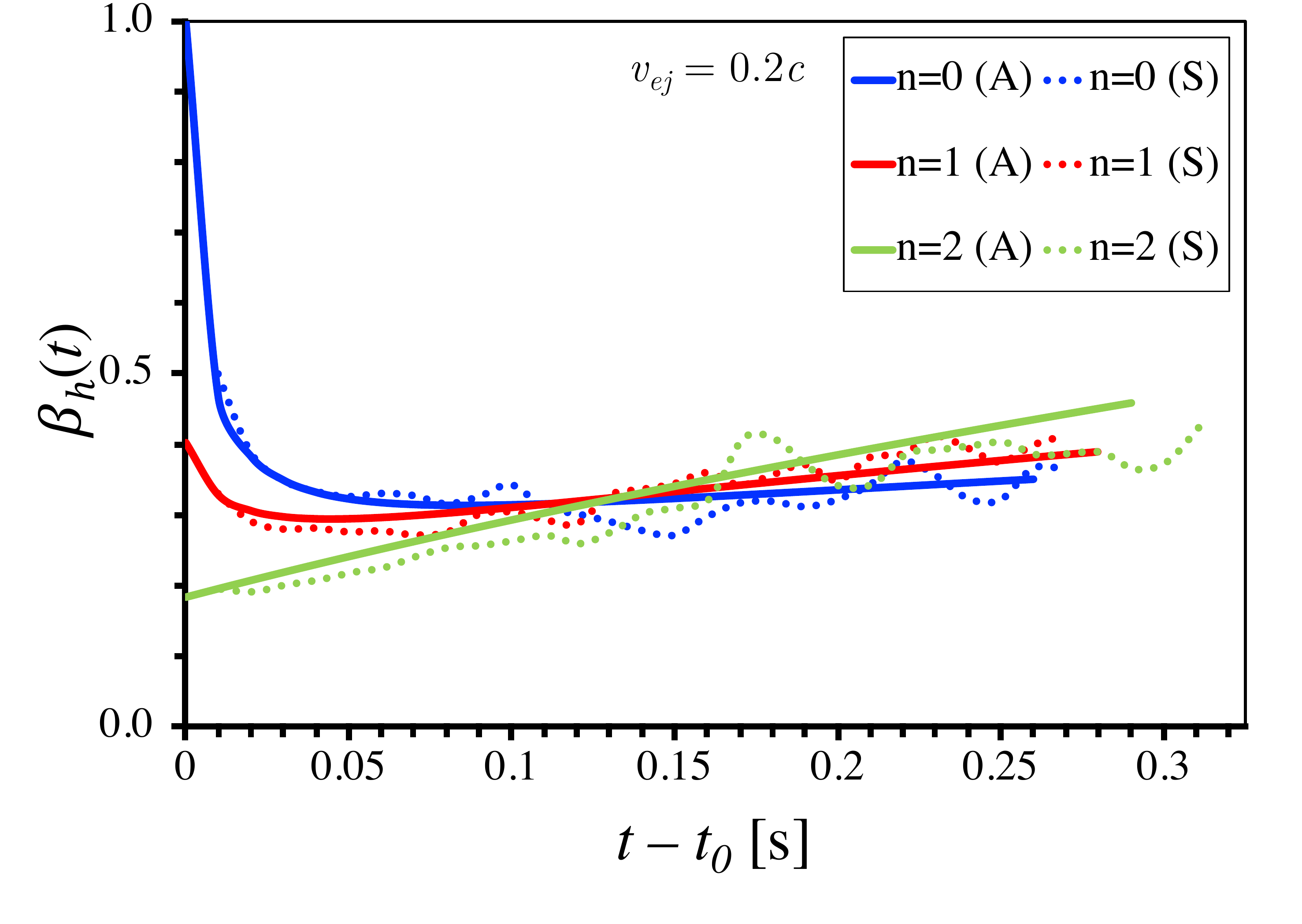}
  \end{subfigure}
  \caption{Jet head radius (left) and velocity (right) in mediums with different density profiles and different expansion velocities (``V'' models, see table \ref{table:models}). Density profile power-law indices are $n=0$ (blue), $n=1$ (red) and $n=2$ (green). Expansion velocities are $v_{ej}=0$ (top two), $v_{ej}=0.1\:c$ (middle two) and $v_{ej}=0.2\:c$ (bottom two). Solid lines [denoted with ``(A)''] show the analytic solution, colored dotted lines [denoted with ``(S)''] show results from 2D numerical hydrodynamic simulations. The black line shows the outer radius of the ejecta. Analytic lines in the top two panels are calculated using equations (\ref{eq:r stat}) and (\ref{eq:v stat}). Analytic lines in the middle and bottom panels are calculated using equations (\ref{eq:r dyn}) and (\ref{eq:v dyn}).}
  \label{fig:P4a-v012s} 
\end{figure*}

Figure \ref{fig:P4a-v012s} illustrates both analytical and numerical results for jet head motion in different environments. Analytical jet head radial position $r_h(t)$ and velocity $\beta_h(t)$ are shown for the static case ($v_{ej}=0$) and the expanding case ($v_{ej}=0.1\:c$ and $v_{ej}=0.2\:c$) and for different density profiles $n=0$, $n=1$ and $n=2$. As it can be seen in figure \ref{fig:P4a-v012s}, the agreement is good. The agreement is good for the case $v_{ej}=0$, where $r_h(t)$ shows a difference within $10-20\%$. For the other expanding cases, the agreement is very good, showing a difference of the order of a few percent. 

At late times, just before the breakout, the gap between analytical and numerical results widens in most models (see figure \ref{fig:P4a-v012s}). Our interpretation is that this difference is mainly due to the combination of two effects; the first effect leads to an underestimation and the second leads to an overestimation (of the jet head position). First, numerically, as the ejecta's outer shell expands in the very low density CSM, the gap in density results in an additional acceleration and a widening of this outermost shell of the ejecta, which affects the initial density profile around the outer radius, giving a steeper density profile. The analytical model overlooks this effect. Second, in the analytical model, from equation (\ref{eq:beta_h_1}) to equation (\ref{eq:beta_h_2}), the term $\beta_j-\beta_a$ has been approximated as $\beta_j\simeq1$ (as $\beta_j \gg \beta_a$). 
However, in the case of significantly fast ejecta ($v_{ej}\gtrsim 0.2\:c$), the approximation $\beta_j-\beta_a \simeq 1$ starts to break down, especially when the jet head gets closer to the highly expanding outer medium for $r_h(t)/r_m(t) \simeq1$. 
In addition to these two effects, there is the approximation of a constant jet opening angle (as $\theta_j\approx\theta_0/f_j$ with $f_j\approx 5$), which, although reasonable, can include an error of up to a factor of $\sim 2$ in velocity in extreme conditions (i.e. a relativistic jet case). 

Although the analytical model is well-defined for $n\sim0-3$, the models with a much steeper density profiles ($n=3$, $n=4$ and $n=5$) are compared to simulations (breakout times can be found in table \ref{table:models}). In these cases, the gap between analytical and numerical results gets to about a factor of 2, in particular for the extremely steep cases ($n=4$ and $n=5$). 
This is mainly because of the high level of collimation in the inner region, so that $\theta_j\approx\theta_0/f_j$ (with $f_j\approx 5$) is no longer a reasonable approximation. Also, in steep density profiles, it is common that the analytic model gives unphysical jet head velocities ($\beta_h\gtrsim 1$). 
Note that simulations show that collimation happens in the inner (very dense) region, and as soon as the jet head reaches outer regions, the collimated jet opens up; and in some cases the jet head accelerates and loses contact with the rear jet. Overall, we get good agreements for $n\simeq3$, with a difference within $\sim20\%$. For the other steeper models, results are questionable, especially in terms of jet head velocity.

\subsubsection{the breakout}
Here we compare the analytic breakout time [equations (\ref{eq:tb stat 1}) and (\ref{eq:tb dyn})] with the breakout time inferred from numerical simulations. 
The breakout time in numerical simulations is defined as the time when the jet head reaches the ejecta's region expanding with the maximum ejecta velocity $v_{ej}$\footnote{Note that, in numerical simulations -- as the CSM is assumed static and having a very low in density relative to the ejecta -- the outer most region of the ejecta initially expanding with the maximum velocity $v_{ej}$ accelerates slightly to exceed $v_{ej}$. This can be seen in figure \ref{fig:P4a-snap}. However, as this has little effect on the jet propagation and its breakout time, we stick to defining the breakout time as the time when the jet head reaches the ejecta's fluid expanding with $v_{ej}$.}. That is, $r_h(t_b)=r_{m}(t_b)=v_{ej}\:(t_b-t_0)+r_{m,0}$.

\begin{figure}
 \includegraphics[width=\columnwidth]{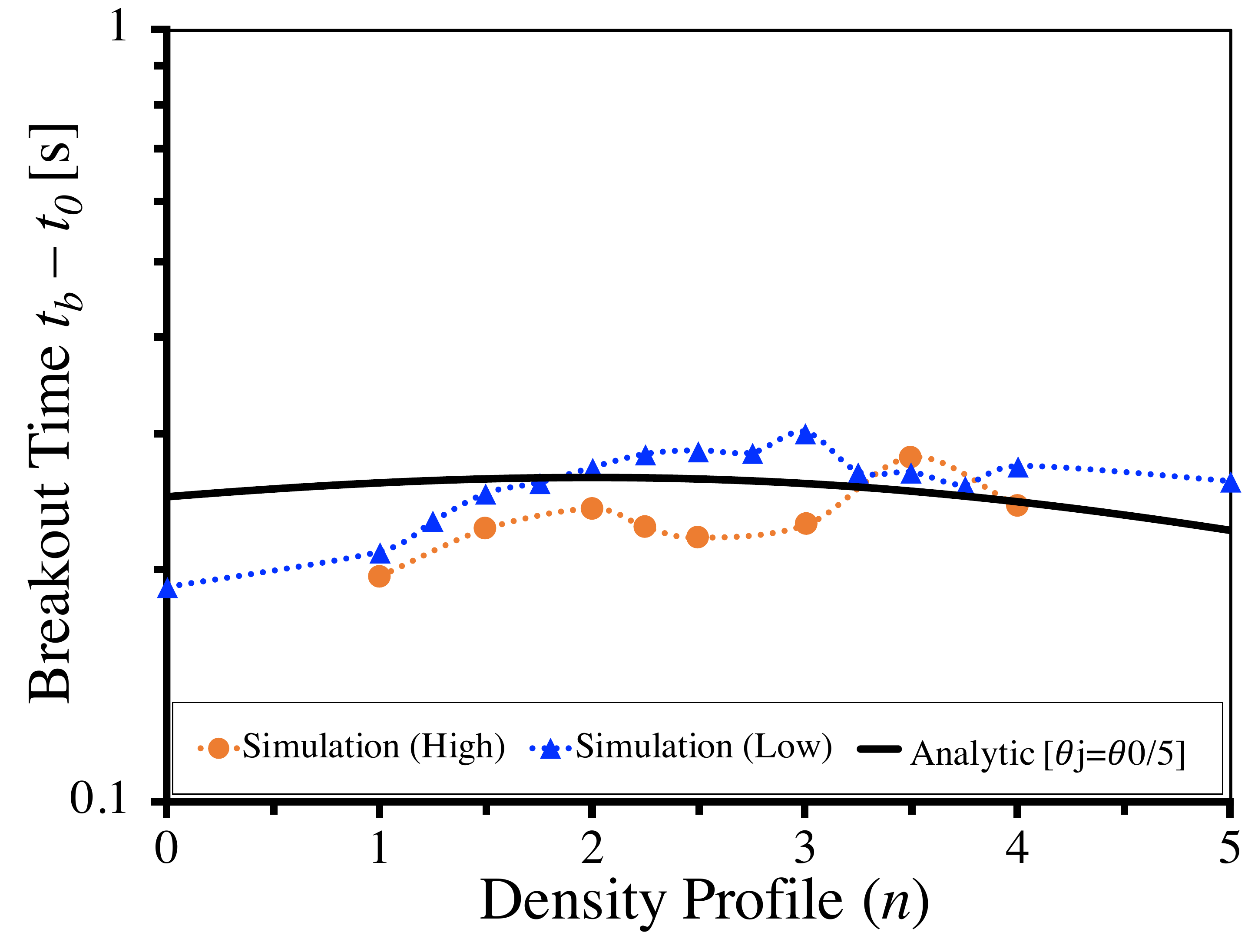}
 \caption{The breakout time for the different models (of the group ``N'') varying in the density profile index $n$. The red circles, and blue triangles are the breakout times from 2D hydrodynamical simulations, using high and low-resolutions, respectively. Dotted lines are smooth interpolations. The black solid line shows the analytic breakout time as a function of the density profile's index $n$, for $\theta_j\simeq  \theta_0/5$ [using equation (\ref{eq:tb dyn})].}
 \label{fig:P4a-tb-Ns.pdf}
\end{figure}

\begin{figure}
 \includegraphics[width=\columnwidth]{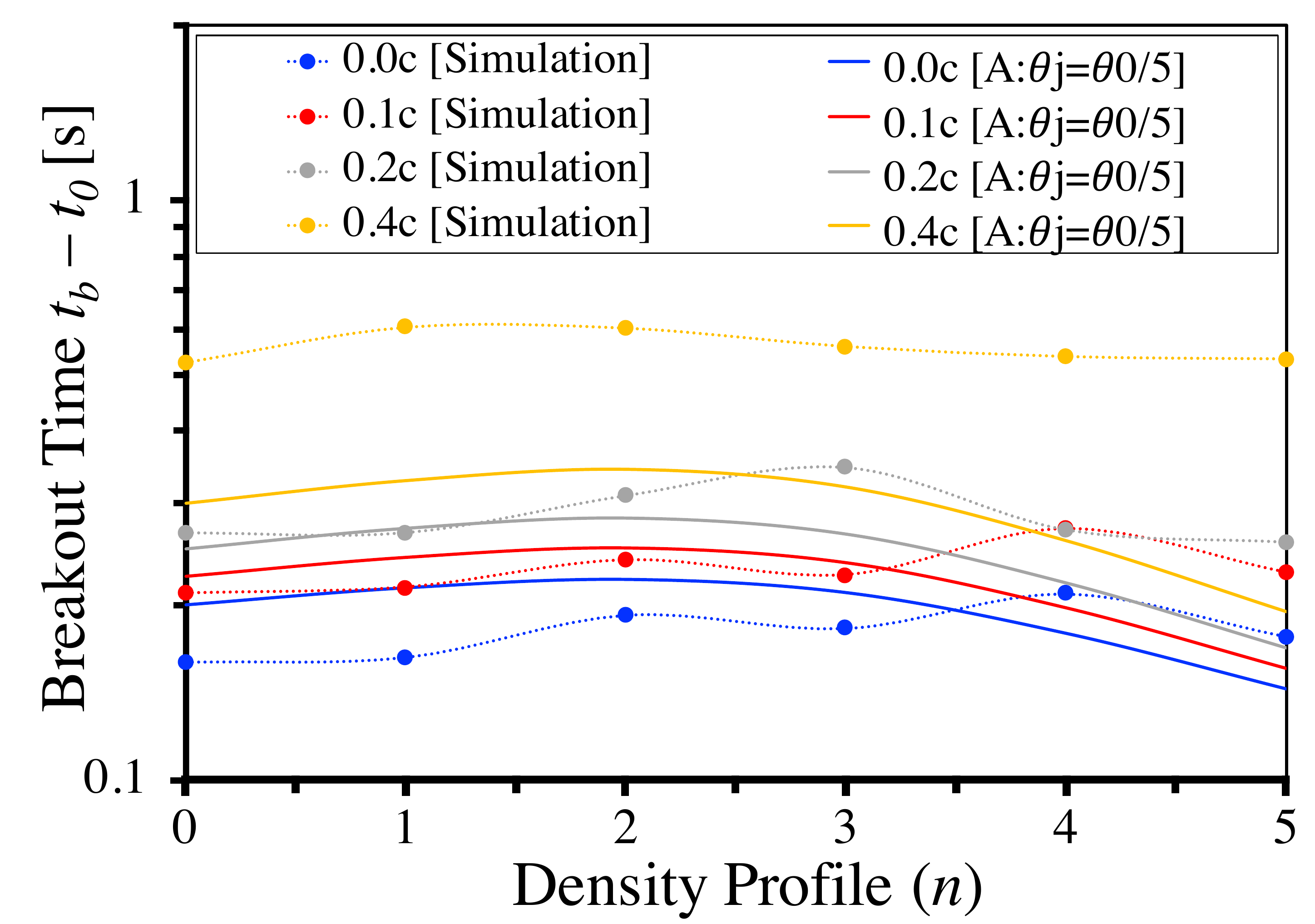}
 \caption{The breakout time for the different models (of the group ``V'') varying in the ejecta's maximum velocity $v_{ej}$ and the density profile index $n$. Circles show breakout times from the 2D hydrodynamical simulations, smoothly interpolated with dotted lines. Solid lines are analytic breakout times [using equation (\ref{eq:tb dyn}) and taking $\theta_j = \theta_0/5$]. Colors show models with different maximum velocities, blue, red, grey and orange, for $v_{ej}=$ $0$, $0.1c$, $0.2c$ and $0.4c$, respectively.}
 \label{fig:P4a-tb-v0124s.pdf}
\end{figure}

Figures \ref{fig:P4a-tb-Ns.pdf} and \ref{fig:P4a-tb-v0124s.pdf} show the results for analytical and numerical breakout times, for ``N'' models (varying in $n$) and ``V'' models (varying in $v_{ej}$), respectively (see also table \ref{table:models}). In figure \ref{fig:P4a-tb-Ns.pdf}, we see a very good match between the simulations and analytical calculations.
The agreement is along the whole domain of density profiles $0 - 5$, although both low and high $n$ limits show less impressive agreement. Also, the close $t_b$ values found in low and high-resolution calculations suggest that the resolutions in our simulations should be acceptable. Also, in this parameter space, both analytical and numerical breakout times show small variations for different $n$. 

Figure \ref{fig:P4a-tb-v0124s.pdf} shows breakout times for a variety of expansion velocities (group ``V''; using $\theta_j\approx \theta_0/5$). 
First, let's consider models with $v_{ej}=0$, $0.1c$, and $0.2c$. In the domain $n=0-3$, we see a very good agreement (overall a difference of $\sim15-20\%$). However, for $n=4-5$ the gap between analytic and numerical values widens (overall a difference of $\sim25\%$ but even up to $\sim50\%$). This is due to the same effects discussed in $\S$ \ref{subsec:jet head motion} which tends to give analytically high jet head velocities. 
Second, for models with $v_{ej}=0.4c$, there is a large gap (factor $\sim 2$ difference). As previously discussed in $\S$ \ref{subsec:jet head motion}, unless $v_{ej} \lesssim 0.4\:c$, the approximation of assuming $\beta_j-\beta_a \simeq 1$ starts to break down (see Appendix \ref{app:exact solution n=2} for a more rigorous calculation).

\begin{figure}
 \includegraphics[width=0.99\linewidth]{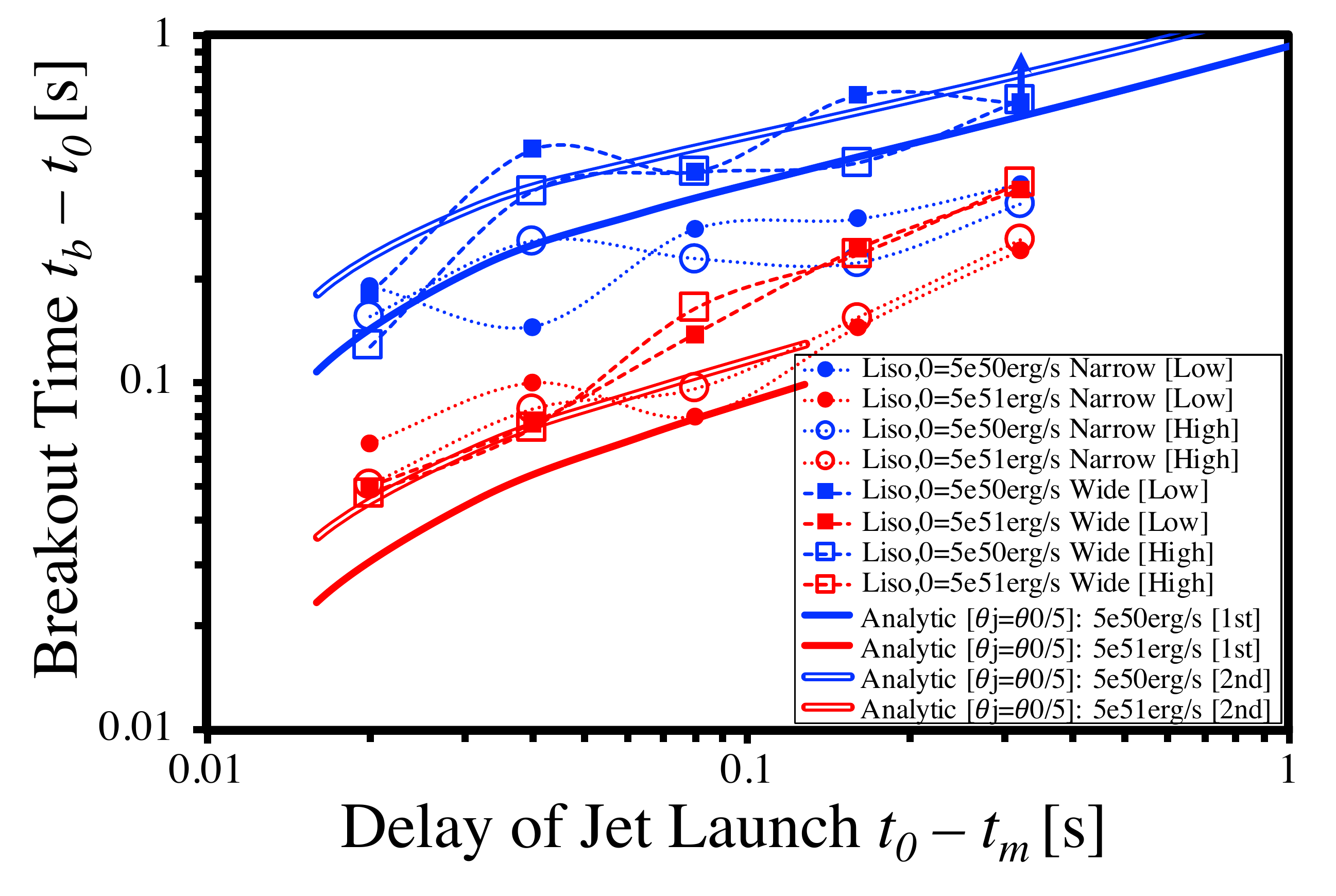}
 \caption{The breakout time for the different models (of the group ``T'' in table \ref{table:models}) varying in the delay time $t_0-t_m$, and also in the jet isotropic luminosity $L_{iso,0}$ and the jet opening angle $\theta_0$. Circles (blue, red, filled and open) and Squares (blue, red, filled and open) are for models with a narrow and wide jet, respectively ($\theta_0=6.8^\circ$ and $\theta_0=18^\circ$, respectively). Blue and red are for models with $L_{iso,0}=5\times10^{50}$ erg s$^{-1}$ and $L_{iso,0}=5\times10^{51}$ erg s$^{-1}$, respectively. Filled and open symbols are for calculations using low and high-resolutions, respectively. Dotted and dashed lines show smooth interpolations. Solid lines are analytic breakout times for $\theta_j=\theta_0/5$ [using equation (\ref{eq:tb dyn})]. The double solid lines show more accurate analytic breakout times [using equation (\ref{eq:tb exact})].}
 \label{fig:P4a-tb-Ts.pdf}
\end{figure}

We also compare the analytic and numerical breakout times for the ``T'' group. Figure \ref{fig:P4a-tb-Ts.pdf} shows the breakout times for all models, different not only in $t_0-t_m$, but also in jet power, opening angles, and resolution. We notice that: i) numerical breakout times vary around the analytic breakout times calculated using $\theta_j\approx\theta_0/5$ (by $\sim \pm30\%$ overall and up to $\sim100\%$ in the extreme case); ii) our low and high-resolution calculations give similar results suggesting that the resolution is high enough; and iii) the more accurate analytic solution [calculations are presented in Appendix \ref{app:exact solution n=2}, equation (\ref{eq:tb exact})] gives a better agreement, especially for wide jet models (where the difference is reduced from $\sim100\%$ to $\sim80-90\%$). 
Note that in simulations with $L_{iso,0}=5\times10^{51}$ erg s$^{-1}$, wide opening angle, and a large $t_0-t_m$, we notice accumulation of heavy ejecta matter in the jet head [a plug; see: \citet{2001ApJ...550..410M}; \citet{2004ApJ...608..365Z}; \citet{2013ApJ...767...19L}]. This artificial numerical issue exclusive of 2D simulations is one reason for the long breakout times.

Finally, an additional test of our analytic breakout times is to make a comparison with other breakout times in simulations found in the literature. We compare with breakout time in \citet{2014ApJ...784L..28N}. Table \ref{table:Comparison to other works.} show the comparison. For most calculations, our simple analytic breakout time assuming $\theta_j \approx \theta_0 /5$ (with a calibration coefficient of $N_s= 2/5$) is very reasonable, and gives very similar results to independently carried out simulations (difference of  $\sim 20 - 50\%$ for most models).

\begin{table*}
 \caption{Comparison of analytic breakout times $t_b-t_0$ [derived using equation (\ref{eq:tb dyn})] with breakout times from simulations in \citet{2014ApJ...784L..28N}.}
 \label{table:Comparison to other works.}
 \begin{tabular}{llllllllll}
  \hline
         & $M_{ej}$    &  $n$  & $L_j$ & $\theta_0^*$ &  $r_0$  & $r_{m,0}$ & $v_{ej}$ &  Simulation: & Analytic: ($\theta_j=\theta_0/5$)\\ 
  Models & [$M_\odot$] &       &  [erg s$^{-1}$] &[deg]& [cm] & [cm]    & [$c$]      & $t_b-t_0$ [s] & $t_b-t_0$ [s]\\
    \hline
    M-ref & $10^{-2}$          & 3.5 & $2\times10^{50}$ &	27	&	$1.2\times10^8$	& $6.1\times10^8$ & 0.4 &	0.231 & 0.222\\
    M-L4 & $10^{-2}$        & 3.5 & $4\times10^{50}$ &	27	&	$1.2\times10^8$	& $6.1\times10^8$ & 0.4 & 0.195 & 0.139\\
    M-th30 & $10^{-2}$        & 3.5 & $2\times10^{50}$ &  42	&	$1.2\times10^8$	& $6.1\times10^8$ & 0.4 & 0.626 &	0.422\\
    M-th45 & $10^{-2}$        & 3.5 & $2\times10^{50}$ &	57	&	$1.2\times10^8$	& $6.1\times10^8$ & 0.4 &  -    &  0.676\\
    M-ti500	& $10^{-2}$        & 3.5 & $2\times10^{50}$ &	27	&	$5.6\times10^8$	& $60.1\times10^8$& 0.4 & 0.899 &	0.506\\
    M-M3	& $10^{-3}$        & 3.5 & $2\times10^{50}$ &	27	&	$1.2\times10^8$	& $6.1\times10^8$ & 0.4 &	0.105 & 0.051\\
    M-M2-2 & $2\times10^{-2}$ & 3.5 & $2\times10^{50}$ &	27	&	$1.2\times10^8$	& $6.1\times10^8$ & 0.4 & 0.320 &	0.366\\
    M-M1 & $10^{-1}$        & 3.5 & $2\times10^{50}$ &	27	&	$1.2\times10^8$	& $6.1\times10^8$ & 0.4 &	0.750 & 1.306\\
     \hline
 \end{tabular}
 \vspace{1ex}
 
     \raggedright {\footnotesize
     
     $^*$ The initial opening angle is defined as: $\theta_0\approx\theta_{inj}+1/\Gamma_0$, not to be confused with $\theta_{inj}$.
     }
\end{table*}

\section{Numerical relativity simulations and the dynamical ejecta}
\label{sec:4.NR}

\begin{figure*}
    \vspace{4ex}
  \begin{subfigure}
    \centering
    \includegraphics[width=0.495\linewidth]{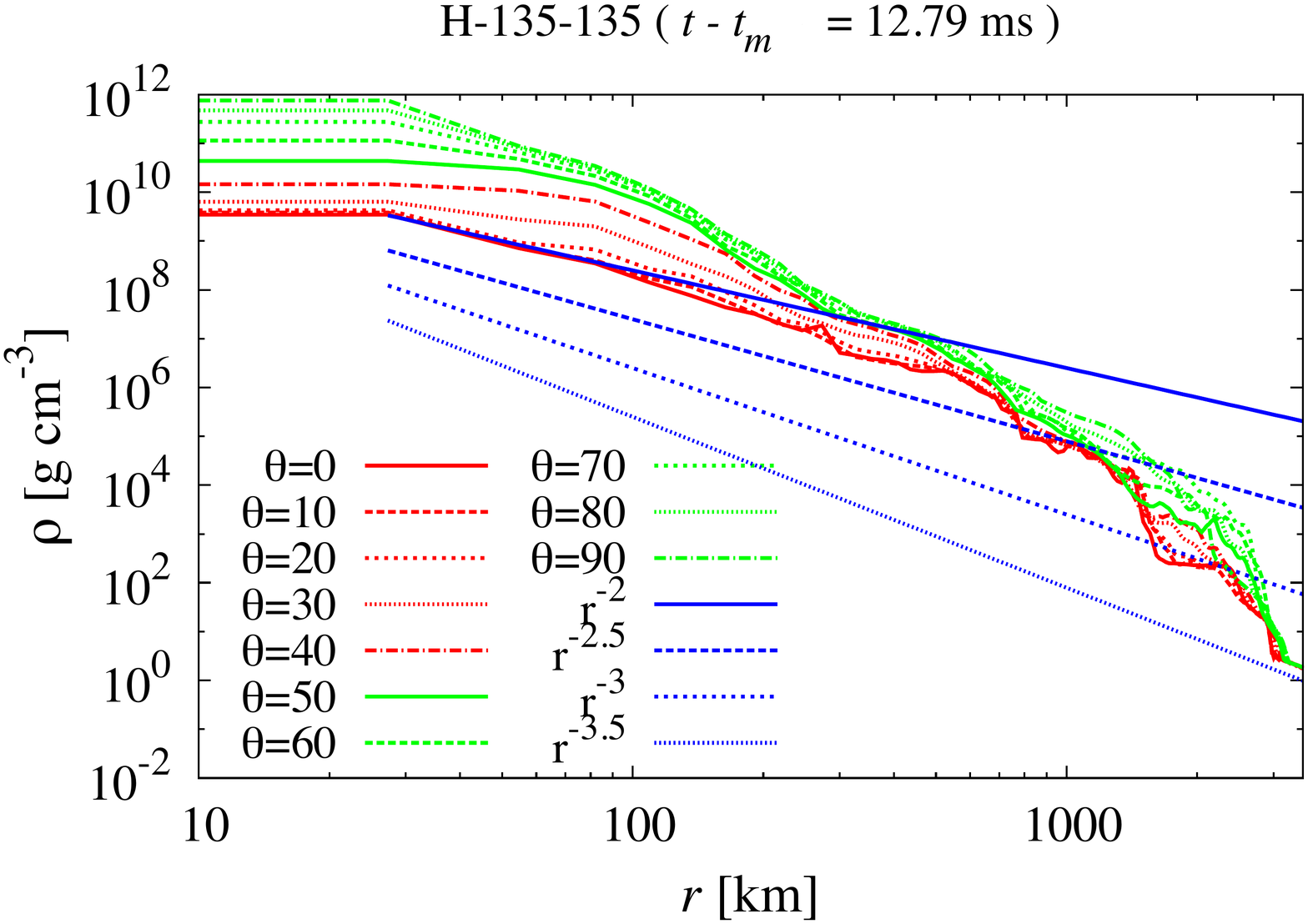} 
  \end{subfigure}
  \begin{subfigure}
    \centering
    \includegraphics[width=0.495\linewidth]{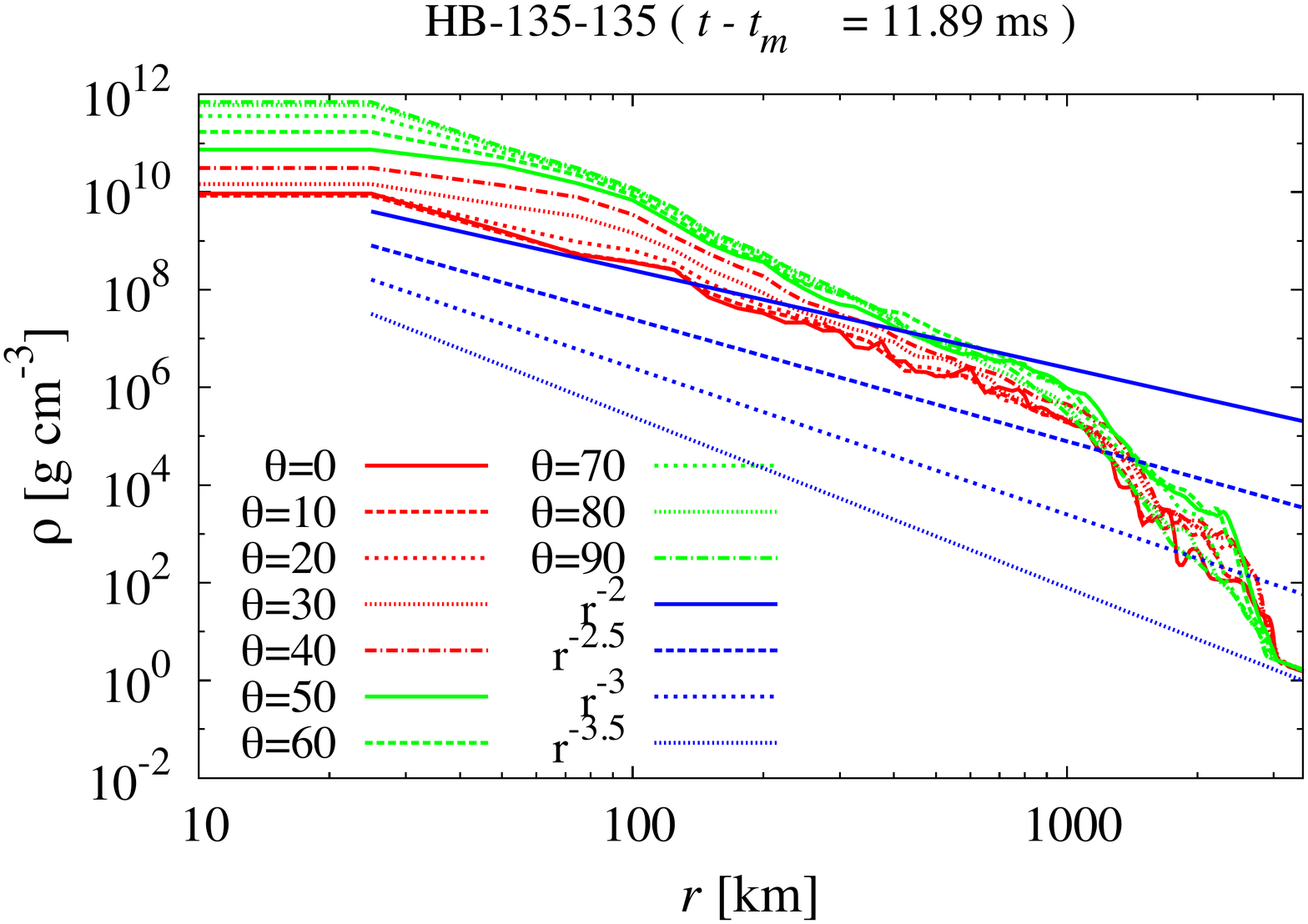} 
  \end{subfigure} 
  \begin{subfigure}
    \centering
    \includegraphics[width=0.495\linewidth]{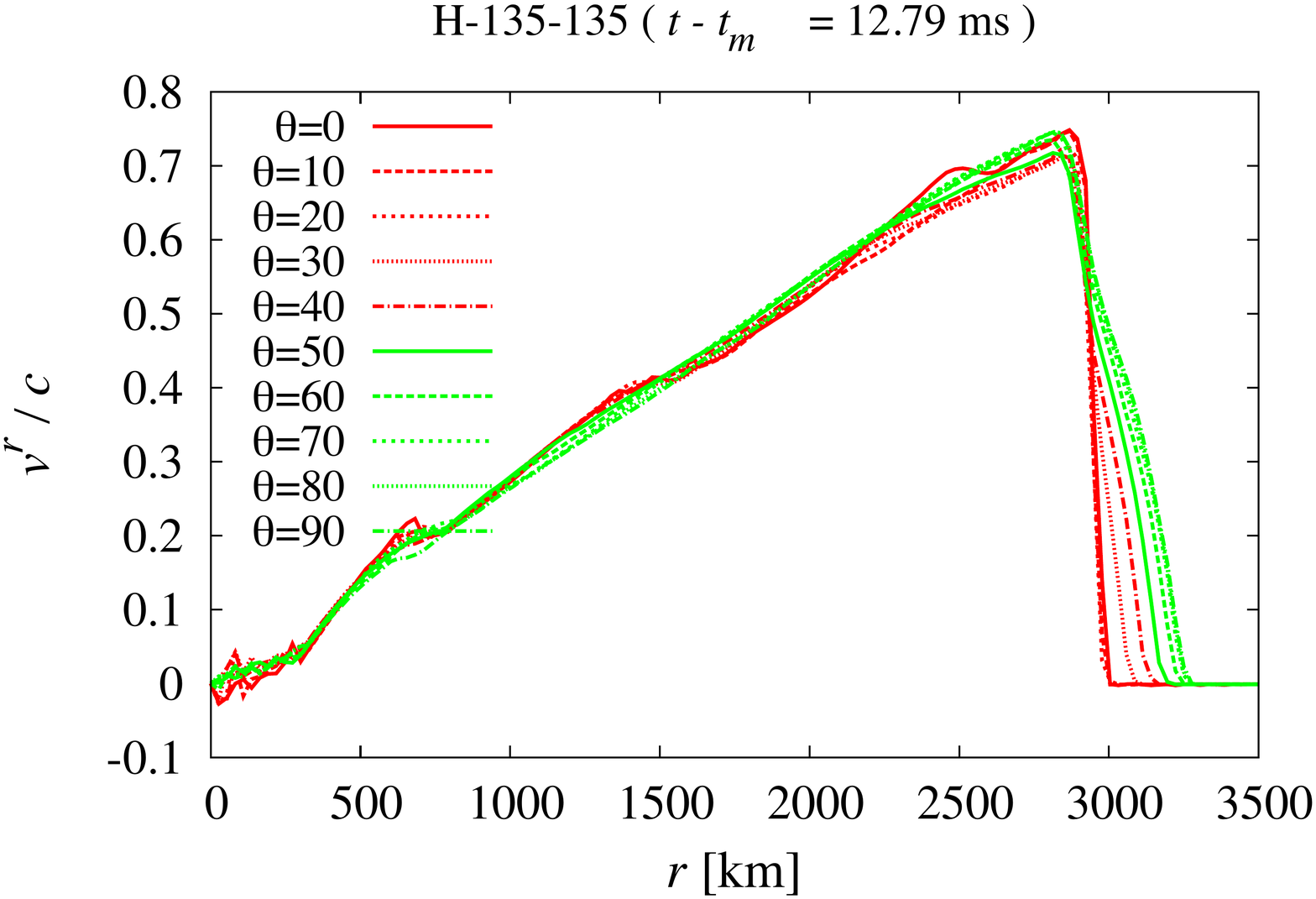} 
  \end{subfigure}
  \begin{subfigure}
    \centering
    \includegraphics[width=0.495\linewidth]{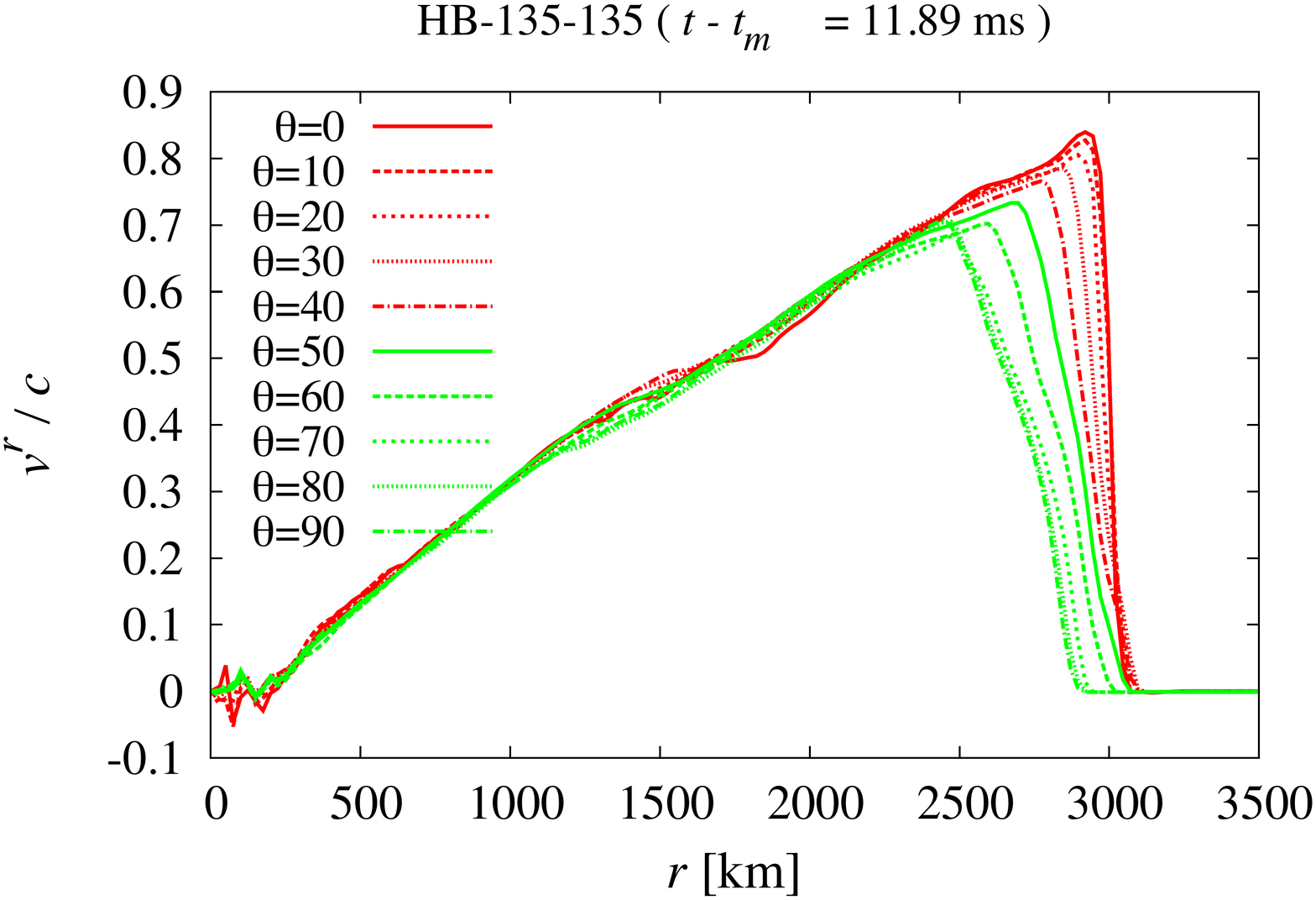} 
  \end{subfigure} 
  \caption{Results from very high-resolution numerical relativity simulations of a BNS merger by \citet{2017PhRvD..96h4060K}. The BNS system is of $1.35M_\odot-1.35M_\odot$. Two models with different equation of state are presented: H-135-135 (at $t-t_{m}=12.79 $ ms; where $t_{m}$ is the time when the gravitational-wave amplitude reaches its peak) and HB-135-135 (at $t-t_{m}=11.89 $ ms). Top two panels show the density profiles and the bottom two panels show the velocity profiles. For more information about the parameters of the two models H-135-135 and HB-135-135, please refer to tables 1 and 2 in \citet{2017PhRvD..96h4060K}.}
  \label{fig:H_HB_135_135} 
\end{figure*}

The assumed ejecta profiles for density and velocity vary in the different studies on jet propagation. One very commonly adopted density profile is, as in \citet{2014ApJ...784L..28N}, $\rho_a(r)\propto r^{-3.5}$. 
In others studies, the adopted index differs (e.g. $n=2$ in \citealt{2018MNRAS.479..588G}; \citealt{2018ApJ...863...58X}; \citealt{2017ApJ...848L...6L}). For the velocity profile, the adopted maximum velocity $v_{ej}$ in different studies varies. \citet{2014ApJ...784L..28N} took the maximum ejecta velocity as $v_{ej} = 0.4$c. In most other studies, this value is typically taken as $0.2c$

Here we determine these important parameters by going back to numerical relativity simulations.
We present results from numerical relativity simulations with the highest resolution to date \citep{2017PhRvD..96h4060K}.
We focus on a merger event similar to GW170817, where the total mass of the binary is $\sim 2.7M_\odot$ \citep{2017PhRvL.119p1101A}.
Considering GW170817, there is uncertainty on the mass ratio of the two NSs of the binary ($q$), and an asymmetric BNS system is possible (i.e. $q<1$).
However, for simplicity, we only consider the case of an equal mass BNS system (i.e. $q=1$), where each NS has a mass of $1.35 M_\odot$.
Note that, depending on this mass ratio, the dynamical ejecta can be largely different \citep{2019arXiv190800655V}.
Hence, we stress that the density profile presented here is limited to the particular case of a symmetric BNS system.

Also, another simplification is that we consider jet propagation in a medium composed of the ``dynamically'' ejected mass, only.
In reality, other mass ejection mechanisms are possible, even within the short timescale ($<1.7$ s in the case of GW170817) before the jet launch.
In particular, the viscous outflow from the central region may contribute substantially to the total mass of the ejecta, its density profile, and its angular distribution (\citealt{2018ApJ...860...64F}).

We analyze the produced dynamical ejecta during the first few milliseconds after the merger. 
We present the results in figure \ref{fig:H_HB_135_135}.
We show two models, each with a different EOS. 
With EOS H, the radius of each of the two NSs is $12.27$km; while with EOS HB, the radius is $11.61$km (see \citealt{2017PhRvD..96h4060K} for more details). 

\subsection{Density profile of the ejecta}
\label{sec:Density profile of the ejecta}
The top two panels in figure \ref{fig:H_HB_135_135} show the density profile at $t-t_m \approx 12$ ms, for the models H-135-135 and HB-135-135 (see \citealt{2017PhRvD..96h4060K}). Both models show an overall similar density profile. In the inner region of the ejecta, the density profile can be approximated to a power-law with an index $n\approx2$ (i.e., $\rho(r)\propto r^{-n}$). In the outer region of the ejecta expanding at $\sim 0.2c-0.3c$, the density profile steepens to $n\approx3.5$. And for the fast expanding outer region -- the fast tail -- the density profile is even steeper with $n\sim 5 - 6$. However, since the mass of the fast tail is very small, it can be neglected as long as we focus on jet propagation. 

Ideally, the density profile of BNS merger ejecta should be a combination of two (or more) power-law functions, one power-law with an index $n_1\sim 2$ to fit the inner region and a steeper power-law with an index $n_2\sim 3.5$ to fit the outer region. However, we know from the analytic results that the overall jet propagation (i.e. the breakout time) is not dramatically affected by variation in the density profile's index $n$ [see equation (\ref{eq:tb dyn}) or equations (\ref{eq:tb dyn sim n<3}) and (\ref{eq:tb dyn sim n>3})]; simulations confirm this point by showing that the breakout time varies weakly for models with different $n$ (see figure \ref{fig:P4a-tb-Ns.pdf}, or the breakout times for group ``N'' in table \ref{table:models}; also see figure \ref{fig:P4a-tb-v0124s.pdf}). Hence, the whole density profile of the ejecta can be approximated to a single power-law function with a power-law index $n\sim 2 - 3.5$. Therefore, for the case of GW170817, we will consider the ejecta's density profile as a power-law with an index $n=2$. 

\subsection{Velocity profile of the ejecta}
\label{sec:Velocity profile of the ejecta}
The velocity profile in figure \ref{fig:H_HB_135_135} (bottom two panels) at $t-t_m\approx12$ ms after the merger shows a homologous behavior $v_a(r)\propto r$. 
Other numerical relativity simulations in the literature present the average velocity of the ejecta (see table 4 in \citealt{2013PhRvD..87b4001H}; table 1 in \citealt{2013ApJ...773...78B}; table 1 in \citealt{2015MNRAS.448..541J}; table 1 in \citealt{2017CQGra..34j5014D}; table 2 in \citealt{2018ApJ...869..130R}; and for a summary see figure 1 in \citealt{2019arXiv190109044S}). Overall, the average velocity of the ejecta (including the bound part) is found as $\sim0.2$c. Hence, we adopt a linear velocity profile throughout the ejecta, with $(0.2{c})^2=(1/M_{ej})\int v_a(r)^2\rho_a(r) dV$; hence, for $n=2$, the maximum velocity is adopted as $v_{ej}=\sqrt{3} \times 0.2c \simeq 0.346c$.

As pointed out in $\S$ \ref{sec:Density profile of the ejecta}, a fast tail is visible in the outer part of the ejecta. 
The fast tail is very sensitive to the EOS, hence there are large uncertainties on its properties. 
Note that, in comparison to the whole dynamical ejecta, in terms of mass and energy, the fast tail is several orders of magnitude weaker than the ejecta. 
Also, the fast tail is about one order of magnitude less energetic than the cocoon ($E_c\sim10^{49}-10^{50}$ erg, Hamidani et al. 2019 in preparation), and its very low mass gives a very short diffusion time. 
Hence, its presence would not dramatically affect the jet propagation, nor the cocoon and its EM counterparts at later times; although it can be relevant to the prompt emission \citep{2018MNRAS.475.2971B}, and to the early afterglow emission (\citealt{2014MNRAS.437L...6K}; \citealt{2018ApJ...867...95H}). 
Therefore, we neglect the entire fast tail part of the ejecta.

\subsection{Angular dependence and the mass of the ejecta $M_{ej}$} 
\label{sec:Angular dependance}
Numerical relativity simulations' results, as in figure \ref{fig:H_HB_135_135}, show clearly that the dynamical ejecta is not spherical; the density in the equatorial region is higher than the density along the polar axis.  
We know from follow-up observations of GW170817 that the jet opening angle is small ($\theta_c \approx 4^\circ$; \citealt{2018Natur.561..355M}; \citealt{2018arXiv180806617T}; \citealt{2019Sci...363..968G}), this implies that the jet propagated in the polar region of the ejecta.
Since the density in the equatorial region is not relevant to jet propagation, and since the scope of this study is limited to jet propagation, we do not include the excess in density in the equatorial region (relative to the polar region) in our calculation for GW170817's jet propagation. 
In other words, we assume a spherically symmetric polar density profile, for simplicity.
Therefore, for an application to GW170817, and by considering a fiducial value for the dynamical ejecta mass, as $0.01 M_\odot$, we adopt in our calculations an effective ejecta mass $M_{ej} = 0.002 M_\odot$\footnote{
The reduction in mass is by a factor $\sim 4-5$, which is determined by the difference between the polar density and the equatorial density, in the region where most of the ejecta mass is contained. In the case of the ejecta results presented in figure \ref{fig:H_HB_135_135}, this region is located within a radius of $\sim 1000$ km (which is the radius at which the density profile steepens); and the difference between the polar density and the equatorial density is roughly one order-of-magnitude. }.

Note that, in the case of a highly asymmetric BNS system the angular distribution is much different, because the mass contribution of the tidal component of the dynamical ejecta is substantially higher (\citealt{2019arXiv190703790K}; \citealt{2019arXiv190800655V}). Hence, our estimation of $M_{ej}$ here should be understood as limited to the case of a symmetric BNS system.

\section{Application to GW170817}
\label{sec:GW170817}

Here we apply our numerical modeling to GW170817/\textit{s}GRB 170817A. 
Our strategy here is to combine all robust information on GW170817/\textit{s}GRB 170817A from observations and numerical relativity simulations. 
Then we use our analytic model in order to isolate key parameters of the central engine which are not well-known: the engine power $L_{iso,0}$ and the timescale of the engine activation relative to the merger time $t_0-t_m$.

\subsection{Key information from numerical relativity simulations}

We presented data from numerical relativity simulations in $\S$ \ref{sec:4.NR}. Accordingly, the key parameters are fixed as follows: $n=2$, $v_{ej} = 0.346\:c$, and $M_{ej}=0.002 M_{\odot}$ (refer to $\S$ \ref{sec:Density profile of the ejecta}; $\S$ \ref{sec:Velocity profile of the ejecta}; and $\S$ \ref{sec:Angular dependance}; respectively).   
With $r_0\approx10^6$ cm around the vicinity of the event horizon of a BH with $2.7 M_\odot$, the only remaining unknowns for GW170817/\textit{s}GRB 170817A are the engine parameters: $L_j$, $\theta_0$, and $t_0-t_m$ (or $L_{iso,0}$ and $t_0-t_m$). 

\subsection{Key information from observations}
\subsubsection{Prompt GRB observations: Delay time}
Electromagnetic detection by Fermi shows a delay of $\sim1.7$s later than GW observations by LIGO \citep{2017ApJ...848L..13A}. 
This delay should be the results of three delays: delay due to the engine activation $t_0-t_m$, delay in the jet breakout $t_b-t_0$, and delay in the release of EM radiation from the jet. Therefore, this piece of information can be useful to constrain $t_0-t_m$ and $L_{iso,0}$ (to which $t_b-t_0$ depends). 

\subsubsection{Afterglow observations: Jet energy and opening angle}
\label{subsec:Late observations}
Late afterglow observations of GW170817 provided an estimation of the jet (and cocoon) energy and its final opening angle (\citealt{2018Natur.561..355M}; \citealt{2018arXiv180806617T}; \citealt{2019Sci...363..968G}). Radio observations also determined the afterglow's peak time at $\sim 150$ days, and the flux at this time. Lorentz factor could be inferred from the superluminal motion as $\Gamma\gtrsim4$. 
Then, the isotropic equivalent energy can be found as $E_{iso}\approx3\times10^{51}-10^{53}$erg for reasonable ranges of $\epsilon_e$ and $\epsilon_B$ (for more details, please refer to \citealt{2018Natur.561..355M}). As far as our study is concerned, the conclusion is that the late time jet energy should satisfy $E_{iso}\leqslant 10^{53}$ erg.\footnote{We do not consider the lower limit on energy because it is possible that, after the prompt emission, later engine activity did inject energy, which could have contributed to the afterglows.}

From the afterglow observations, the final opening angle of the jet was constrained as $\theta_f=3.38^{+0.974}_{-0.974}$ degree (table 2 in \citealt{2018arXiv180806617T}). Other studies suggested similar results: $\theta_f<5^\circ$ in \citet{2018Natur.561..355M}, $\theta_f=3.4^\circ\pm 1^\circ$ in \citet{2019Sci...363..968G}, and $\theta_f=6.3^{1.1}_{0.6}$ degree for GW170817 (and $\theta_f=6.9^{2.3}_{2.3}$ degree for cosmological \textit{s}GRBs) in \citet{2019ApJ...880L..23W}. Taking into account these estimations, we set the jet opening angle for GW170817, where $\theta_f$ varies within two extremes: i) narrow jet case $\theta_{f,narrow}=3.4^{\circ}$, and ii) wide jet case $\theta_{f,wide}=9.0^\circ$. 

Given our assumption $\theta_f\approx\theta_0/2$ (see $\S$ \ref{sec:Simulated models}), the initial opening angle of the jet takes values between two extremes: $\theta_{0,narrow}\approx6.8^{\circ}$ and $\theta_{0,wide}\approx18^{\circ}$; hence our choice of $\theta_0$ for the ``T'' group models\footnote{Note that considering much narrower jets $\theta_0<6.8^\circ$ is numerically challenging.}.
\subsection{Results}
\label{sec:Results}
Figure \ref{fig:P4a-L=dt} presents our constraints on the central engine of GW170817. The engine isotropic luminosity $L_{iso,0}\approx 4L_j/\theta_0^2$, and the delay between the merger (at $t_m$) and the engine activation (at $t_0$), are unknown parameters. 

If we assume that the jet launch is triggered by the collapse of the HMNS, the timescale $t_0-t_m$ should be comparable to the lifetime of the HMNS. 
The lifetime of the HMNS has been constrained in several studies (\citealt{2018ApJ...856..101M}; \citealt{2018ApJ...860...64F}; etc.), and with the above assumption such constraints could be translated into the timescale $t_0-t_m$ (see the green lines in figure \ref{fig:P4a-L=dt}). 
\citet{2018ApJ...856..101M} proposed a scenario to explain the ejecta composition (in particular the electron faction $Y_e$) as inferred from the blue macronova. 
In this scenario a fast rotating and highly magnetized HMNS (i.e., a magnetar) survives for a timescale $\sim 0.1 - 1$s  (see figure 2 and $\S$ 4 in \citealt{2018ApJ...856..101M}).
\citet{2018ApJ...860...64F} considering the viscous wind ejecta, explained that a long-lived HMNS is needed to assist the launch of ejecta high in electron fraction ($Y_e$) in the polar region, so that the blue macronova of GW170817 can be explained. 
With a stiff EOS, it is possible that the HMNS survives for longer timescale, comparable to the timescale of neutrino cooling $\sim 10$s (see figure 11; table 3 and $\S$ 4 in \citealt{2018ApJ...860...64F}; also see \citealt{2017PhRvD..96l3012S}). 

Other constraints on the timescale $t_0-t_m$ can be found in the literature (\citealt{2018ApJ...866L..16M}; \citealt{2019ApJ...877L..40G}; etc.).

\subsubsection{Maximum $L_{iso,0}$ as inferred from afterglow observations [in Orange]}
\label{sec:orange}
First, we show the upper limit for the engine isotropic luminosity $L_{iso,0}$ (orange solid line). 
There are three luminosities involved: the luminosity of the engine $L_{iso,0}\simeq 4L_j/\theta_0^2$; the luminosity of the jet after the breakout (which has a reduced opening angle from $\theta_0$ to $\theta_f\approx \theta_0/2$; see $\S$ \ref{sec:Simulated models}) giving $L_{iso,f}\simeq 4L_j/\theta_f^2\approx 4L_{iso,0}$; and the luminosity of jet after radiation has been released $L_{iso}\approx L_{iso,f}\times(1-\epsilon_{rad})$. Afterglow observations did put a limit on the late time jet energy as $E_{iso}\leqslant 10^{53}$ erg ($1\sigma$ upper limit; see $\S$ \ref{subsec:Late observations}; for more details, refer to \citealt{2018Natur.561..355M}; \citealt{2018arXiv180806617T}; \citealt{2019Sci...363..968G}). With the duration of GW170817 being $\sim 2$ s, and assuming a radiative efficiency of $\epsilon_{rad} \sim 50 \%$ \citep{2015ApJ...815..102F} to account for all the energy lost in radiation, the luminosity of the jet after the breakout $L_{iso,f}$ satisfying $L_{iso,f}\times T_{90} (1- \epsilon_{rad})=E_{iso} \leqslant 10^{53}$ erg, gives $L_{iso,f} \leqslant 10^{53}$ erg/s. Hence, the engine luminosity should satisfy $L_{iso,0} \leqslant 2.5\times 10^{52}$ erg/s.

\subsubsection{Breakout time $t_0-t_b$ and the 1.7s delay [in blue]}
Second, as shown in figure \ref{fig:P4a-L=dt}, we present different analytic breakout times, $t_b-t_0$, using the equation (\ref{eq:tb dyn}). As discussed in $\S$ \ref{sec:Comparison with Numerical Simulations}, comparison with simulations shows that our analytic breakout times are very reliable.

The key line in figure \ref{fig:P4a-L=dt} is the solid blue line. 
In the region below this line, the delay between the GW signal and the EM signal (prompt emission) is strictly larger than the measured 1.7s, which excludes this region for GW170817. 
Let's derive the time delay between GW signal and EM signal for an observer at $\theta_v$. 
The jet breaks out at a radius $R_b$ at a time $t_b$, $t_b -t_m$ seconds after the merger. 
After the breakout, it takes a certain timescale $x$ for prompt emission photons to be released. This timescale, $x$, is not well-understood, as is the mechanism of the prompt emission. It has been argued that $x$ can take milliseconds to seconds (see \citealt{2019arXiv190500781Z}; in particular table 1).
At the breakout time, $t_b$, the jet should have gained a distance $R_b \cos\theta_v$ toward the observer with a viewing angle $\theta_v$, while GW signal should have gained a distance $c(t_b-t_m)$. In other words, the jet is already late by a timescale: $(t_b-t_m) - \frac{R_b}{c} \cos\theta_v$ seconds. Adding the timescale $x$ for the prompt EM signal to be released from the jet, the timescale delay between GW and EM signals should be: $t_{delay} = (t_b -t_m) - R_b \cos\theta_v/c + x$. Then, with $x > 0$ the equality becomes inequality as follows: 

\begin{eqnarray}
t_{delay} \geqslant (t_b -t_m) - R_b \cos\theta_v/c ,
\end{eqnarray}
with $R_b \approx (t_b-t_m)v_{ej}$. For a given delay time ($t_{delay}$), the breakout time should satisfy the following equation:

\begin{eqnarray}
t_b - t_m \leqslant \frac{t_{delay}}{1-(v_{ej}/c)\:\cos\theta_v} .
\end{eqnarray}
For the case of GW170817, where the observer is at a viewing angle $\theta_v\approx 20^\circ$\footnote{$\theta_v \sim 20^\circ - 30^\circ$ (\citealt{2017PhRvL.119p1101A}; \citealt{2018arXiv180806617T}). 
Difference in the delay time is not significant but since  $\theta_v = 20^\circ$ gives a more conservative constraint on GW170817 (a larger parameter space for $L_{iso,0}$ and $t_0-t_m$), we adopt $\theta_v = 20^\circ$ for a more solid constraint.}, and $t_{delay}=1.7$ s, we get: 

\begin{eqnarray}
t_b - t_0 \leqslant \frac{1.7\text{s}}{1-(v_{ej}/c)\:\cos 20^\circ} - (t_0-t_m) .
\label{eq:1.7s}
\end{eqnarray}
Using the analytic expression for the breakout time $t_b-t_0$ [equation (\ref{eq:tb dyn})], we can write:
\begin{eqnarray}
A_c= \frac{1-(r_0/r_{m,0})^\frac{4-n}{2}}{(t_b-t_0)(4-n)}\left[\sqrt{r_{m,0}}+\sqrt{r_{m,0}+v_{ej}(t_b-t_0)}\right] ,
\label{eq:A_c from t_b}
\end{eqnarray}
and with the expression of $A_c$ in equation (\ref{eq:A_c dyn}), the engine luminosity can be written as:
\begin{eqnarray}
L_{iso,0}= \frac{A_c^2(3-n)M_{ej}c}{4[1-(r_0/r_{m,0})^{3-n}]},
\label{eq:Liso from A_c}
\end{eqnarray}
where $r_{m,0}$ can be written as a function of $t_0-t_m$ as: $r_{m,0} \approx v_{ej}(t_0-t_m) + r_0$. With all the ejecta parameters already fixed, and using equations (\ref{eq:A_c from t_b}) and (\ref{eq:Liso from A_c}), the condition in equation (\ref{eq:1.7s}) is equivalent to an upper limit on $t_b-t_0$ (or a lower limit on $L_{iso,0}$) which varies as a function of $t_0-t_m$.
This constraint can be plotted analytically as a function of $L_{iso,0}$ and $t_0-t_m$ as shown in figure \ref{fig:P4a-L=dt} (solid blue line). 

Our constraints show that, based on the delay time between GW and EM signals and the afterglow observations, the allowed values for the breakout time (of GW170817's jet) is wide; it can take values as $t_b-t_0\sim 0.01 - 2$ s.
Although long breakout times $\gtrsim 1$s are theoretically allowed, such cases requires a long accretion time ($\gtrsim1$ s), which might be difficult in a typical BNS post-merger scenario.

Finally, as a reference, the dashed blue line shows a jet breakout time ($t_b-t_0$) equal to the median of the prompt emission duration ($T_{90}$) of Swift \textit{s}GRBs: $t_b-t_0=\text{Median}(T_{90})=0.36$ s\footnote{$\text{Median}(T_{90})$ for \textit{s}GRBs was calculated using Swift data, available at: \url{https://swift.gsfc.nasa.gov/}.}. 
This line is a reference to the typical timescale of the prompt emission, and shows the corresponding engine parameters if the breakout time is comparable to this timescale \citep{2013ApJ...764..179B}. 
The isotropic equivalent luminosity of the jet on the on-axis after the breakout ($\theta_v < \theta_f$) around this dashed line is $L_{iso,f} = 4 L_{iso,0} \sim 10^{51}$ erg/s, which is comparable to the observed range of isotropic luminosity for \textit{s}GRBs (see figure \ref{fig:P4a-Lgrbv2}).

\subsubsection{Jet head breakout velocity $v_b$ [in grey]}
\begin{figure*}
 \includegraphics[width=0.99\linewidth]{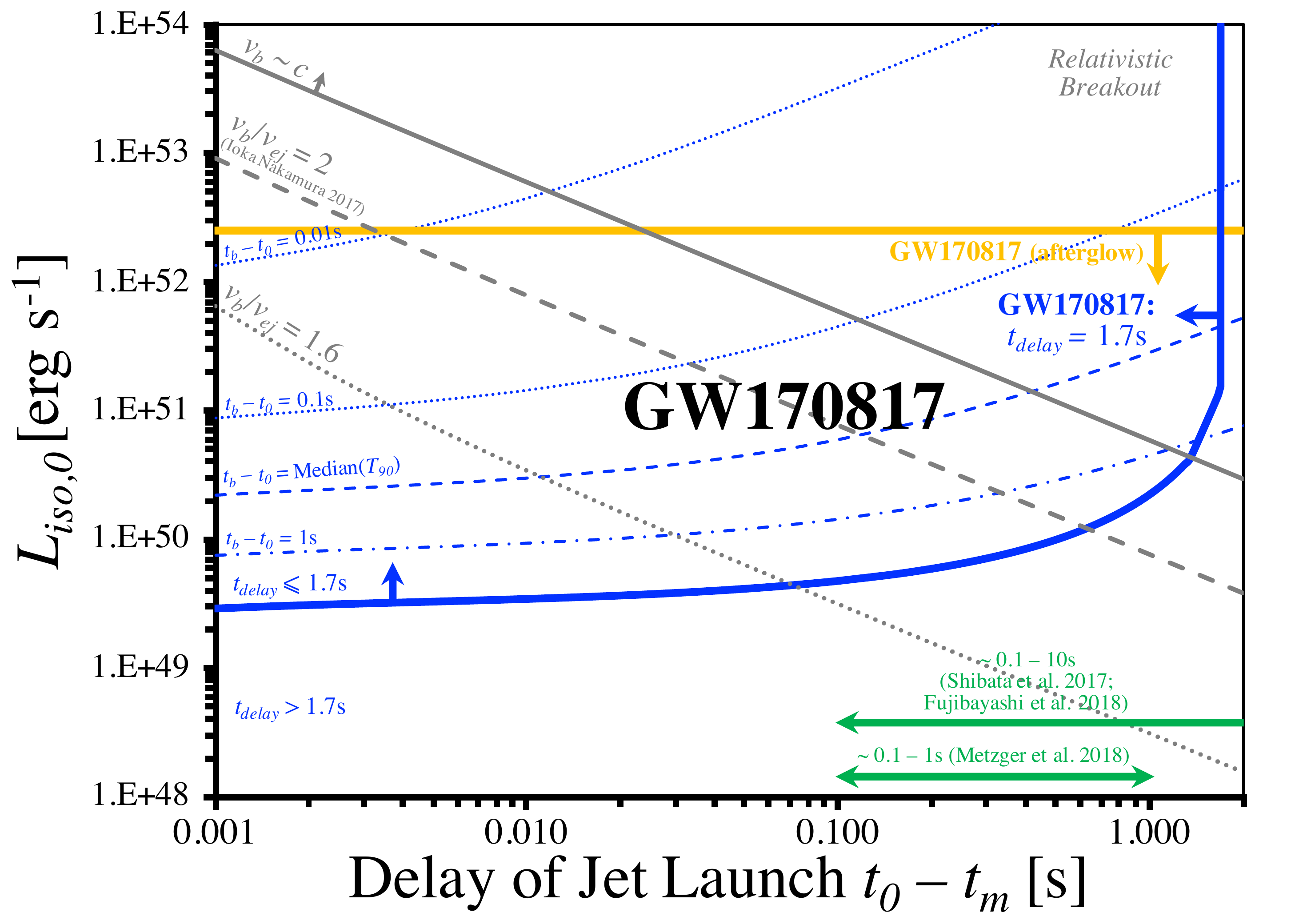}
 \caption{The allowed parameter space for GW170817's central engine in terms of its isotropic equivalent engine luminosity ($L_{iso,0}$) and the delay between the merger and its activation time ($t_0-t_m$). The blue solid line is an analytic constraint based on the 1.7s delay time between the gravitational wave and the electromagnetic signal [using equation (\ref{eq:1.7s})]. Other blue lines (dotted, dashed, and dotted dashed) show analytic breakout times [using equation (\ref{eq:tb dyn})]. Grey lines show analytic breakout velocities of the jet head [using equation (\ref{eq:v_b dyn})]. In green constraints on $t_0-t_m$ from previous studies are shown (see $\S$ \ref{sec:Results}, and for more details refer to: \citealt{2017PhRvD..96l3012S}; \citealt{2018ApJ...860...64F}; and \citealt{2018ApJ...856..101M}). The line in orange is based on late time radio observations ($1 \sigma$ upper limit, see $\S$ \ref{sec:orange}; also refer to: \citealt{2018Natur.561..355M}; \citealt{2018arXiv180806617T}; and \citealt{2019Sci...363..968G}). The following parameters of the ejecta have been used: $n=2$, $v_{ej} = 0.2\sqrt{3}c$, and $M_{ej} =0.002 M_\odot$ (refer to $\S$ \ref{sec:Density profile of the ejecta}; $\S$ \ref{sec:Velocity profile of the ejecta} and $\S$ \ref{sec:Angular dependance}, respectively). } 
 \label{fig:P4a-L=dt}
\end{figure*}

Figure \ref{fig:P4a-L=dt} also shows lines for a constant breakout velocity of the jet head $v_b/v_{ej}$ (grey lines). Previously \citet{2018PTEP.2018d3E02I} presented an estimation of this velocity as $v_b/v_{ej} \sim 2$. Using our analytic modeling [in particular equation \ref{eq:v_b dyn}], we get $A_c$ as a function of the jet head breakout velocity as follows:
\begin{eqnarray*}
A_c= \frac{1}{\sqrt{r_{m,0}}}\left[v_b - v_{ej}\left(1+\frac{1-(r_0/r_{m,0})^\frac{4-n}{2}}{4-n}\right) \right] .
\end{eqnarray*}
Equation (\ref{eq:Liso from A_c}) can be used to find the engine luminosity $L_{iso,0}$ as a function of $t_0-t_m$ for different breakout velocities. For the case of GW170817 (i.e. the ejecta parameters assumed for GW170817), we find that the breakout velocity can be constrained as:
\begin{eqnarray}
1.5 < v_b/v_{ej} < 5/\sqrt{3} ,
\end{eqnarray}
with $v_b/v_{ej}\approx 2$ as a central value for GW170817's parameter space.

Note that there is a parameter space where we predict a relativistic breakout $v_b \sim c$. Such a relativistic breakout may produce a bright emission. However, the luminosity depends on the size of the fast tail of the ejecta, and hence it is not clear whether this region should be excluded for GW170817, or not. Also, note that our analytical modeling is limited for a non-relativistic jet head.

\section{Discussion: The nature of \textit{s}GRB 170817A}
\label{sec:GRB 170817A}

\begin{figure}
 \includegraphics[width=0.99\linewidth]{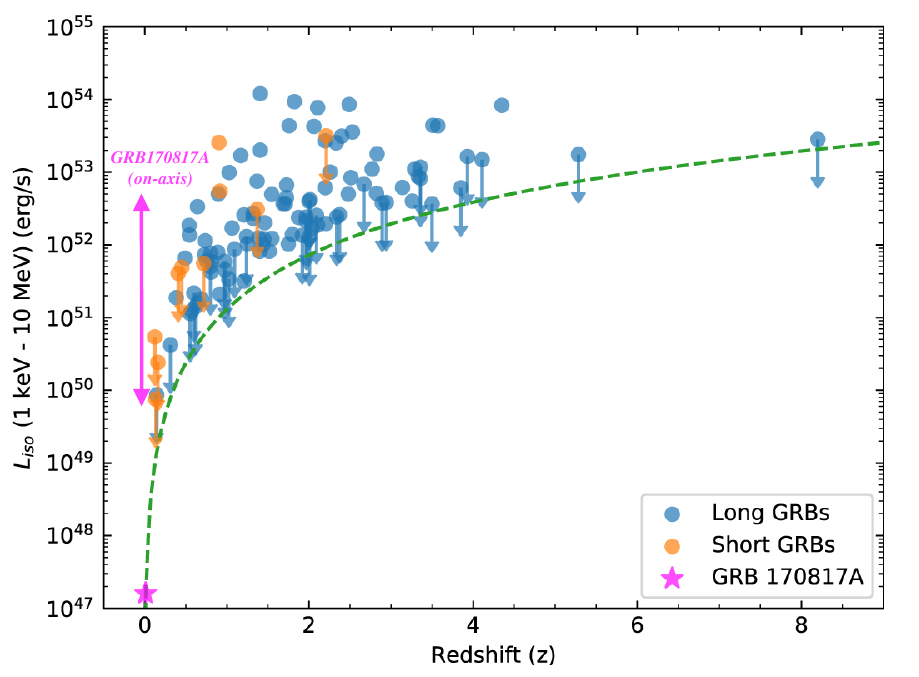}
 \caption{Isotropic equivalent luminosity for \textit{short} and \textit{long} GRBs, and \textit{s}GRB 170817A for a comparison (credit: \citealt{2017ApJ...848L..13A}). The constraint on the luminosity of the prompt emission of \textit{s}GRB 170817A (if observed with a line of sight near the jet on-axis) is shown in magenta. The luminosity of the prompt emission of \textit{s}GRB 170817A is calculated from the jet luminosity, assuming a radiative efficiency parameter $\epsilon_{rad}\sim 50\%$. }
 \label{fig:P4a-Lgrbv2}
\end{figure}

\textit{s}GRB 170817A is a unique \textit{s}GRB, in particular with its extremely low isotropic luminosity $L_{iso}$ (as it can be seen in figure \ref{fig:P4a-Lgrbv2}; see \citealt{2017ApJ...848L..13A}). This faintness has been interpreted as due to one of two main scenarios: i) \textit{s}GRB 170817A is apparently different from typical \textit{s}GRBs due to the observer's large viewing angle, a so-called off-axis model (e.g., \citealt{2018PTEP.2018d3E02I}; \citealt{2019MNRAS.487.4884I}); ii) \textit{s}GRB 170817A is intrinsically different and a unique \textit{s}GRB. 

Based on the inferred engine isotropic luminosity in the on-axis $\sim 3\times10^{49} - 2.5\times10^{52}$ erg/s, we can estimate the observed luminosity of \textit{s}GRB 170817A for an observer with a line of sight near the on-axis. A rough estimation, using a radiative efficiency $\epsilon_{rad} \sim 50 \%$\footnote{Taking a lower radiative efficiency, even $\epsilon_{rad}\sim 10\%$ would not change the conclusion.} \citep{2015ApJ...815..102F}, and assuming a top-hat jet with the jet opening angle $\theta_f\approx \theta_0/2$ (see $\S$ \ref{sec:A roughly constant cross section}), one can deduce the isotropic equivalent luminosity for the prompt emission in the range $\sim 6\times10^{49} - 5\times10^{52}$ erg s$^{-1}$, which is typical for a \textit{s}GRB as it can be seen in figure \ref{fig:P4a-Lgrbv2}.

Hence, based on the close similarity -- in luminosity -- 
on-axis observers would most likely observe a typical \textit{s}GRB; although the possibility that \textit{s}GRB 170817A is produced by a different mechanism than ordinary \textit{s}GRBs cannot be entirely excluded.
Hence, GW170817/\textit{s}GRB 170817A and its energetics support the NS merger scenario for typical \textit{s}GRBs.

\section{Conclusion}
\label{sec:Conclusion}

We investigated jet propagation in BNS merger ejecta. 
Our study combines hydrodynamical simulations of jet propagation and analytic modeling. 
We constructed the post-merger ejecta based on high resolution numerical relativity simulations' results. Then, we analytically solved jet propagation, from the vicinity of the central engine $r_0$, to the edge of the expanding BNS merger ejecta. 
Our analytic model gives the jet head position as a function of time, which allows us to derive crucial quantities such as the breakout time and the breakout velocity (see $\S$ \ref{subsubsec:Case II. Expanding medium} or Appendix \ref{app subsec:Dynamic ambient medium: BNS merger ejecta}).

Although analytic modeling of the jet head motion in an expanding medium has been the subject of several recent studies (\citealt{2018ApJ...866....3D}; \citealt{2018ApJ...866L..16M}; \citealt{2019ApJ...876..139G}; etc.), the analytical model presented here is more extensive and gives much more reliable results. This is because the expansion of the ejecta is properly taken into account, and none of the key parameters have been overlooked (in particular, the delay time of the jet launch $t_0-t_m$). 
As a result, we identify a new term contributing to the breakout time for an expanding medium, which does not exist for a static medium [see equations (\ref{eq:tb dyn sim n<3}) and (\ref{eq:tb dyn sim n>3})].

We carried out hydrodynamical simulations of jet propagation for a wide variety of BNS merger ejecta and a wide variety of engine models. 
We showed that our analytic solutions for jet head motion, breakout time, etc., do agree with numerical simulations. 
We discussed the limitations of our analytical model; in particular, for steep density profiles $n>3$, or high ejecta velocities $v_{ej}\gtrsim 0.4c$, our analytical model reaches its limit and is less reliable.
We also showed the parameter space where the jet head becomes relativistic, and for which our non-relativistic analytic model cannot be applied.

Then, we applied our analytic model to GW170817/\textit{s}GRB 170817A. 
We considered the following key facts:
\begin{enumerate}
    \item GW observation of GW170817 (e.g., $\sim 2.7 M_\odot$), and the merger ejecta properties as revealed by numerical relativity simulations. 
    \item Late time afterglow observations constraining the jet's energy as $E_{iso}\approx3\times10^{51}-10^{53}$ erg \citep{2018Natur.561..355M}, and jet's opening angle as $\theta_f\approx3.4^\circ-9.0^\circ$ \citep{2018arXiv180806617T}.
    \item Observations showing a $\sim1.7$ s delay between the GW merger signal and the prompt EM signal \citep{2017ApJ...848L..13A}.
\end{enumerate}
With the above facts, we reached the following conclusions for GW170817/\textit{s}GRB 170817A, in particular concerning the central engine:
\begin{enumerate}
    \item The time delay between the merger time and the jet launch time $t_0-t_m$ was constrained as $t_0-t_m \lesssim 1.3$s.\label{i}
    \item The central engine's isotropic equivalent luminosity was constrained as $L_{iso,0}\approx 3\times10^{49} - 2.5\times10^{52}$ erg s$^{-1}$. Higher power contradicts afterglow observations, while lower power results in a delay larger than 1.7s.\label{ii}
    \item The jet head breakout velocity was constrained as $0.52 < v_b/c < 1$. This result is in agreement with \citet{2018PTEP.2018d3E02I}.
\end{enumerate}

This new constraint is more robust but in an overall agreement with previous studies (in agreement with \citealt{2018ApJ...856..101M}; partially in agreement with theoretical estimations based on numerical relativity simulations: \citealt{2017PhRvD..96l3012S} and \citealt{2018ApJ...860...64F}). 
The relatively tighter constraint found in \citet{2019ApJ...876..139G} ($t_0-t_m = 0.98_{-0.26}^{+0.31}$ s) seems questionable; in particular due to the fact that the analytic modeling of the jet propagation in \citet{2019ApJ...876..139G} did not properly account for the expansion of the ejecta, the fact that the assumed density of the ejecta did not take into consideration the angular dependence of the density, and also the fact that no comparison with numerical simulation was available.   

One implication of our constraint on the central engine of GW170817 is on the nature of \textit{s}GRB 170817A, and on the origin of its extreme faintness. We argue that based on the engine power, the prompt emission's luminosity for an on-axis observer would be in the range $6\times10^{49}-5\times10^{52}$ erg s$^{-1}$ (assuming $\epsilon_{rad} \approx50\%$). 
First, this suggests that \textit{s}GRB 170817A is most likely a \textit{s}GRB by nature, and that its faintness is due to nothing intrinsic; rather it is most likely due to a large viewing angle effect (e.g., \citealt{2018PTEP.2018d3E02I}; \citealt{2019MNRAS.487.4884I}). 
Second, based on the similarity in luminosity with other \textit{s}GRBs, this would imply that the many other \textit{s}GRBs are most likely similar events to \textit{s}GRB 170817A; that is, BNS merger events at larger distances but viewed with much smaller viewing angles. Hence, the BNS merger model is a promising model for \textit{s}GRBs.

These conclusions and constraints are based on the first and the only GW/EM detection for a \textit{s}GRB so far; with more GW/EM observations expected in the near future, the method presented here has the potential to improve the quality of the constraints. 
Also, the analytical modeling presented here can be applied to investigate other closely related topics (e.g. the cocoon, the macronova, and the EM counterparts; Hamidani et al. 2019 in preparation).

As a final note, it should be pointed that our work overlooks some complex physics: magnetic field, neutrinos, GR effects, etc. There are many other numerical limitations as we use 2D hydrodynamical simulations, rather than the ideal 3D simulations. Also, due to limitations in the computational resources, simulations ideally with inner boundaries $\sim10^6$ cm, were not possible. These are interesting future perspectives. 

\section*{Acknowledgements}
\addcontentsline{toc}{section}{Acknowledgements}

    We wish to record our sense of gratitude to Amir Levinson, Hiroki Nagakura, Kazuya Takahashi, Koutarou Kyutoku, Masaomi Tanaka, Masaru Shibata, Ore Gottlieb, Sho Fujibayashi, Takashi Hosokawa, Tomoki Wada, Toshikazu Shigeyama, Takahiro Tanaka, Tsvi Piran and Yudai Suwa, for their very fruitful comments. 
    We thank the participants and the organizers of the workshops with the identification number YITP-T-19-04, YITP-W-18-11 and YITP-T-18-06, for their generous support and helpful comments. 
    
    Numerical computations were achieved thanks to the following: Cray XC50 of the Center for Computational Astrophysics at the National Astronomical Observatory of Japan, Cray XC40 at the Yukawa Institute Computer Facility, Oakforest-PACKS at Information Technology Center of the University of Tokyo, K computer on AICS (project number hp19016), and post-K computer project (Priority issue No. 9) of MEXT Japan.
    
    This work is partly supported by JSPS KAKENHI nos. 18H01213, 18H01215, 17H06357, 17H06362, 17H06131 (KI).




\bibliographystyle{mnras}
\bibliography{04a-mnras} 




\appendix

\section{Setup for the analytic model}
\label{app:A}
Here we present the analytical model for the motion of a jet head, powered by the central engine, and expanding in a certain medium (or in certain ejecta). The shock jump conditions can be written as (\citealt{1989ApJ...345L..21B}; \citealt{1997ApJ...479..151M}; \citealt{2003MNRAS.345..575M}; \citealt{2013ApJ...777..162M}; \citealt{2018PTEP.2018d3E02I}):
\begin{eqnarray}
h_j \rho_j c^2 (\Gamma\beta)_{jh}^2 + P_j = h_a \rho_a c^2 (\Gamma\beta)_{ha}^2 + P_a ,
\end{eqnarray}
where $h$, $\rho$, and $P$ are enthalpy, density, and pressure of each fluid element, all measured in the fluid's rest frame. The subscripts $j$, $h$, and $a$ refer to the three domains: the jet, jet head, and ambient medium, respectively. Both $P_a$ and $P_j$ can be neglected. Hence, we can write the jet head velocity as:
\begin{eqnarray}
\beta_h  =  \frac{\beta_j - \beta_a}{1 + \tilde{L}^{-1/2}} +  \beta_a ,
\label{eq:general case}
\end{eqnarray}
where $\tilde{L}$ is the ratio of energy density between the jet and the ejecta:
\begin{eqnarray}
\tilde{L}  =  \frac{h_j \rho_j \Gamma_j^2}{h_a \rho_a \Gamma_a^2} \simeq \frac{L_j}{\Sigma_j \rho_a c^3} ,
\label{eq:L expression Ap}
\end{eqnarray}
where $\Sigma_j$ is the jet head cross section $\Sigma_j=\pi\theta_j^2 r_h^2(t)$, and $\theta_j$ is the jet head opening angle.

We take the following set of assumptions (for more depth refer to $\S$ \ref{sec:model assumptions}):
\begin{enumerate}
    \item The velocity of the ejecta $\beta_a$ is neglected in comparison to the highly relativistic jet outflow $\beta_j\simeq 1$. Hence, in equation (\ref{eq:general case}), we can take $\beta_j -\beta_a \simeq 1$.
    \item We consider the case of a non-relativistic jet head $\tilde{L}\ll (1-\beta_a)^2$. Hence, we can simplify equation (\ref{eq:general case}) to the following form:
\begin{eqnarray}
\beta_h \simeq \tilde{L}^{1/2} + \beta_a .
\label{eq:general case2}
\end{eqnarray}
    \item The ejecta expands in a homologous manner: $v_a(r,t)=(r/r_m(t))v_{ej}$, with $r_m(t)$ as the outer radius of the ejecta.
    \item The mass of the ejecta is defined based on the ambient medium's density through which the jet head propagates: $M_{ej}=\int_{r_0}^{r_{m}(t)}4\pi r^2 \rho_a(r,t)dr$.
    \item At the jet launch time $t_0$, the density of the ambient medium is assumed to follow a power-law with an index $n$: $\rho_a(r) = \rho_0 (r_{0}/r)^n$; where $\rho_0 = (M_{ej}/4\pi r_0^n)(3-n)/(r_{m,0}^{3-n} - r_0^{3-n})$. Note that, for the expanding medium case, the ejecta's expansion results in the density being time dependent as: $\rho_a(r,t) =\rho_0 (r_{0}/r)^n(r_{m,0}/r_m(t))^{3-n}$.
\end{enumerate}

Finally, by assuming that the jet launch takes place at the vicinity of the central engine, $r_0$ is constrained as: $r_0\approx10^6-10^7$ cm. Also, for a small opening angle $\theta_0$, we can write the isotropic equivalent luminosity at the base of the jet as $L_{iso,0}\simeq4L_j/\theta_0^2$.

\section{Analytic modeling using a constant jet opening angle}
\label{app:B}
Here, two additional approximations are considered:
\begin{enumerate}
    \item The jet head opening angle $\theta_j$ is approximated as constant over time [see $\S$ \ref{sec:A roughly constant cross section}; figure \ref{fig:P4a-rh2}; and equation (\ref{eq:f_j})].
    \item We approximate the jet opening angle as $\theta_j=\theta_0/f_j$ where $f_j$ is a constant and $\theta_0$ is the jet initial opening angle. For the case of expanding ejecta, based on results within our parameter space, we get $f_j\approx 5$ (see $\S$ \ref{sec:f_j}).
\end{enumerate}
\subsection{Static ambient medium}
\label{app subsec:Static ambient medium}
For a static medium case, $\beta_a \simeq 0$, equation (\ref{eq:general case2}) can be simplified to the following:
\begin{eqnarray}
\beta_h  \simeq \tilde{L}^{1/2} .
\end{eqnarray}
Hence, the jet head velocity is determined by the following equation:
\begin{eqnarray}
\frac{dr_h(t)}{dt}  =  c\:\tilde{L}^{1/2} .
\end{eqnarray}
With the assumption of a power-law density profile, and the expression of $\tilde{L}$ in equation (\ref{eq:L expression Ap}), the result is the following equation of motion (a differential equation of order one):
\begin{eqnarray}
\frac{dr_h(t)}{dt}  =  A\:r_h(t)^\frac{n-2}{2} ,
\label{eq:diff static case}
\end{eqnarray}
where $A$ is a constant $\propto \tilde{L}^{1/2}$ and can be written as:
\begin{equation*}
A=\sqrt{ \left(\frac{r_m^{3-n}-r_0^{3-n}}{3-n}\right)\left(\frac{4\:L_j}{\theta_j^2\:M_{ej}\:c}\right)    } .
\end{equation*}
With the approximation of constant jet opening angle $\theta_j$ over time [see $\S$ \ref{sec:A roughly constant cross section}; figure \ref{fig:P4a-rh2}; and equation (\ref{eq:f_j})], $A$ is also constant over time. Hence, with the boundary condition $r_h(t=t_0)=r_0$, the solution of equation (\ref{eq:diff static case}) is:
\begin{eqnarray}
r_h(t)  =  \left[\left(\frac{4-n}{2}\right)A\:(t-t_0) +  r_0^\frac{4-n}{2}\right]^{\frac{2}{4-n}} .
\end{eqnarray}
Returning to the differential equation (\ref{eq:diff static case}), we obtain the jet head velocity as:
\begin{eqnarray}
v_h(t)  =  A\:\left[ \left(\frac{4-n}{2}\right)\:A\:(t-t_0)    + r_0^\frac{4-n}{2}            \right]^\frac{n-2}{4-n} .
\end{eqnarray}

With $t_0$ as the time when the jet is launched, and $t=t_b$ as the time when the jet head reaches the outer radius $r_h(t_b)=r_m$, the breakout time $t_b-t_0$ and the breakout velocity $v_b$ can be expressed as:

\begin{align}
\label{eq:tb stat}
t_b - t_0 =& \frac{2}{A\:(4-n)}\left[r_m^\frac{4-n}{2} - r_0^\frac{4-n}{2} \right], \\
v_b =& A\:r_m^\frac{n-2}{2}.
\end{align}

In the static medium case (i.e. collapsar case), the collimation factor takes values roughly as $f_j\sim$ a few $\times 10$, depending on the parameters of the engine and the medium.

For a non-relativistic jet head ($\tilde{L}\ll 1$), analytic results needs to be corrected using the calibration factor $N_s$ \citep{2018MNRAS.477.2128H}.
Hence, $A$ in the above equations should be replaced by the following $A_c$:
\begin{equation*}
A\to A_c = N_s\sqrt{ \left(\frac{r_m^{3-n}-r_0^{3-n}}{3-n}\right)\left(\frac{4\:L_j}{\theta_j^2\:M_{ej}\:c}\right)    } ,
\end{equation*}
where $N_s\approx 2/5$ [see equation (\ref{eq:N_s})].

\subsection{Dynamic ambient medium: BNS merger ejecta}
\label{app subsec:Dynamic ambient medium: BNS merger ejecta}
Here we consider the case where the ejecta velocity is non-negligible to the jet head velocity. The start of the ejecta expansion (i.e. the merger time) is assumed at $t=t_m$, and the jet is assumed to launch later at $t=t_0$. Hence, the velocity and density profiles of ejecta can be written as:
\begin{eqnarray}
    v_a(r)  =& v_{ej}\left(\frac{r}{r_{m,0}}\right) ,\\
    \rho_a(r) =& \rho_0 \left(\frac{r_{0}}{r}\right)^n ,
\end{eqnarray}
where $r_0$ is the jet head location at $t=t_0$, $r_{m,0}$ is the ejecta's outer radius at $t=t_0$, $\rho_0$ is the density at the inner boundary $r_0$ at $t=t_0$, and $n$ is the density profile's power-law index. With the ejecta constantly expanding, both the velocity and the density profile are dependent on the time $t$. The ejecta's maximum radius can be written as a function of time as $r_m(t) = v_{ej}\: (t-t_0) + r_{m,0}$. Hence, $v_a(r,t)$ and $\rho_a(r,t)$ can be written as:

\begin{eqnarray}
    v_a(r,t)  =& v_{ej}\left(\frac{r}{r_{m}(t)}\right) , \\
    \rho_a(r,t) =& \rho_0 \left(\frac{r_{0}}{r}\right)^n\left(\frac{r_{m,0}}{r_m(t)}\right)^{3-n} .
\end{eqnarray}
From equation (\ref{eq:general case2}), the jet head dynamics can be described by the following equation:

\begin{eqnarray}
    \frac{dr_h(t)}{dt}  = v_{ej}\left[\frac{r_h(t)}{r_m(t)}\right] + c\:{\tilde{L}}^{1/2} .
\label{eq:diff eq case 2}
\end{eqnarray}

By substituting $\rho_a(r,t)$ and $\Sigma_j=\pi \theta_j^2 r_h^2(t)$ in equation (\ref{eq:L expression Ap}), $c\:{\tilde{L}}^{1/2}$ can be written as:
\begin{eqnarray}
    c\:{\tilde{L}}^{1/2} = {A}\:{r_h(t)}^\frac{n-2}{2}\:{r_m(t)}^\frac{3-n}{2} ,
    \label{eq:B13}
\end{eqnarray}
where $A$ is a constant $\propto \tilde{L}^{1/2}$ and can be written as:
\begin{equation*}
A=\sqrt{ \left(\frac{r_{m,0}^{3-n}-r_0^{3-n}}{(3-n)\:r_{m,0}^{3-n}}\right)\left(\frac{4\:L_j}{\theta_j^2\:M_{ej}\:c}\right)    } .
\end{equation*}
With equations (\ref{eq:diff eq case 2}) and (\ref{eq:B13}), the equation of motion of the jet head can be found as:
\begin{eqnarray}
    \frac{dr_h(t)}{dt}  + \left( -\frac{v_{ej}}{r_{m}(t)}\right)r_h(t) = {A}\:{r_m(t)}^\frac{3-n}{2}{r_h(t)}^\frac{n-2}{2} .
    \label{eq:dif dynamic}
\end{eqnarray}

With the boundary condition $r_h(t_0)=r_0$, we can solve equation (\ref{eq:dif dynamic}) as:
\begin{equation}
\begin{split}
    r_h(t) = &\left[ (4-n){\frac{A}{v_{ej}}}[\sqrt{r_m(t)} - \sqrt{r_{m,0}}] + \left(\frac{r_0}{r_{m,0}}\right)^{\frac{4-n}{2}} \right]^\frac{2}{4-n} \\ 
    & \times r_m(t) .
\end{split} 
\end{equation}
Using back equation (\ref{eq:dif dynamic}), the jet head velocity $v_h(t)$ can be found as:
\begin{eqnarray}
v_h(t) = A \left[\frac{r_h(t)}{r_m(t)}\right]^\frac{n-2}{2}\sqrt{r_m(t)} + v_{ej}\:\left[ \frac{r_h(t)}{r_m(t)}\right] .
\label{eq:v_h app}
\end{eqnarray}

The breakout time is by definition the time at which we have $r_h(t=t_b)=r_m(t=t_b)=r_b$. 
Hence, the breakout time $t_b-t_0$ is:
\begin{equation}
    t_b - t_0 = \left[ \frac{ r_{m,0}^\frac{4-n}{2} - r_{0}^\frac{4-n}{2}  }{ r_{m,0}^\frac{4-n}{2}} \frac{\sqrt{v_{ej}}}{(4-n)A} + \sqrt{\frac{r_{m,0}}{v_{ej}}}         \right]^2 - \frac{r_{m,0}}{v_{ej}} .
    \label{eq:tb app}
\end{equation}
Also, the breakout velocity $v_h(t=t_b)=v_b$ can be expressed as:
\begin{eqnarray}
v_b = A \:\sqrt{R_b} + v_{ej} ,
\label{eq:vb app}
\end{eqnarray}
with $R_b = v_{ej}\:(t_b-t_0) + r_{m,0}$ as the breakout radius. 

For a non-relativistic jet head propagating in an expanding medium ($\tilde{L}\ll (1-\beta_a)^2$), we find that this analytic solution needs to be corrected in the same way as the analytic solution of a non-relativistic jet head in a static medium (see $\S$ \ref{sec:calibration}). Hence, with $A\propto\tilde{L}^{1/2}$, the above $A$ should be corrected to $A_c$ as follows:
\begin{equation}
A \to A_c=N_s\sqrt{ \left(\frac{r_{m,0}^{3-n}-r_0^{3-n}}{(3-n)\:r_{m,0}^{3-n}}\right)\left(\frac{4\:L_j}{\theta_j^2\:M_{ej}\:c}\right)    } ,
\label{eq:A_c app}
\end{equation}
where we take $N_s\approx 2/5$ [see equation (\ref{eq:N_s})], and $A$ in the above equations should be replaced by $A_c$.

\subsubsection{A more rigorous solution (for $n=2$)}
\label{app:exact solution n=2}
For $n=2$, $\beta_a$ in equation (\ref{eq:general case}) is not approximated and is calculated rigorously. For $\tilde{L}\ll (1-\beta_a)^2$, and with $\beta_j\simeq 1$, equation (\ref{eq:general case}) gives:
\begin{eqnarray}
\beta_h=\tilde{L}^{1/2}(1-\beta_a) + \beta_a .
\end{eqnarray}
From equation (\ref{eq:L expression}) we can write $\tilde{L}$ as follows:
\begin{eqnarray*}
c\:\tilde{L}^{1/2}=A\left[\frac{r_h(t)}{r_m(t)}\right]^\frac{n-2}{2}\sqrt{r_m(t)} , \\
\end{eqnarray*}
with:
\begin{eqnarray*}
A=\sqrt{ \left(\frac{r_{m,0}^{3-n}-r_0^{3-n}}{(3-n)\:r_{m,0}^{3-n}}\right)\left(\frac{4\:L_j}{\theta_j^2\:M_{ej}\:c}\right)    } .
\end{eqnarray*}
Replacing, $\beta_h$ and $\beta_a$ gives an equation of motion as follows:
\begin{equation}
\begin{split}
    \frac{dr_h(t)}{dt}  + &\left( -\frac{v_{ej}}{r_m(t)}\right)r_h(t) =\\
    &{A}\:\left[\frac{r_h(t)}{r_m(t)}\right]^\frac{n-2}{2}\sqrt{r_m(t)}\:\left(1-\frac{v_{ej}}{c}\frac{r_h(t)}{r_m(t)}\right) .
    \label{eq:dif dynamic exact}
\end{split}
\end{equation}
Again, we use the approximation of a constant jet opening angle (see $\S$ \ref{sec:A roughly constant cross section}). The solution for this equation of motion is complex (hypergeometric function), unless $n=2$.
For $n=2$, the term with the power $\frac{n-2}{2}$ is equal to  unity. 
With the boundary condition $\frac{r_h(t=t_0)}{r_m(t=t_0)}=\frac{r_0}{r_{m,0}}$, the solution of equation (\ref{eq:dif dynamic exact}) is:
\begin{equation}
    \frac{r_h(t)}{r_m(t)} = \frac{c}{v_{ej}}\left[1-\left(\frac{r_{m,0}\:c-r_0\:v_{ej}}{r_{m,0}\:c}\right)e^{\frac{-2A}{c}[\sqrt{r_m(t)}-\sqrt{r_{m,0}}]   } \right] .
\end{equation}
Using back the equation of motion (\ref{eq:dif dynamic exact}), the velocity can be written as:
\begin{eqnarray}
v_h(t) = v_{ej}\left[\frac{r_h(t)}{r_m(t)}\right]\left[1-\frac{A\sqrt{r_m(t)}}{c}\right] + A\sqrt{r_m(t)} .
\end{eqnarray}
This equation is very similar to the previous case where $\beta_j-\beta_a\approx1$ [see equation (\ref{eq:v_h app})]. The main difference is the additional term $[1-\frac{A\sqrt{r_m(t)}}{c}]$.

The breakout quantities can be found by setting $r_h(t_b)=r_m(t_b)$, which gives the breakout time as:
\begin{equation}
\begin{split}
    t_b - t_0 = \frac{1}{v_{ej}}\left[ \left(\frac{c}{2A}\ln\left(\frac{c r_{m,0} - v_{ej} r_0}{c r_{m,0}- v_{ej} r_{m,0}}\right) + \sqrt{r_{m,0}}\right)^2 - r_{m,0}                    \right] ,
\end{split}
\label{eq:tb exact}
\end{equation}
and the breakout velocity as: 
\begin{eqnarray}
v_b = A \:\sqrt{R_b}\left[1- \frac{v_{ej}}{c}\right] + v_{ej}  ,
\end{eqnarray}
with $R_b = v_{ej}\:(t_b-t_0) + r_{m,0}$ as the breakout radius. 

Note that, here too, $A$ in the above equations needs to be corrected to $A_c=N_s A$ [see equation (\ref{eq:A_c app}); with $N_s\approx2/5$].

\section{Analytic modeling of the jet propagation with the cocoon collimation}
\label{sec:app C}
Here we do not approximate the jet's opening angle as constant over time, as taken in previous sections. We consider a cocoon collimating a non-relativistic jet ($\tilde{L}<\theta_0^{-4/3}$) as in \citet{2011ApJ...740..100B} for the collapsar case; although here we consider the case of an expanding medium.

The unshocked jet's height $\hat{z}$ can be written as a function of the jet luminosity $L_j$ and the cocoon's pressure $P_c$ (\citealt{2011ApJ...740..100B}):
\begin{eqnarray}
  \hat{z}(t) = \sqrt{\frac{L_j}{\pi c P_c}} .
\end{eqnarray}
The jet is uncollimated below $\hat{z}/2$, and collimated beyond  $\hat{z}/2$ \citep{2011ApJ...740..100B}. Hence, the jet cross-section can be found for the two modes as follows (see \citealt{2011ApJ...740..100B}):
\begin{equation}
  \Sigma_j(t) =
    \begin{cases}
      \pi r_h^2(t)\theta_0^2 & \text{if $r_h(t)<\hat{z}(t)/2$ (uncollimated jet)} ,\\
      \pi r_h^2(t)\theta_j^2(t) & \text{if $r_h(t)>\hat{z}(t)/2$ (collimated jet)} ,
    \end{cases}       
\end{equation}
where $r_h(t)$ is the jet head radius, $\theta_0$ is the jet initial opening angle, and $\theta_j(t)$ is the opening angle of the collimated jet. The system of equations for the cocoon collimating a jet gives (see \citealt{2011ApJ...740..100B}):
\begin{align}
\label{eq:c1}
r_c \approx& \chi c\langle{\beta_\perp}\rangle (t-t_0) ,\\
\label{eq:c2}
    \langle{\beta_{\perp}}\rangle =&
    \sqrt{\frac{P_{c}}{\langle{\rho}_{a}(t)\rangle c^{2}}} , \\
    \label{eq:c3}
    P_{{c}} =& \:\:\:\:\:\: \frac{E_{in}}{3\:V_{{c}}} \:\:\:\:\:\:= \eta \frac{L_j\left(1-\langle{\beta_h}\rangle \right) \:(t-t_0)}{2 \pi r_c^{2} r_{{h}}(t)} , \\
    \label{eq:c4}
    \Sigma_j(t) =& \pi r_h^2(t) \theta_j^2(t) = \frac{L_j \theta_0^2}{4 c P_c} ,
\end{align}
where $t_0$ is time at which the jet is launched at $r_0$, $t-t_0$ is the time since the jet launch, $r_c$ is the lateral width of the cocoon, $\langle{\beta_{\perp}}\rangle$ is the lateral average velocity of the expanding cocoon, $\left<\beta_h\right>=\frac{r_h(t)-r_0}{c(t-t_0)}$ is the average jet head velocity, and $V_c$ is the cocoon volume (we approximate the cocoon shape to an ellipsoidal: $V_c=(2\pi/3)r_c^2 r_h(t)$).
$\eta$ is a parameter that theoretically takes values between 0 and 1 (\citealt{2011ApJ...740..100B}; \citealt{2019arXiv190707599S}). 
$\eta$ accounts for ignored effects such as the adiabatic expansion. 
$r_c$ is defined as $r_c \approx c\langle{\beta_\perp}\rangle (t-t_0)$ in the collapsar case, where the medium is static (\citealt{2011ApJ...740..100B}; \citealt{2018MNRAS.477.2128H}).
However, $r_c$ is more complex in the expanding medium case, therefore $\chi$ is an additional parameter ($>1$ in the expanding medium case) which accounts for enhancement of $r_c$ due to the comoving expansion of the medium. 
Taking into account these two parameters ($\eta$ and $\chi$) and deducing them based on results from numerical simulations, we make this analytic model much more reliable in comparison to previous models (\citealt{2018MNRAS.475.2659M}; \citealt{2018ApJ...866L..16M}; \citealt{2019ApJ...876..139G}; etc.). 
Finally, $\left<\rho_a(t)\right>$ is the average density in the medium surrounding the cocoon. We can write $\left<\rho_a(t)\right>$ as:
\begin{eqnarray*}
\langle{\rho_a(t)}\rangle = \int_{r_0}^{r_h(t)} \rho_a(r,t)dV/({4\pi(r_h^3(t)-r_0^3)}/{3}) ,
\end{eqnarray*}
where $r_m(t) = r_{m,0} + v_{ej}\:(t-t_0)$ is the ejecta outer most radius. Assuming $r_h(t)\gg r_0$ gives: 
\begin{eqnarray*}
\langle{\rho_a(t)}\rangle \approx \frac{3 M_{ej}}{4\pi r_m^3(t)} \left[ \frac{r_m(t)}{r_h(t)}\right]^n .
\end{eqnarray*}
Hence, using the equations (\ref{eq:c1}), (\ref{eq:c2}) and (\ref{eq:c3}), $r_c$ and $P_c$ can be found as:
\begin{align}
    \label{eq:r_c app}
    r_c \approx& \chi \sqrt{ \frac{P_c 4 \pi r_m^3(t) (t-t_0)^2  }{3 M_{ej}}\left[\frac{r_h(t)}{r_m(t)}\right]^n} , \\
    \label{eq:P_c app}
    P_c =& \sqrt{\frac{\eta}{\chi^2}}\sqrt{ \frac{3 L_j(1-\frac{r_h(t)-r_0}{c(t-t_0)})M_{ej}}{8 \pi^2\: (t-t_0)\: r_m^3(t)\: r_h(t)} \left[\frac{r_m(t)}{r_h(t)}\right]^n} .
\end{align}
Finally, replacing equation (\ref{eq:P_c app}) in equation (\ref{eq:c4}) gives:
\begin{equation}
    \left[\frac{\theta_j(t)}{\theta_0}\right]^2 = 
    \left[\frac{\chi^2}{\eta} \frac{L_j}{6 M_{ej}c^2(1-\frac{r_h(t)-r_0}{c(t-t_0)})}        \right]^{\frac{1}{2}} r_h^{\frac{n-3}{2}}(t) r_m^{\frac{3-n}{2}}(t)(t-t_0)^{\frac{1}{2}} .
    \label{eq:theta_j^2}
\end{equation}
For simplicity, let's take $(1-\frac{r_h(t)-r_0}{c(t-t_0)})\approx 1$\footnote{In reality, $(1 - \frac{r_h(t)-r_0}{c(t-t_0)})$ gives a factor $<1$ for a non-relativistic jet head. It varies over time and depends on the parameters of the engine and the ejecta. In our calculations, this factor is implicitly absorbed into $\eta$.}. Hence, with $r_m(t)=r_{m,0}+v_{ej} (t-t_0)$ we obtain:
\begin{equation}
    \frac{\theta_j(t)}{\theta_0} =
    \left[\frac{\chi^2}{\eta} \frac{L_j}{6 M_{ej}c^2 v_{ej}}        \right]^{\frac{1}{4}}\: \left[\frac{r_h(t)}{r_m(t)} \right]^{\frac{n-3}{4}} \: [r_m(t)-r_{m,0}]^{\frac{1}{4}} .
    \label{eq:theta_j/theta_0 app}
\end{equation}
Notice the weak dependence of the opening angle of the jet on time. 
This is consistent with our numerical simulations (see figure \ref{fig:P4a-rh2}) and justifies our approximation of a roughly constant opening angle for the jet (in $\S$ \ref{sec:A roughly constant cross section}).

On the other hand, the equation of the motion is the following differential equation [see equation (\ref{eq:dif dynamic})]:
\begin{eqnarray}
    \frac{dr_h(t)}{dt}  + \left( -\frac{v_{ej}}{r_{m}(t)}\right)r_h(t) = {A(t)}\:{r_m(t)}^\frac{3-n}{2}{r_h(t)}^\frac{n-2}{2} ,
    \label{eq:dif dynamic coll}
\end{eqnarray}
where the only difference from Appendix \ref{app:B} is that $A$ is time-dependent here: $A \equiv A(t)$. It can be written as:
\begin{equation*}
A(t)=\sqrt{ \left(\frac{r_{m,0}^{3-n}-r_0^{3-n}}{(3-n)\:r_{m,0}^{3-n}}\right)\left(\frac{4\:L_j}{\theta_0^2 M_{ej}\:c}\right)    } \times\left[\frac{\theta_0}{\theta_j(t)}\right] .
\end{equation*}
Let's write $A(t)$ as $A(t)=A_0[\theta_0/\theta_j(t)]$, with:
\begin{equation*}
A_0=\sqrt{ \left(\frac{r_{m,0}^{3-n}-r_0^{3-n}}{(3-n)\:r_{m,0}^{3-n}}\right)\left(\frac{4\:L_j}{\theta_0^2 M_{ej}\:c}\right)    } .
\end{equation*}

For an always collimated jet, we solve the equation of motion [equation (\ref{eq:dif dynamic coll})]. 
The solution is given by the following integration:
\begin{equation}
    \left[\frac{r_h(t)}{r_m(t)}\right]^{\frac{5-n}{4}} =  A_1 \frac{5-n}{4} \int {r_m^{-3/4}(t)[1-r_{m,0}/r_m(t)]^{-1/4} dt},
\end{equation}
where $A_1$ is a constant:
\begin{equation}
A_1=\left[\frac{\eta}{\chi^2}\right]^\frac{1}{4}\left[ \left(\frac{r_{m,0}^{3-n}-r_0^{3-n}}{(3-n)\:r_{m,0}^{3-n}}\right)^2\left(\frac{96\:L_j\:v_{ej}}{\theta_0^4 M_{ej}}\right)        \right]^{1/4} .
\end{equation}

The system of equations (\ref{eq:c1}), (\ref{eq:c2}), (\ref{eq:c3}) and (\ref{eq:c4}); which is equivalent to the above integration; can be solved numerically (\citealt{2011ApJ...740..100B}; \citealt{2018MNRAS.477.2128H}; \citealt{2018ApJ...866L..16M}). Still, some approximations can allow us to solve it analytically. 
One possible approximation is, as in the case of short delay between the merger time, $t_m$, and the jet launch time, $t_0$, in comparison to the breakout time $t_b-t_m$: $t_0-t_m\ll t_b - t_m$; we get $r_m(t) \gg r_{m,0}$ giving:
\begin{equation}
    \int {r_m^{-3/4}(t)[1-r_{m,0}/r_m(t)]^{-1/4} dt}  \simeq \int{ r_m^{-3/4}(t)dt} + C .
    \label{eq:r_m>>r_m,0 app}
\end{equation}
With the boundary condition at $t=t_0$, $r_m(t_0)=r_{m,0}$ and $r_h(t_0)=r_0$, we can get:
\begin{equation}
    \left[\frac{r_h(t)}{r_m(t)}\right]^\frac{5-n}{4} =  \left[\frac{A_1 (5-n)}{v_{ej}}\right]\: \left[ r_m^{\frac{1}{4}}(t) - r_{m,0}^\frac{1}{4}\right] +\left[\frac{r_0}{r_{m,0}}\right]^\frac{5-n}{4} .
    \label{eq:r_h/r_m step 1 app c}
\end{equation}
Jet head velocity can be deduced by using equation (\ref{eq:r_h/r_m step 1 app c}) as:
\begin{equation}
v_h(t)=v_{ej}\left[ \frac{r_h(t)}{r_m(t)}\right] +A(t)\left[ \frac{r_h(t)}{r_m(t)}\right]^\frac{n-2}{2}[r_m(t)^\frac{1}{4}(r_m(t)-r_{m,0})^\frac{1}{4}] .
\end{equation}

Finally, the breakout time and the breakout velocity, when $r_h(t_b)/r_m(t_b)=1$, can be derived as:
\begin{equation}
t_b - t_0 =\frac{1}{v_{ej}}\left[ \frac{v_{ej}}{A_1(5-n)}\left(1-\left(\frac{r_0}{r_{m,0}}\right)^\frac{5-n}{4}\right)+ r_{m,0}^\frac{1}{4}\right]^{4} - \frac{r_{m,0}}{v_{ej}} ,
\end{equation}
and,
\begin{equation}
\:\:\:\:\:\:\: v_b = v_{ej} + A(t_b)R_b^\frac{1}{4}\left[R_b-r_{m,0}\right]^\frac{1}{4},
\label{eq:v_b app c}
\end{equation}
with: 
\begin{equation*}
A(t)=A_1 \left[\frac{r_h(t)}{r_m(t)}\right]^\frac{3-n}{4}[r_m(t)-r_{m,0}]^{-\frac{1}{4}},
\end{equation*}
These breakout quantities are very similar to those found previously by using the approximation of a constant jet opening angle [see Appendix \ref{app:B}, in particular equations (\ref{eq:tb app}) and (\ref{eq:vb app})], although a bit more complex; in particular, equation (\ref{eq:v_b app c}) where $R_b\gg r_{m,0}$ gives $v_b \approx v_{ej} + A(t_b)\sqrt{R_b}$.

As previously explained, in the non-relativistic domain ($\tilde{L}\ll (1-\beta_a)^2$), the analytic solution gives a jet head velocity (i.e., $\tilde{L}^{1/2}$, which is $\propto A(t)$) which is $\sim 2.5 - 3$ times higher than numerical simulations (see $\S$ \ref{sec:calibration}). Therefore, the above $A(t)$ should be corrected to $A_c(t)=N_s A(t)$. Hence, $A_1$ in the above equations is corrected to $A_{1,c}$:
\begin{equation*}
A_1 \to A_{1,c}=N_s \left[\frac{\eta}{\chi^2}\right]^\frac{1}{4}\left[ \left(\frac{r_{m,0}^{3-n}-r_0^{3-n}}{(3-n)\:r_{m,0}^{3-n}}\right)^2\left(\frac{96\:L_j\:v_{ej}}{\theta_0^4 M_{ej}}\right)        \right]^{1/4} .
\end{equation*}
where we take $N_s\approx 2/5$ [see equation (\ref{eq:N_s})].

Finally, $\chi$ and $\eta$ are poorly known parameters. $\chi$ accounts for the expansion of the ejecta and the cocoon. Hence, $\chi\gtrsim1$. $\eta$ here accounts for several effects, in particular the adiabatic expansion of the cocoon, and the approximation of a complete thermalization. Also, note that the component $1-\left<\beta_h\right> < 1$ is absorbed in $\eta$. With $\chi\gtrsim1$ and $\eta\lesssim1$, the ratio $\eta/\chi^2$ should be smaller than unity. 
Comparison with simulations show that $\left[ \eta/\chi^2\right]^{\frac{1}{4}}\approx 1/2$ (Hamidani et al. 2019 in preparation). 

\section{Glossary of main mathematical symbols} 
\label{sec:app D}
\subsection{General variables}
$t$: time.\\
$r$: radius.\\
$t-t_m$: time since the merger.\\
$t-t_0$: time since the jet launch.\\
$t_0-t_m$: time delay between the merger and the jet launch.
$t_b - t_0$: jet breakout time, since the jet launch.\\
$r_h(t)$: jet head radius.\\
$R_b$: breakout radius.\\
$v_b$: jet head's breakout velocity.\\
$\beta_h$ [$=v_h/c$]: jet head velocity in units of the speed of light.\\
$L_{iso,0}$: isotropic equivalent luminosity of the jet at its base (i.e., of the central engine).\\
$\tilde{L}$: the ratio of the energy density between the jet and the ambient medium.\\
$A$ [as well as $A_0$ and $A_1$]: a constant $\propto \tilde{L}$ which depends on the parameters of the ejecta and the engine, and reflects how easily penetrable the ejecta is (for the jet).\\
$A_c$ [and $A_{1,c}$]: the corrected form of $A$, using the calibration coefficient $N_s$.\\

\subsection{Symbols used in the analytic model [Appendix \ref{app:A} and Appendix \ref{app:B}]}
$r_0$: the jet head radius at $t=t_0$.\\
$r_{m,0}$: the ejecta's outer radius at $t=t_0$.\\
$r_m(t)$: radius of the outer ejecta (expanding medium case).\\
$r_m$: the outer radius of the ambient medium (static medium case).\\
$\beta_j$: velocity of the unshocked jet-outflow in units of the speed of light ($\simeq 1$).\\
$\beta_a$ [=$v_a/c$]: velocity of the ambient medium in units of the speed of light.\\
$v_{ej}$: maximum velocity of the ejecta.\\
$\Sigma_j$: cross section of the jet.\\
$\theta_0$: initial jet opening angle.\\
$\theta_j$: opening angle of the collimated jet.\\
$f_j$: the degree of collimation, relative to the initial opening angle.\\
$\theta_f$: opening angle of the jet after the breakout.\\
$\rho_a$: density of the ambient medium.\\
$\rho_0$: density of the ambient medium at $r=r_0$ and $t=t_0$.\\
$L_j$ [$=L_{iso,0}\theta_0^2/4$]: luminosity of the jet (one sided).\\
$M_{ej}$: mass of the ambient medium (or the ejecta).\\
$n$: index of the ejecta's density profile.\\
$N_s$: calibration coefficient for the analytic model, based on numerical simulations.\\

\subsection{Symbols used in the analytic model [Appendix \ref{sec:app C}]}
$\tilde{z}$: the unshocked jet's height.\\
$P_c$: cocoon pressure.\\
$r_c$: maximum lateral width of the cocoon.\\
$\chi$: degree of enhancement in the width of the cocoon due to the expansion of the ejecta (>1 in the case of an expanding medium).\\
$\langle{\beta_\perp}\rangle$: average lateral velocity of the cocoon.\\
$V_c$: volume of the cocoon.\\
$E_{in}$: internal energy deposited by the jet into the cocoon.\\
$\eta$: parameter for how efficiently the energy deposited by the jet into the cocoon is converted into thermal energy (varies between 0 and 1).\\

\subsection{Symbols used in numerical simulations}
$\rho$: density.\\
$P$: pressure.\\
$h$: enthalpy.\\
$\rho_{CSM}$: density of the CSM.\\
$\Gamma_0$: initial Lorentz factor.\\
$r_{in}$: jet injection nozzle.\\
$\theta_{inj}$: opening angle of the injected jet at the injection nozzle.\\
$\gamma$: adiabatic index.\\
$K_{ej}$: scaling factor for the ejecta pressure to density.\\
$\Delta r_{min}$: highest radial resolution.\\
$\Delta \theta_{min}$: highest angular resolution.\\
$\theta_j(r)$: average opening angle of the jet.\\
$\theta_{j,av}(r)$: opening angle of the jet outflow at the radius $r$.\\

\subsection{Other symbols}
$E_{iso}$: total isotropic equivalent energy of the jet.\\
$L_{iso,f}$: isotropic equivalent luminosity of the jet after the breakout.\\
$L_{iso}$: isotropic equivalent luminosity of the jet after radiation has been released.\\
$\epsilon_{rad}$: radiation efficiency for a GRB jet.\\
$\theta_v$: viewing angle, relative to the jet axis (polar region).\\
$x$: timescale for the radiation to be released after the jet breakout.\\
$t_{delay}$: delay timescale between the GW and the EM signal.\\


\bsp	
\label{lastpage}
\end{document}